%% file: journal_paper.tex
\documentclass[preprint,3p]{elsarticle}

\PassOptionsToPackage{hyphens}{url}\usepackage{hyperref}
\usepackage[T1]{fontenc}
\usepackage[usenames,dvipsnames]{xcolor}
\usepackage[breakable, theorems, skins]{tcolorbox}
\tcbset{enhanced}
\usepackage{graphicx}
\usepackage{amssymb}
\usepackage{amsmath}
\usepackage{amsthm}
\usepackage{paralist}
\usepackage{todonotes}
\usepackage[algoruled,vlined,boxed,linesnumbered]{algorithm2e}
\SetKwRepeat{Do}{do}{while}
\usepackage{algpseudocode}
\usepackage{multirow, makecell}
\usepackage{enumitem}
\usepackage[framemethod=default]{mdframed}
\usetikzlibrary{automata,arrows}
\usetikzlibrary{datavisualization}
\usetikzlibrary{datavisualization.formats.functions}
\usetikzlibrary{positioning, calc, backgrounds, arrows}
\usetikzlibrary{shapes.geometric}
\usetikzlibrary{decorations.pathreplacing}
\usetikzlibrary{patterns}

\journal{Information Systems}

\usepackage[figuresright]{rotating}

\makeatletter
\newtheorem*{rep@theorem}{\rep@title}
\newcommand{\newreptheorem}[2]{%
\newenvironment{rep#1}[1]{%
 \def\rep@title{#2 \ref{##1}}%
 \begin{rep@theorem}}%
 {\end{rep@theorem}}}
\makeatother

\newreptheorem{theorem}{Theorem}

\newtheorem{definition}{Definition}[section]
\newtheorem{lemma}{Lemma}[section]
\newreptheorem{lemma}{Lemma}

\makeatletter
\newcommand{\shorteq}{%
    \mkern3mu
  \settowidth{\@tempdima}{--}
  \resizebox{\@tempdima}{\height}{=}%
  \mkern3mu
}
\makeatother
\newcommand{\fsm}{\mathit{FSM}}
\newcommand{\labels}{\Sigma}
\newcommand{\nodes}{\mathit{N}}
\newcommand{\arcs}{\mathit{A}}
\newcommand{\source}{s}
\newcommand{\finStates}{\mathit{R}}
\newcommand{\src}{\mathit{src}}
\newcommand{\lbl}{\lambda}
\newcommand{\tgt}{\mathit{tgt}}
\newcommand{\shift}{\blacktriangleright}
\newcommand{\logL}{\mathit{L}}
\newcommand{\dafsa}{\mathit{D}}

\newcommand{\lnet}{\mathit{PN}}
\newcommand{\places}{\mathit{P}}
\newcommand{\netTransitions}{\mathit{T}}
\newcommand{\netArcs}{\mathit{F}}
\newcommand{\netLabel}{\lambda}

\newcommand{\inTr}[1]{\bullet{#1}}
\newcommand{\outTr}[1]{#1\bullet}
\newcommand{\wNet}{\mathit{WN}}

\newcommand{\sysNet}{\mathit{SN}}
\newcommand{\initMarking}{m_0}
\newcommand{\finalMarkings}{M_R}

\newcommand{\reachGraph}{\mathit{RG}}

\newcommand{\rg}{\reachGraph}
\newcommand{\rgNodes}{\nodes_{\reachGraph}}
\newcommand{\rgArcs}{\arcs_{\reachGraph}}
\newcommand{\rgSource}{\source_{\reachGraph}}
\newcommand{\rgFinStates}{\finStates_{\reachGraph}}

\newcommand{\dfa}{\dafsa}
\newcommand{\dfaNodes}{\nodes_{\dafsa}}
\newcommand{\dfaArcs}{\arcs_{\dafsa}}
\newcommand{\dfaSource}{s_{\dafsa}}
\newcommand{\dfaFinStates}{\finStates_{\dafsa}}

\newcommand{\match}{\mathit{MT}}
\newcommand{\lhide}{\mathit{LH}}
\newcommand{\rhide}{\mathit{RH}}

\newcommand{\syncs}{\mathit{S}}
\newcommand{\sync}{\beta}
\newcommand{\op}{\mathit{op}}
\newcommand{\fnSyncDFA}{a_{\dafsa}}
\newcommand{\fnSyncRG}{a_{\reachGraph}}
\newcommand{\filterSyncDFA}{\alignment|_{\dafsa}}
\newcommand{\filterSyncRG}{\alignment|_{\reachGraph}}
\newcommand{\alignment}{\epsilon}
\newcommand{\properAlignments}{\xi}

\newcommand{\costF}{\mathit{g}}

\newcommand{\start}{\mathit{s}}
\newcommand{\TRtype}{\alpha}
\newcommand{\reps}{\mathit{k}}
\newcommand{\TRs}{\mathit{TR}}

\newcommand{\redLog}{\mathit{RL}}
\newcommand{\redTr}{\mathit{rt}}
\newcommand{\redTrace}{\redTr}
\newcommand{\additionalCostF}{\mathit{p}}
\newcommand\mydots{..}
\newcommand{\seq}[2]{#1\mydots#2}
\newcommand{\val}{\mathit{val}}
\newcommand{\trPositions}{\mathit{TR}_l}
\newcommand{\position}{\mathit{i}}
\newcommand{\secondPos}{\mathit{j}}

\newcommand{\redCostF}{\rho}
\newcommand{\alignpos}{\mathit{pos}_{\alignment}}
\newcommand{\redReps}{k_{red}}

\newcommand{\alignments}{\mathcal{A}}
\newcommand{\reductions}{\mathit{Reductions}}
\newcommand{\trReductions}{\mathit{rTrace}}
\newcommand{\reduce}{\gamma}
\newcommand{\pairs}{\mathit{Pairs}}
\newcommand{\redDafsa}{\fsm_{\dafsa,red}}

\newcommand{\open}{\mathit{o}}
\newcommand{\closed}{\mathit{c}}
\newcommand{\actN}{\mathit{n}_{act}}
\newcommand{\out}{\mathit{a}}
\newcommand{\tieconcat}{\oplus}

\newcommand{\pref}{\mathit{Prefix}}
\newcommand{\suffix}{\mathit{Suffix}}
\newcommand{\dom}{\mathit{dom}}

\newcommand{\tpos}{\mathit{pos}_t}

\newcommand{\extendedAlignments}{\alignments_{ext}}
\newcommand{\extendedAlignment}{\alignment_{ext}}

\newcommand{\firstCopy}{\alignment_{1}}

\newcommand{\secondCopy}{\alignment_{2}}

\newcommand{\mPath}{\mathit{path}}

\begin{document}

\begin{frontmatter}



\title{Efficient Conformance Checking\\using  Approximate Alignment Computation with Tandem Repeats}


\author{Daniel Rei{\ss}ner\corref{cor1}\fnref{Melb}}
\ead{dreissner@student.unimelb.edu.au}
\cortext[cor1]{Corresponding author}
\author{Abel Armas-Cervantes\fnref{Melb}}
\ead{abel.armas@unimelb.edu.au}
\author{Marcello La Rosa\fnref{Melb}}
\ead{marcello.larosa@unimelb.edu.au}

\affiliation[Melb]{organization={University of Melbourne},
            country={Australia}}

\begin{abstract}
Conformance checking encompasses a body of process mining techniques which aim to find and describe the differences between a process model capturing the expected process behavior and a corresponding event log recording the observed behavior. Alignments are an established technique to compute the distance between a trace in the event log and the closest execution trace of a corresponding process model. Given a cost function, an alignment is optimal when it contains the least number of mismatches between a log trace and a model trace. Determining optimal alignments, however, is computationally expensive, especially in light of the growing size and complexity of event logs from practice, which can easily exceed one million events with traces of several hundred activities. A common limitation of existing alignment techniques is the inability to exploit repetitions in the log.  
By exploiting a specific form of sequential pattern in traces, namely \emph{tandem repeats}, we propose a novel approximate technique that uses pre- and post-processing steps to compress the length of a trace and recomputes the alignment cost while guaranteeing that the cost result never under-approximates the optimal cost. 
In an extensive empirical evaluation with 50 real-life model-log pairs and against six state-of-the-art alignment techniques, we show that the proposed compression approach systematically outperforms the baselines by up to an order of magnitude in the presence of traces with repetitions, and that the cost over-approximation, when it occurs, is negligible.
\end{abstract}


\begin{highlights}
  \item The method reduces primitive and maximal tandem repeats in traces of an event log to only two copies reducing the length of the input traces for computing alignments.
  \item The technique then computes alignments with an adjusted cost function to prioritize relating the collapsed tandem repeats to loop structures in the process model.
  \item The technique can extend reduced alignments to proper alignments that fully represent the original traces and form a path through the process model.
  \item The technique further reduces computation time by applying a binary search for original traces that map to the same reduced trace.
  \item The cost over-approximation of the technique is at most less than two repetitions of each reduced tandem repeat. The evaluation shows that cost over-approximations rarely occur and are much lower than the worst case cost.
\end{highlights}


\end{frontmatter}


\input{tex/intro}
\input{tex/background}
\input{tex/Preliminaries}
\input{tex/Approach}
\input{tex/evaluation}
\input{tex/conclusion}


\newpage
\bibliographystyle{elsarticle-num}
\bibliography{lit}
\appendix
\input{tex/Appendix.tex}






\end{document}

%% file: tex/intro.tex
\vspace{-1\baselineskip}
\section{Introduction}

Business processes are the backbone of modern organizations \cite{fBPM2}. Processes such as order-to-cash or procure-to-pay are executed hundreds of times in sales and retail organizations, as claims handling or loan origination processes are core to the success of financial companies such as insurances and banks. These processes are supported by one or more enterprise systems. For example, sales processes are typically supported by an enterprise resource planning system while claims handling processes are supported by claims management systems. These systems maintain detailed execution traces of the business processes they support, in the form of so-called \emph{event logs}. An event log contains sequences of events (called \emph{traces})  that are performed within a given process case, e.g.\ for a given order or claim application. In turn, each event refers to the execution of a particular process activity, such as ``Check purchase order'' or ``Assess credit risk'' and is timestamped based on the activity completion time.  

Process mining techniques aim to extract insights from event logs, in order to assist organizations in their operational excellence or digital transformation programs \cite{ProcessMiningBook,fBPM2}. Conformance checking is a specific family of process mining techniques whose goal is to identify and describe the differences between an event log and a corresponding process model \cite{ProcessMiningBook,fBPM2}. While the event log captures the observed business process behavior (the as-is process), the process model used as input by conformance checking techniques captures the expected behavior of the process (the to-be or prescriptive process). 

A common approach for conformance checking is by computing \emph{alignments} between traces in the log and execution traces that may be generated by the process model. In this context, a trace alignment is a data structure that describes the differences between a log trace and a possible model trace. These differences are captured as a sequence of moves, including \emph{synchronous moves} (moving forward both in the log trace and in the model trace) and \emph{asynchronous moves} (moving forward either only in the log trace or only in the model trace). A desirable feature of a conformance checking technique is that it should identify a minimal (yet complete) set of behavioral differences. In trace alignments this means that the computed alignments should have a minimal length, or more generally, a \emph{minimal cost}. Existing techniques that fulfill these properties, e.g.\ \cite{AdriansyahDA11,BVD-Alignment}, exhibit scalability limitations in the context of large and complex real-life logs. In fact, the sheer number of events in a log and the length of each trace are rapidly increasing, as logging mechanisms of modern enterprise systems become more fine-grained, as well as business processes become more complex to comply with more stringent regulations. For example, the BPI Challenge 2018 \cite{BPIC18}, one of the logs used in the evaluation of this paper, features around 2.5M events with traces up to 3K events in length. State-of-the-art alignment techniques are worst-case exponential in time on the length of the log traces and the size of the process model. This lack of scalability hampers the use of such techniques in interactive settings as well as in use cases where it is necessary to apply conformance checking repeatedly, for example in the context of automated process discovery \cite{PD-Discovery-BM}, where several candidate models need to be compared by their conformance.

This paper starts from the observation that activities are often repeated within the same process case, e.g.\ the amendment of a purchase request may be performed several times in the context of a procure-to-pay process, due to errors in the request. In the case of the BPI Challenge 2018 log, nearly half of the 3,000 activities in the longest trace are in fact repeated. When computing alignments, the events corresponding to these repeated activities are aligned with the same loop structure in the process model.
Based on this, we use \emph{tandem repeats} \cite{FindingTRs,BoseTRinProcessMining}, a type of sequential pattern, to encode repeated sequences of events in the log and collapse them into two occurrences per sequence, effectively reducing the number of times the repeated sequence needs to be aligned with a loop structure in the process model. When computing alignments, we use an adjusted cost function to prioritize repeatable sequences in the process model for the collapsed sequences of events in the log. Later, we extend these collapsed sequences to form alignments that fully represent the events in the original traces, and form a valid path in the process model. 
The alignments computed by the proposed technique may over-approximate the minimal cost of an alignment by up to a little bit less than two repetitions of a reduced tandem repeat in a worst case. However, we show in our experiments that, in most cases, the proposed technique computes alignments with minimal cost and that the observed cost over-approximations are very low.
Collapsing the traces of an event log also allows us to reduce the number of unique traces
, since two different traces may differ only in the number of occurrences of a given sequence of events, so when reduced, these two traces may map to the same unique trace. We can then use a binary search to find if the reduction of different sequence of repeated events leads to the same reduced alignment for a unique trace. If that is the case, we can reuse these alignments for several original traces, leading to a further improvement in computational performance. 

We apply this technique to a specific class of Petri nets, namely concurrency-free workflow nets, \emph{a.k.a.} state machine workflow nets, with unique activity labels. In order to relax the concurrency-free requirement, the presented technique can be integrated into a decomposition framework for computing alignments -- as described in~\cite{s-comps} --, thus the technique can be applied to sound and free-choice workflow nets . Free choice workflow nets have been shown to be a versatile class of Petri nets, as they map directly to BPMN models with core elements, which are widely used in practice.


We implemented our technique as an open-source tool as part of the Apromore software ecosystem. Using this tool, we extensively evaluated the efficiency and accuracy of the technique via a battery of 50 real-life model-log pairs, against five baseline approaches for alignment computation.

The rest of this paper is organized as follows. Section \ref{sec:related} discusses existing conformance checking and string compression techniques. Section \ref{sec:preliminary} introduces preliminary definitions and notations related to Automata-based conformance checking, alignments and tandem repeats.  Section \ref{sec:approach} presents the new alignment technique based on tandem repeats. Throughout Sections~\ref{sec:preliminary} and~\ref{sec:approach} we also demonstrate each step of the proposed technique with a running example. Section \ref{sec:evaluation} discusses the results of the empirical evaluation. Finally, Section \ref{sec:conclusion} summarizes the contributions and discusses avenues for future work.

%% file: tex/background.tex
\section{Related Work}\label{sec:related}

This section reviews techniques for computing alignments used in conformance checking, and techniques for string compression.

\subsection{Alignment approaches}

Conformance checking in process mining aims at relating the behavior captured in a process model with the behavior observed in an event log. In this article, we specifically focus on identifying behavior observed in the log that is disallowed by the model (a.k.a.\ unfitting behavior).  
\emph{Trace alignments} are central artifacts in process mining for measuring unfitting behavior. 
While many other conformance checking techniques exist in process mining such as token-based replay~\cite{TokenBasedReplay,TokenExtension}, fast approximation techniques that only compute a fitness score~\cite{leemans18}, techniques producing natural language statements~\cite{GarciaL17}, etc., 
this article solely focuses on techniques that produce trace alignments.
Hereafter, we introduce the concept of trace alignments and then review existing techniques for their computation. 

\paragraph{Trace alignment} Trace alignments, first introduced in~\cite{AdriansyahDA11,Adriansyah14},  relate each trace in the event log to its closest execution in the process model in terms of its Levenshtein distance. 
In this context, an alignment of two traces is a sequence of \emph{moves} (or \emph{edit operations}) that describes how two cursors can move from the start of the two traces to their end. In a nutshell, there are two types of edit operations. A \emph{match} operation indicates that the next event is the same in both traces. Hence, both cursors can move forward synchronously by one position along both traces. 
Meanwhile, a \emph{hide} operation (deletion of an element in one of the traces) indicates that the next events are different in each of the two traces. Alternatively, one of the cursors has reached the end of its trace while the other has not reached its end yet. Hence, one cursor advances along its traces by one position while the other cursor does not move. An alignment is optimal if it contains a minimal number of hide operations. 
Given that a process model can contain a possibly infinite set of traces due to loop structures, several traces can have alignments with minimal distance for the same trace of the event log. In this article, we focus on techniques that compute only one minimal distance alignment for each trace of the event log. 

In the following, we first review approaches that compute (exact) trace alignments with minimal distance. These techniques have a worst-case exponential time complexity in terms of the length of the input trace and the size of the process model. Hence, several approaches have been proposed to compute trace alignments with \emph{approximate} cost or that employ \emph{divide-and-conquer} strategies. These latter two categories of approaches are reviewed afterwards.

\paragraph{Exact techniques}
The idea of computing alignments between a process model (captured as a Petri net) and an event log was developed in Adriansyah et al.~\cite{AdriansyahDA11,Adriansyah14}. This proposal maps each trace in the log into a (perfectly sequential) Petri net, a trace net. It then constructs a synchronous Petri net as a product out of the model and the trace net. Finally, it applies an $A^*$ algorithm to find the shortest path through the synchronous net which represents an optimal alignment.
Van Dongen~\cite{BVD-Alignment} extends Adriansyah et al.'s approach by strengthening the underlying heuristic function. This latter approach was shown to outperform~\cite{AdriansyahDA11,Adriansyah14} 
on an artificial dataset and a handful of real-life event log-model pairs. In the evaluation reported later in this article, we use both~\cite{AdriansyahDA11,Adriansyah14} and \cite{BVD-Alignment} as baselines.

In previous work~\cite{ReissnerCDRA17}, we translate both the event log and the process model into automata structures. Then, we use an $A^*$ algorithm to compute minimal distance trace alignments by bi-simulating each trace of the event log on both automata structures allowing for asynchronous moves, i.e.\ the edit operations. This approach utilizes the structure of the automata to define prefix and suffix memoization tables in order to avoid re-computing partial alignments for common trace prefixes and suffixes. This approach was shown to outperform~\cite{AdriansyahDA11, Adriansyah14} on some real-life and synthetic datasets. We also retain this technique as a baseline approach for the evaluation section.

\newpage
De Leoni et al.~\cite{leoni2017} translate the trace alignment problem into an automated planning problem. Their argument is that a standard automated planner provides a more standardized implementation and more configuration possibilities from the route planning domain. Depending on the planner implementation, this approach can either provide optimal or approximate solutions. In their evaluation, De Leoni et al. showed that their approach can outperform~\cite{AdriansyahDA11} only on very large process models. Subsequently, \cite{BVD-Alignment} empirically showed that trace alignment techniques based on the $A^*$ heuristics outperform the technique of De Leoni et al. in the general case. Thus, in this article we do not retain the technique by De Leoni et al. as a baseline.


In the above approaches, each trace is aligned to the process model separately. An alternative approach, explored in~\cite{GarciaL17}, is to align the entire log against the process model, rather than aligning each trace separately. 
Concretely, this approach transforms both the event log and the process model into \emph{event structures}~\cite{NielsenPW1981}. It then computes a synchronized product of these two event structures. Based on this product, a set of natural-language statements are derived, which characterize all behavioral relations between activities captured in the model but not observed in the log and vice-versa. The emphasis of this \emph{behavioral alignment} is on the completeness and interpretability of the set of difference statements that it produces. As shown in~\cite{GarciaL17}, the technique is less scalable than that of~\cite{AdriansyahDA11, Adriansyah14}, in part due to the complexity of the algorithms used to derive an event structure from a process model. Since the emphasis of the present article is on scalability, we do not retain~\cite{GarciaL17} as a baseline. On the other hand, the technique proposed in this article outputs the same data structure as~\cite{GarciaL17} -- a so-called Partially Synchronised Product (PSP). Hence, the output of this paper's technique can be used to derive the same natural-language difference statements produced by~\cite{GarciaL17}.

\paragraph{Approximate techniques} In order to cope with the inherent complexity of the problem of computing optimal alignments, several authors have proposed algorithms to compute approximate alignments. We review the main approaches next.
\emph{Sequential alignments} \cite{Boudewijn17} is an approximate approach that implements an incremental method  
to calculate alignments. It uses an ILP program to find the cheapest edit operations for a fixed number of steps (e.g.\ three events) taking into account an estimate of the cost of the remaining alignment. The approach then recursively extends the found solution with another fixed number of steps until a full alignment is computed. 
This approach is not considered as a baseline in our empirical evaluation since its core idea was used in the extended marking equation alignment approach in~\cite{BVD-Alignment}, which derives optimal alignments and exhibits better performance than sequential alignments. In other words, \cite{BVD-Alignment} subsumes \cite{Boudewijn17}.

\emph{Alignments of Large Instances} or ALI~\cite{ALI_TOSEM} is another approximate alignment approach. It starts by finding an initial candidate alignment using a replay technique and then improves it using a local search algorithm until no further improvements can be found. This approach has shown promising results in terms of scalability when compared to the exact trace alignment approaches presented in~\cite{AdriansyahDA11, Adriansyah14, BVD-Alignment}. Accordingly, we use this technique as a further baseline in our evaluation.

Another approach 
is the \emph{evolutionary approximate alignments}~\cite{EvolutionaryAllOptimal}. It encodes the computation of alignments as a genetic algorithm. Tailored crossover and mutation operators are applied to an initial population of model mismatches to derive a set of alignments for each trace. In this article, we focus on computing one alignment per trace (not all possible alignments) and thus we do not consider approaches like \cite{EvolutionaryAllOptimal} as baselines in our evaluation. Approaches that compute all-optimal alignments are slower than those that compute a single optimal alignment per trace, and hence the comparison would be unfair.

Bauer et al.~\cite{TraceSampling} propose to use \emph{trace sampling} to approximately measure the amount of unfitting behavior between an event log and a process model. The authors use a measure of trace similarity in order to identify subsets of traces that may be left out without substantially affecting the resulting measure of unfitting behavior. This approach does not address the problem of computing trace alignments, but rather the problem of (approximately) measuring the level of fitness between an event log and a process model. Trace sampling is orthogonal to the contribution of this article. Thus, it can be applied as a pre-processing step prior to any other trace alignment approach, including the techniques presented in this article.


Last, Burattin et al.~\cite{OnlineConformance} propose an approximate approach to find alignments in an online setting. In this approach, the input is an event stream instead of an event log. Since traces are not complete in such an online setting, the approach computes alignments of trace prefixes and estimates the remaining cost of a possible suffix. The emphasis is on the quality of the alignments made for trace prefixes, and as such, this approach is not directly comparable to trace alignment techniques that take full traces as input. 



\paragraph{Divide-and-conquer approaches} In divide-and-conquer approaches, the process model is split into smaller parts to speed up the computation of alignments by reducing the size of the search space. Van der Aalst~\cite{van2013decomposing} propose a set of criteria for a valid decomposition of a process model in the context of conformance checking. One decomposition approach that fulfills these criteria is the \emph{single-entry-single-exit} (SESE) process model decomposition approach. Mu{\~n}oz-Gama et al.~\cite{Munoz-GamaCA14} present a trace alignment approach based on SESE decomposition. The idea is to compute an alignment between each SESE fragment of a process model and the event log projected onto this model fragment. An advantage of this approach is that it can pinpoint mismatches to specific fragments of the process model. However, it does not compute alignments at the level of the full traces of the log -- it only produces partial alignments between a given trace and each SESE fragment. A similar approach is presented in~\cite{wang2017aligning}. 

Verbeek et al.~\cite{verbeek2016merging}, an extension of the approach in~\cite{Munoz-GamaCA14}, merges the partial trace alignments produced for each SESE fragment in order to obtain a full alignment of a trace.~\cite{verbeek2016merging} sometimes computes optimal alignments, but other times it produces so-called \emph{pseudo-alignments} -- i.e., alignments that correspond to a trace in the log but not necessarily to a trace in the process model. 
In this article, the goal is to produce actual alignments (not pseudo-alignments). Therefore, we do not retain \cite{verbeek2016merging} as a baseline.

Lee et al.~\cite{lee2018recomposing} present another approach for recomposing partial alignments, which does not produce pseudo-alignments. Specifically, if the merging algorithm in~\cite{verbeek2016merging} cannot recompose two partial alignments into an optimal combined alignment, the algorithm merges the corresponding model fragments and re-computes a partial alignment for the merged fragment. 
This procedure is repeated until the re-composition yields an optimal alignment. In the worst case, this may require computing an alignment between the trace and the entire process model. 
However, the recomposition technique from~\cite{lee2018recomposing} has only been tested with a manual decomposition of the model.
The goal of the present article is to compute alignments between a log and a process model automatically, and hence we do not retain~\cite{lee2018recomposing} as a baseline.
In the article, it was suggested to use the SESE decomposition proposed in~\cite{Munoz-GamaCA14} as an automatic decomposition technique, but this was has not been tested so far.

Last, in~\cite{s-comps} we extend the Automata-based approach from~\cite{ReissnerCDRA17} to a decomposition-recomposition approach based on \emph{S-Components}. This approach first decomposes the input process model into concurrency-free sub-models, i.e. its S-Components, based on the place invariants of the process model. Then, Automata-based approach is applied to each pair of S-Component and a sub-log derived by trace projection. Next, the approach recomposes the decomposed alignments of each S-Component to form proper alignments for full traces of the event log. This approach was shown to outperform both~\cite{AdriansyahDA11, Adriansyah14} and \cite{BVD-Alignment} on process models with concurrency on a set of real-life log-model pairs. Therefore, we keep the S-Components approach as a baseline in the evaluation section.

\subsection{String compression techniques}

The technique presented in this article relies on a particular type of sequential pattern mining, namely \emph{tandem repeats}, and specifically on string compression techniques to detect and collapse repeated sequences of events, so as to reduce the length of the traces in a log. In the rest of this section we review  different types of string compression techniques, and the types of repetitive patterns that can be compressed. Last, we review the usage of string compression techniques in process mining.

\newpage
\paragraph{Lossless vs. loss-prone compression approaches}
String or text compression techniques can be broken down into two families of approaches: \emph{dictionary based-approaches} and \emph{statistical approaches}~\cite{textCompressionSurvey}. Dictionary-based approaches aim at achieving a \emph{lossless} representation of the input data by recording all reduced versions of repetitive patterns in the data source in a dictionary to be able to later reconstruct an exact representation of the original data. Statistical approaches on the other hand rely on statistical models such as alphabet or probability distributions to compress the input data. This type of approaches can achieve a better degree of compression, but can only reconstruct an approximate representation of the original data, i.e.\ the compression is \emph{prone to the loss of information}. As such, this latter approach is more applicable when a small loss of information is tolerable and the amount of information is very large, e.g.\ in the field of image compression. In the context of trace alignment, this is not suitable because any loss of information may result in further (spurious) differences between the log and the model. Hence, our focus is on lossless compression techniques.

Dictionary based approaches can be further sub-divided into approaches that implicitly represent compressed sequences as tuples, i.e.\ approaches based on ``Lempel Ziv 77''~\cite{LZ77}, or explicitly record compressions in a dictionary, i.e.\ approaches based on ``Lempel Ziv 78''~\cite{LZ78}. The former approaches aims at identifying the longest match of repetitive patterns in a sliding window and compresses the repeated pattern with a tuple consisting of an offset to the previous repetition, the length of the pattern and the first symbol after the pattern. Several approaches improved on this idea by reducing the information of the tuple or by improving the identification of repetitions~\cite{LZ77Family}.

Approaches based on Lempel Ziv 78 build up a dictionary for compressed repetitive sequences, where each compressed pattern is linked to an index of its extended form in the dictionary. When the input source is large, the dictionary grows extensively and leads to a lower compression rate. Several approaches tackle this issue by using different dictionary types, for example with static length~\cite{LZ78static} or over a rolling window~\cite{LZ78dynamic}. These two latter approaches are faster in decoding repetitive patterns than in compressing them. This is because they need to constantly identify repetitive patterns during the compression. However, they can decode the patterns faster since all necessary information is stored either in the tuples or in the dictionary.

In this article, we will define a reduction of an event log based on the ideas of \cite{LZ77} representing repetitive patterns as tuples. We will use the additional information of the tuples about the reduced pattern, i.e.\ reduced number of repetitions, to guide the computation of compressed alignments that can then be decoded into proper alignments for the process model.

\vspace{-0.75\baselineskip}
\paragraph{Types of repetitive patterns}
A repetitive pattern~\cite{RepeatTypes} is a sequence of symbols that is repeated in a given period or context, i.e.\ in this work the context is a given trace of an event log. The repeating sequence (a.k.a.\ the \emph{repeat type}) can either be \emph{full}, i.e.\ all symbols of the repeat type are repeated, or \emph{partial}, i.e.\ only some symbols of the repeat type are repeated. A repeat type can either be repeated \emph{consecutively}, i.e.\ all repetitions follow one another, or \emph{gapped}, i.e.\ the repetitions of the repeat type occur at different positions within a given trace. In addition, a repetitive pattern can also be \emph{approximate} with a Levenshtein distance of $k$ symbols, i.e.\ the pattern allows up to $k$ symbols disrupting the repeating sequence. In the context of conformance checking, we aim at relating a repetitive pattern to the process model to find if it can be repeated in a loop structure of the process model. For that purpose, we will rely on a restrictive class of repeat patterns, i.e.\ full repeat types with consecutive repetitions (a.k.a. \emph{tandem repeats}). If the pattern were partial, approximate or gapped, the execution context of the process model would be lost and hence no cyclic behavior of the process model could be extended when decoding the repetitive patterns later on.

\vspace{-0.75\baselineskip}
\paragraph{Repetitive patterns in process mining}
In the context of process mining, repetitive patterns have been used to define trace abstractions in~\cite{BoseTRinProcessMining}. These trace abstractions haven then been used to discover hierarchical process models. In this context, tandem repeats have been considered for discovering loop structures and full repeat types with gapped repetitions 
have been used for discovering subprocesses. The properties of tandem repeats have been further explored in~\cite{BoseTRinProcessMining}. Specifically, a tandem repeat is called maximal, if the repeat type cannot be extended by another consecutive repetition before its starting position or after the last repetition of the tandem repeat. Conversely, a tandem repeat is called primitive, if the repeat type in itself is not another tandem repeat. These categorizations were made to discourage redundant discoveries of similar repeat types. In this article, we will hence use maximal and primitive tandem repeats to reduce the event log for the purpose of speeding up the computation of trace alignments.

%% file: tex/Preliminaries.tex
\newcommand{\rightshift}{\blacktriangleright}
\newcommand{\stateMachineWNet}{\mathit{WN_{SM}}}
\newcommand{\syncCostStandard}{\mathit{cost}}

\newcommand{\head}[1]{\mathit{head}~#1}
\newcommand{\tail}[1]{\mathit{tail}~#1}
\newcommand{\Closed}{\Theta}
\newcommand{\insertF}{\mathit{INSERT}}
\newcommand{\push}{\mathit{PUSH}}
\newcommand{\isinvisible}{\mathit{ISINVISBLE}}
\newcommand{\f}{\mathit{f}}
\newcommand{\n}{\mathit{m}}
\newcommand{\inc}{\mathit{inc}}
\newcommand{\replaceTau}{\mathit{replaceTau}}
\newcommand{\sbt}{\,\begin{picture}(-1,1)(-1,-3)\circle{3}\end{picture}\ \ }
\newcommand{\removeArcs}{\mathit{removeArcs}}
\newcommand{\arc}{\mathit{arc}}
\newcommand{\remarcs}{\mathit{remArcs}}
\newcommand{\remnodes}{\Xi}
\newcommand{\INV}{\Psi}
\newcommand{\inv}{\mathit{a}}
\newcommand{\replaceTauBack}{\mathit{replaceTauBackwards}}
\newcommand{\removeUnconnectedMarkings}{\mathit{removeUnconnectedNodes}}
\newpage
\section{Preliminaries}\label{sec:preliminary}

The approach presented in this paper builds on the concepts introduced in this section: finite state machines, Petri nets, event logs, alignments and tandem repeats. 
Observe that throughout this and the next section, a running example is developed to explain relevant concepts.

\vspace{-0.5\baselineskip}
\subsection{Finite State Machines (FSM).}

Our technique represents the behavior of a process model and the event log as \emph{Finite State Machines (FSM)}. A FSM captures the execution of a process by means of edges representing activity occurrences and nodes representing execution states. Activities and their occurrences are identified by their name. Hereinafter, $\labels$ denotes the set of labels (activity names) in both the model and the log. 

\begin{definition}[Finite state machine]\label{FSM}
Given a set of labels $\labels$, a \emph{FSM} is a tuple $(\nodes,\arcs,\source,\finStates)$, where $\nodes$ is a set of nodes, $\arcs \subseteq \nodes\times\labels\times\nodes$ is a set of arcs, $\source \in \nodes$ is the initial node and $\finStates \subseteq \nodes$ is a set of final nodes.  The sets $\nodes,\arcs$ and $\finStates$ are non-empty and finite.
\end{definition}

An arc  $a=(n_s,l,n_t) \in \arcs$ represents the occurrence of an activity $l\in\labels$ at the (source) node $n_s$ that leads to the (target) node $n_t$.
The functions $\src(a)=n_s$, $\lbl(a)=l$ and $\tgt(a)=n_t$ retrieve the source node, label and target node of $a$, respectively. 
Given an arc $a$ and a node $n$, we define a function $n\shift a$ to traverse arc $a$ and retrieve its target node $n_t$ if node $n$ is the source node of $a$, i.e. $n\shift a = n_t$ if $n=n_s$. If we cannot traverse arc $a$ from node $n$, the function will return $n$, i.e. $n\shift a=n$ if $n\neq n_s$.
The incoming and outgoing arcs for a node $n$ are retrieved as $\shift n = \{a\in\arcs\mid\tgt(a)=n\}$ and $n\shift=\{a\in\arcs\mid\src(a)=n\}$, respectively.

\subsection{Petri net and reachability graph.}

Process models can be represented in various modelling languages, in this work we use Petri nets due to its well-defined execution semantics. This modelling language has two types of nodes, transitions, which in our case represent activities, and places, which represent execution states. The formal definition for Petri nets is presented next.

\begin{definition}[(Labelled) Petri net]
A (labelled) \emph{Petri net}, or simply a \emph{net}, is the tuple $\lnet = (\places, \netTransitions, \netArcs, \netLabel)$, where $\places$ and $\netTransitions$ are disjoint sets of \emph{nodes}, \emph{places} and \emph{transitions}, respectively; $\netArcs \subseteq (\places \times \netTransitions) \cup (\netTransitions \times \places)$ is the flow relation, and $\netLabel : \netTransitions \to \labels\cup\tau$ is a labelling function mapping transitions to labels $\labels \cup \{\tau\}$, where $\tau$ is a special label representing an unobservable action.
\end{definition}

Transitions with label $\tau$ represent silent steps whose execution leaves no footprint but are necessary for capturing certain behavior in the net (e.g., optional execution of activities or loops). In a net, we will often refer to the preset or postset of a node, the preset of a node $y$ is the set $\inTr{y} = \{x \in P \cup T \mid (x,y) \in F\}$ and the postset of $y$ is the set $\outTr{y} = \{z \in P \cup T \mid (y,z) \in F\}$.

The work presented in this paper considers a sub-family of Petri nets: uniquely-labeled free-choice workflow nets~\cite{WFNets,FreeChoice}. It is uniquely labelled in the sense that every label is assigned to at most one transition. Given that these nets are workflow and free choice nets, they have two special places: an initial and a final place and, whenever two transitions $t_1$ and $t_2$ share a common place $s \in \inTr{t_1} \cap \inTr{t_2}$, then all places in the preset are common for both transitions $\inTr{t_1} = \inTr{t_2}$. The formal definitions are given below.

\begin{definition}[Uniquely-Labelled, free-choice, workflow net]\label{def:wNet}
A (labelled) \emph{workflow net} is a triplet $\wNet=(\lnet,i,o)$, where $\lnet = (\places, \netTransitions, \netArcs, \netLabel)$ is a labelled Petri net, $i\in\places$ is the initial and $o\in\places$ is the final place, and the following properties hold:
\begin{compactitem}
    \item $i$ has an empty preset and $o$ has an empty postset, i.e., $\inTr{i}=\outTr{o}=\varnothing$.
    \item If a transition $t^*$ were added from $o$ to $i$, such that $\inTr{i} = \outTr{o} = \{t^*\}$, then the resulting net is strongly connected.
\end{compactitem}
A workflow net $\wNet=(\lnet,i,o)$, where $\lnet = (\places, \netTransitions, \netArcs, \netLabel)$, is \emph{uniquely-labelled} and \emph{free-choice} if the following holds:
\begin{compactitem}
    \item (Uniquely-labelled) For any $t_1,t_2\in \netTransitions,\netLabel(t_1)=\netLabel(t_2)\neq\tau \Rightarrow t_1=t_2$.
    \item (Free-choice) For any $t_1,t_2 \in \netTransitions$: $s \in \inTr{t_1} \cap \inTr{t_2} \implies \inTr{t_1} = \inTr{t_2}$.
\end{compactitem}
\end{definition}

%
This paper treats concurrency and cyclic behaviour of process models separately. In particular, the approach presented in this paper tackles the problem of computing alignments between an event log and concurrency-free Petri nets. Then, in order to obtain concurrency-free Petri nets from a uniquely-labeled free-choice workflow net, any decomposition technique can be used; for this paper, the proposed technique is used in combination with the decomposition/recomposition technique presented in~\cite{s-comps}.
The work in~\cite{s-comps} decomposes a free-choice workflow net into subnets that are concurrency-free and fully represent the behavior of the original workflow net. These subnets belong to a subclass of Petri nets called state machine workflow nets. These workflow nets have the additional restriction that every transition can have at most one place in its pre- and postset.
The sub-family of Petri nets, \emph{uniquely-labelled state machine workflow nets}~\cite{WFNets,FreeChoice}, used in this paper for the computation of alignments in concurrency-free models is presented next.


\begin{definition}[Uniquely-Labelled, state machine workflow net]
A (uniquely-labelled) \emph{state machine workflow net} is a triplet $\stateMachineWNet=(\lnet,i,o)$, where $\lnet = (\places, \netTransitions, \netArcs, \netLabel)$ is a labelled Petri net, $i\in\places$ is the initial and $o\in\places$ is the final place, and the following properties hold:
\begin{compactitem}
    \item $i$ has an empty preset and $o$ has an empty postset, i.e., $\inTr{i}=\outTr{o}=\varnothing$.
    \item If a transition $t^*$ were added from $o$ to $i$, such that $\inTr{i} = \outTr{o} = \{t^*\}$, then the resulting net is strongly connected.
    \item Any transition $t\in\netTransitions$ can only have one place in its pre- and postset, i.e. $\left|\inTr{t}\right|=\left|\outTr{t}\right|=1$.
    \item (Uniquely-labelled) For any $t_1,t_2\in \netTransitions,\netLabel(t_1)=\netLabel(t_2)\neq\tau \Rightarrow t_1=t_2$.
\end{compactitem}
\end{definition}

The execution semantics of a net can be defined by means of \emph{markings} representing its execution states and the \emph{firing rule} describing if an action can occur. 
A marking is a multiset of places, i.e. a function $m : P \rightarrow \mathbb{N}_0$ that relates each place $p \in P$ to a natural number of \emph{tokens}. A transition $t$ is enabled at marking $m$, represented as $m[t\rangle$, if each place of the preset $\inTr{t}$ contains a token, i.e. $\forall p \in \inTr{t} : m(p) \geq 1$. An enabled transition $t$ can fire to reach a new marking $m'$, the firing of $t$ removes a token from each place in the preset $\inTr{t}$ and adds a token to each place in the postset $\outTr{t}$, i.e. $m' = m \setminus \inTr{t} \uplus \outTr{t}$. A fired transition $t$ at a marking $m$ reaching a marking $m'$ is represented as $m[t\rangle m'$. 
A marking $m'$ is \emph{reachable} from $m$, if there exists a sequence of firing transitions $\sigma = \langle t_1,\dots t_n\rangle$, such that $m_{i-1}[t_i\rangle m_i$ holds for all $1\leq i\leq n$, $m_0=m$ and $m_n=m'$. In addition, every marking is reachable by itself, i.e. $m'$ is always reachable from $m$, if $m'=m$.

A net with an initial and a final marking is called a \emph{(Petri) system net}.  
In the case of a 
state machine
workflow net, the initial marking has only one token in the special place $i$ and there is 
only one final marking with a token in the special place $o$.

\begin{definition}[System net]\label{def:SysNet}
A \emph{system net} $\sysNet$ is a triplet $\sysNet = (\stateMachineWNet, \initMarking, \finalMarkings)$, where $\stateMachineWNet=(\lnet,i,o)$ is a labelled %
state machine workflow net
, $\initMarking=\{i\}$ is the initial marking with the special place $i$ and %
$\finalMarkings=\{\{o\}\}$ is the set of final markings containing only one final marking with the special place $o$
.
\end{definition}

A marking is $k$-bounded if every place at a marking $m$ has up to $k$ tokens, i.e., $m(p) \leq k$ for any $p \in P$. 
%

A system net of a state machine workflow net, as defined in Def.~\ref{def:SysNet}, is always sound~\cite{StateMachineWFsound} and safe, i.e. every marking is 1-bounded since every transition of a state machine workflow net can have at most one place in its postset.


All possible markings, as well as the occurrence of observable and invisible activities, of a system net can be captured in a so-called \emph{reachability graph}~\cite{mayr84}. A reachability graph is a non-deterministic FSM, where nodes denote markings, and arcs denote the firing of transitions. The notation for a reachability graph will be the same as the FSM with the subscript $RG$, i.e. $(\rgNodes, \rgArcs, \rgSource, \rgFinStates)$ is a reachability graph.
Observe that for workflow nets, the reachability graph for the corresponding system net has a single final node with no outgoing arcs because the workflow net has one final marking with special place $o$.
The complexity for constructing a reachability graph of a state machine workflow net is $O(\left|\places \cup \netTransitions\right|)$, 
i.e. every place and transition translates to one node and arc, respectively.

\newpage
In order to have a more compact representation of the reachability graph, we remove all arcs labelled with $\tau$ by applying Alg.~\ref{alg:remTau}. The algorithm for removing tau arcs from the reachability graph of system (workflow) net was first proposed in~\cite{ReissnerCDRA17} and was later revised in~\cite{s-comps} to remove $\tau$-arcs targeting final markings. We include a revised version of the algorithm here to address the case when a reachability graph contains a $\tau$-labelled arc from the initial to a final marking and to make the article self contained.

Given a reachability graph of a system net as defined in Def.~\ref{def:SysNet}, Alg.~\ref{alg:remTau} removes all $\tau$ labelled arcs and returns the resulting reachability graph.
Intuitively, for every node $m\in\rgNodes$ reached by a $\tau$-labelled arc $a_1 = (m_1,\tau,m) \in \rgArcs$ and every outgoing arc of $m$, i.e. $a_2 = (m,l,m_2) \in \rgArcs$, the algorithm replaces $a_1$ with new $a_{12} = (m_1,l,m_2)$ (lines \ref{line:begin breadth-first}-\ref{line:end breadth first} and lines \ref{line:begin replaceTau}-\ref{line:end replaceTau}).
This replacement is repeated until all arcs representing $\tau$-transitions are removed. In case all incoming arcs of a node get replaced, the node $m$ is removed together with its outgoing arcs using function $\removeUnconnectedMarkings$ (Lines \ref{line:begin remNodes}-\ref{line:end remNodes}). Function $\emph{replaceTau}$ also handles the case of another outgoing $\tau$-labeled arc $a_2 = (m,\tau,m_2)$ by a depth-first search along $\tau$-transitions in $\rgArcs$ (lines \ref{line:begin replaceTau}-\ref{line:end replaceTau}).
At this point, the remaining $\tau$ arcs target the single final node of the reachability graph.
The algorithm then removes these $\tau$ arcs $a = ( m_1, \tau, m_f )$ targeting a final node $m_f$ while introducing new replacement arcs $a' = ( m_2, l, m_f )$ for each incoming arc of $m_1$, such that $(m_2,l,m_1) \in \rgArcs$ (Lines \ref{line:begin TauRemovelFinalNodes}-\ref{line:end TauRemovelFinalNodes} and function \emph{replaceTauBackwards}).
If a $\tau$ arc from the source to the final node would be removed that way, the source node needs to be added to the set of final nodes of the reachability graph to preserve the behavior of an execution with only $\tau$ arcs.
Finally, after all $\tau$ arcs targeting a final node are removed, any nodes that have become unconnected in the process have to be removed with function $\removeUnconnectedMarkings$. The reachability graph returned by Alg.~\ref{alg:remTau} is now free of $\tau$-labelled arcs.

\begin{figure}[h!]
\centering
\vspace{1\baselineskip}
\includegraphics[width=1\textwidth]{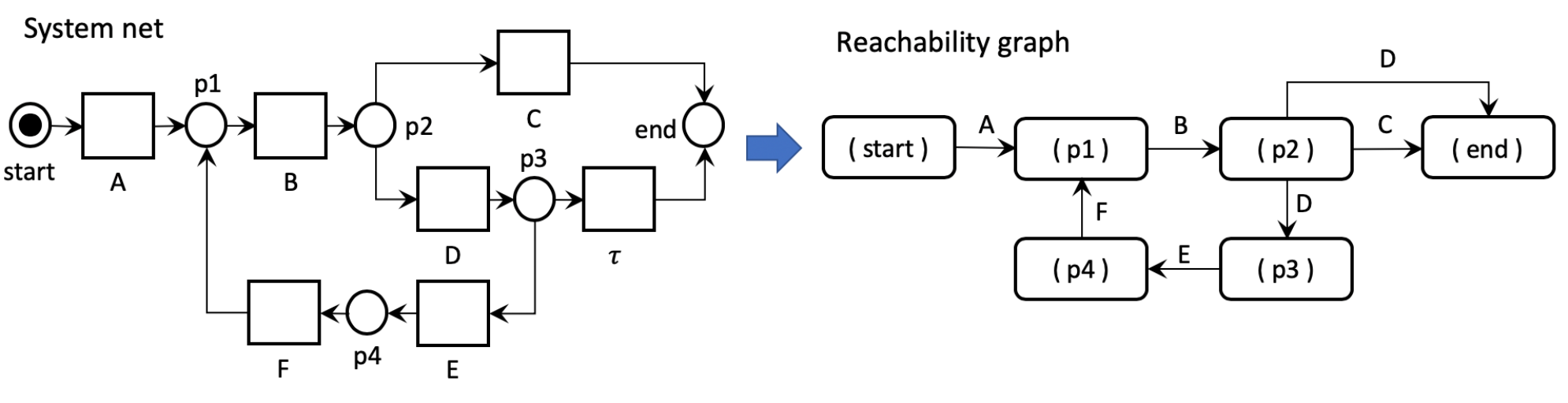}
\vspace{-1\baselineskip}
\caption{System net and reachability graph of the running example.}\label{fig:runningExampleRG}
\vspace{-1\baselineskip}
\end{figure}

The system net and reachability graph shown in Fig.~\ref{fig:runningExampleRG} are going to be used as the input process model in the running example throughout the paper.
The workflow net contains a loop of activities $B$, $D$, $E$ and $F$ that can be executed any number of times after $A$ and, afterwards, the process ends by either choosing activity $C$ or $D$.
Observe that the nodes in the reachability graph represent the markings in the net.


\newblock

\begin{algorithm}[H]{
  \SetKwInOut{Input}{input}
  \SetKwProg{Fn}{Function}{}{}
  \Input{Reachability Graph $\reachGraph=(\rgNodes, \rgArcs, \rgSource, \rgFinStates)$ of a System net;}
  $\open \leftarrow \langle \rgSource \rangle$; \tcp{List of nodes to check}
  $\closed \leftarrow \{\rgSource\}$; \tcp{Nodes checked}
  \While{$\open \neq \langle \rangle$}{\label{line:begin breadth-first}
   $\n \leftarrow \head{\open}$;\tcp{Remove node m from the head the list}
   $\open \leftarrow \tail{\open}$\;
   $\INV \leftarrow \{a = (m_1, l, \n) \in \shift\n \mid l = \tau \land \n \notin \rgFinStates\}$;\tcp{Incoming $\tau$ arcs at m}
  \For{$\inv \in \INV$}{
   $\replaceTau(\inv, \n, \{ \n\})$;\tcp{Replace $\tau$ arcs}
  }
  $\rgArcs \leftarrow \rgArcs \setminus \INV$;\tcp{Remove incoming $\tau$ arcs}
  \For{$(\n, l, m_2) \in \n\shift \mid m_2 \notin \closed$}{\tcp{Insert the target node of each outgoing arc into the auxiliary lists}
   $\open \leftarrow \open \tieconcat m_2$\;
   $\closed \leftarrow \closed \cup \{m_2\}$\;
  }\label{line:end breadth first}
 }
 $\removeUnconnectedMarkings()$\;
 \For{$a = (m_1,l,m_f) \in \shift m_f \mid m_f\in\rgFinStates\land l=\tau$\tcp{Any remaining $\tau$ arc targets a final node}\label{line:begin TauRemovelFinalNodes}}
 {
    $\replaceTauBack(a)$;\tcp{Replace $\tau$ arcs with a backwards replacement}
    \If{$m_1=\rgSource$}
    {
        $\finalMarkings\leftarrow\finalMarkings\cup\initMarking$;\tcp{Handle a $\tau$-arc from the source to $m_f$}
    }
 }
$\removeUnconnectedMarkings()$\;\label{line:end TauRemovelFinalNodes}
 \Return{$\reachGraph$}\;
 \Fn{$\replaceTau((m_1, \tau, \n) \in \rgArcs, \n_t \in \rgNodes, \Closed \in 2^\rgNodes$) 
 }{\label{line:begin replaceTau}
  \For{$(\n_t, l, m_2) \in \n_t\shift$}{
   \uIf{$l \neq \tau \vee m_2 \in \rgFinStates$}{
   \tcp{Replace outgoing arc of $n_t$ that is not $\tau$ and its target is not final}
    $\rgArcs \leftarrow \rgArcs \cup \{(m_1, l, m_2)\}$;  
   }
   \ElseIf{$m_2 \notin \Closed$} {
    $\Closed \leftarrow \Closed \cup \{m_2\}$; 
    $\replaceTau((m_1, \tau, \n), m_2, \Closed)$;\tcp{Try to replace the input $\tau$ arc from the new target marking $m_2$}
   }
  }\label{line:end replaceTau}
 }
 \Fn{$\replaceTauBack((m_1,\tau,m_f\in\rgFinStates) \in \rgArcs$}
 {\label{line:begin replaceTauBackwards}
     \tcp{For each incoming arc of the source node of the input $\tau$ arc}
     \tcp{Replace incoming arcs of $m_1$ with arcs from the predecessors of $m_1$ to $m_f$}
     \For{$(m_2,l,m_1) \in \shift m_1 \mid l \neq \tau$}
     {
        $\rgArcs \leftarrow \rgArcs \cup \{(m_2,l,m_f)\}$; \tcp{Replace arc from the source of the arc, its label and the target marking of the $\tau$ arc}
     }
     $\rgArcs \leftarrow \rgArcs \setminus \{(m_1,\tau,m_f)\}$; \tcp{Remove the $\tau$ arc from the reachability graph}\label{line:end replaceTauBackwards}
 }
 \Fn{$\removeUnconnectedMarkings()$}
 {
     $\remnodes \leftarrow \{\n \in \rgNodes \mid (\shift\n = \varnothing \land \n \neq \rgSource) \lor (\n\shift = \varnothing \land \n \notin \rgFinStates)\}$;\tcp{Non-source nodes with no incoming arcs, and non-final nodes with no outgoing arcs. These nodes result from the deletion of $\tau$ arcs}\label{line:begin remNodes}
 \While{$\remnodes \neq \varnothing$}{
  \For{$\n \in \remnodes$}{
   $\rgArcs \leftarrow \rgArcs \setminus (\shift\n \cup \n\shift)$\tcp{Remove arcs from every node in $\remnodes$}
  }
  $\rgNodes \leftarrow \rgNodes \setminus \remnodes$;\tcp{Remove all disconnected nodes}
  $\remnodes \leftarrow \{\n \in \rgNodes \mid (\shift\n = \varnothing \land \n \neq \rgSource) \lor (\n\shift = \varnothing \land \n \notin \rgFinStates)\}$;\tcp{Determine if any more nodes became disconnected} \label{line:end remNodes}
 }
 }
 \caption{Remove tau-labelled arcs}\label{alg:remTau}
}
\end{algorithm}

\subsection{Event log and DAFSA.}

Event logs record the executions of a business process. These executions are stored as sequences of activity occurrences (a.k.a. events). A sequence of events corresponding to an instance of a process is called a \emph{trace}, where events are represented by the corresponding activity's name. Although event logs are multisets of traces, given that the same trace might have been observed several times, we are only interested in distinct traces and thus an event log is considered as a set of traces. 

\begin{definition}[Trace and Event Log]\label{def:log}
Given a set of labels $\labels$, a \emph{trace} $t$ is a finite sequence of labels $t=\langle l_1,l_2,\dots,l_n\rangle \in \labels^*$ such that $l_i \in \labels$ for any $1\leq i \leq n$. An \emph{event log} $\logL$ is a set of traces.
\end{definition}

The size of a trace $t$ is the number of elements it contains, shorthanded as $\left|t\right|$, and $t[i]$ retrieves the $i$-th element in the trace. %
Further, $t[i,j]$ retrieves the sequence of elements in $t$ from position $i$ up to position $j$. Finally, we use $\oplus$ to denote the concatenation of two sequences, i.e. $t[1]\oplus t[2,\left|t\right|]=t$.

An event log can be represented as a FSM called \emph{Deterministic Acyclic Finite State Automaton (DAFSA)}, as described in \cite{ReissnerCDRA17}.
A DAFSA is a lossless representation of the event log, where every trace in the event log is represented as a path from the initial node to one of the final nodes. It is also an exact representation as it does not contain any paths that are not also traces in the event log. It uses prefix and suffix compression to achieve a minimal size of the FSM in terms of arcs and nodes \cite{daciukJ00}.
The DAFSA of an event log is                                                                                                                        denoted as $\dfa = (\dfaNodes,\dfaArcs,\dfaSource,\dfaFinStates)$, with the elements listed in Def.~\ref{FSM} with subscript $\dafsa$. 
Figure~\ref{fig:runningExampleLog} shows the input event log of the running example, where every trace is annotated with an identifier. 
This identifier will be useful to keep track of its tandem repeats in subsection~\ref{sec:tandemRepeats} and the trace transformations presented in the next section.

\begin{figure*}[htbp]
\centering
\resizebox{0.6\textwidth}{!}{ 
 \tikzstyle{ID} = [draw, rectangle, fill=white, align=center, minimum height=5mm, text width={width("$t  (1)  t$")}, font=\footnotesize]
 \tikzstyle{block} = [draw, rectangle, fill=white, align=left, minimum height=5mm, text width={width("$t  A, B, D, F, B, D, F, B, D, F, B, D, F, B, D, F, B, D  t$")}, font=\footnotesize]
 \begin{tikzpicture}[>=stealth', node distance=-0.3pt]
  \node[block] (log) {\bf{Trace}};
  \node[ID, left=of log] (id) {\bf{ID}};
  \node[block, below=of log] (trace1) {$\langle A, B, C, C, C, C \rangle$};
  \node[ID, left=of trace1] (id1) {1};
  \node[block, below=of trace1] (trace2) {$\langle A, B, D, E, E, F, B, D, E, E, F, B, D, E, E, F, B, C \rangle$};
  \node[ID, left=of trace2] (id2) {2};
  \node[block, below=of trace2] (trace3) {$\langle A, B, D, F, B, D, F, B, D, F, B, D \rangle$};
  \node[ID, left=of trace3] (id3) {3};
  \node[block, below=of trace3] (trace4) {$\langle A, B, D, F, B, D, F, B, D, F, B, D, F, B, D \rangle$};
  \node[ID, left=of trace4] (id4) {4};
  \node[block, below=of trace4] (trace5) {$\langle A, B, D, F, B, D, F, B, D, F, B, D, F, B, D, F, B, D \rangle$};
  \node[ID, left=of trace5] (id5) {5};
  \node[block, below=of trace5] (trace6) {$\langle A,D,B,F,D,B,F,D,B,F,D,B,F,B,C \rangle$};
  \node[ID, left=of trace6] (id6) {6};
  \end{tikzpicture}
  }
 \caption{Input event log for the running example.}\label{fig:runningExampleLog}
\end{figure*}
\vspace{-1\baselineskip}

\subsection{Alignments.}
Alignments capture the common and deviant behavior between a model and a log -- in our case between the FMSs representations for the model and log -- by means of three operations: 
\begin{inparaenum}[1.]
	\item a synchronized move ($\match$) traverses one arc on both FSMs with the same label,
	\item a log operation ($\lhide$) and
	\item a model operation ($\rhide$) that traverse an arc on the log or model FSM, respectively, while the other FSM does not move.
\end{inparaenum}
Note that $\match$ is commonly referred to as \emph{match}, and $\lhide$ and $\rhide$ as \emph{hides}. These operations are applied over a pair of elements that can be either arcs of the two FSMs or $\perp$ (indicating a missing element for $\lhide$ and $\rhide$). These triplets (operation and pair of affected elements) are called \emph{synchronizations}. 

\begin{definition}[Synchronization]
Let $\dfaArcs$ and $\rgArcs$ be the arcs of a DAFSA $\dfa$ and a reachability graph $\rg$, respectively. 
A \emph{synchronization} is a triplet $\sync=(\op,a_{\dafsa},a_{\reachGraph})$, where $\op \in \{\match, \lhide, \rhide\}$ is an operation, $a_{\dafsa} \in \dfaArcs\cup\bot$ is an arc of the DAFSA and $a_{\reachGraph} \in \rgArcs\cup\bot$ is an arc of the reachability graph.
The set of all possible synchronizations is represented as $\syncs(\dfa, \rg) = \{(\lhide, a_{\dafsa}, \perp) \mid a_{\dafsa} \in \dfaArcs\} \cup \{(\rhide, \perp, a_{\reachGraph}) \mid a_{\reachGraph} \in \rgArcs\} \cup \{(\match, a_{\dafsa}, a_{\reachGraph}) \mid a_{\dafsa} \in \dfaArcs \land a_{\reachGraph} \in \rgArcs \land \lbl(a_{\dafsa}) = \lbl(a_{\reachGraph})\}$.
\end{definition}

Given a synchronization $\sync=(\op,a_{\dafsa},a_{\reachGraph})$, the operation, the arc of the DAFSA and the arc of the reachability graph are retrieved by $\op(\sync) = \op$, $\fnSyncDFA(\sync)=a_{\dafsa}$ and $\fnSyncRG(\sync)=a_{\reachGraph}$, respectively. By the abuse of notation, let $\lbl(\sync)$ denote the label of the arc in $\sync$ that is different to $\perp$, i.e., if $\fnSyncDFA(\sync) \neq \perp$ then $\lbl(\sync) = \lbl(\fnSyncDFA(\sync))$, otherwise $\lbl(\sync) = \lbl(\fnSyncRG(\sync))$.

\begin{definition}[Alignment]
An \emph{alignment} is a sequence of synchronizations $\alignment = \langle \sync_1, \sync_2, \dots, \sync_n\rangle$. 
The projection of an alignment $\alignment$ to the DAFSA, shorthanded as $\filterSyncDFA$, retrieves all synchronizations with $\fnSyncDFA(\sync)\neq\bot$
, while the projection to the reachability graph, shorthanded as $\filterSyncRG$, retrieves all synchronizations with $\fnSyncRG(\sync)\neq\bot$.
\end{definition}

As a shorthand, functions $\op$, $\fnSyncDFA$, $\fnSyncRG$ and $\lbl$ can be used for alignments by applying the function to each synchronization wherein. For instance, $\op(\alignment)$ results in the sequence of operations in $\alignment$.

An alignment is \emph{proper} if it 
\begin{inparaenum}[]
	\item represents a trace $t$, this is $\lbl(\filterSyncDFA)=t$, and 
	\item both $\fnSyncDFA(\filterSyncDFA)$ and $\fnSyncRG(\filterSyncRG)$ are paths through the DAFSA and the reachability graph from a source node to one of the final nodes, respectively. 
\end{inparaenum}
We refer to the set of all proper alignments as $\properAlignments(\dfa,\rg)$.




Intuitively, an alignment represents the number of operations to transform a trace (path in the DAFSA) into a path in the reachability graph. Synchronizations in an alignment can be associated with a cost, which can be defined individually for each activity with domain-specific knowledge. For simplicity purposes, we use the standard cost function~\cite{ILP-Alignment,ReissnerCDRA17}, where a weight of 1 is assigned to a synchronization with $\lhide$ and $\rhide$ operations, and 0 to the synchronizations with $\match$ operations. 
The formal definition is given next. 

\begin{definition}[Cost function]\label{def:CostFunction}
Given an alignment $\alignment$ with $1\leq \position\leq\left|\alignment\right|$, we define the cost of a synchronization at position $i$ with function $\syncCostStandard$:
\[
\syncCostStandard(\alignment,\position)=
\begin{cases}
    1,& \text{if } \op(\alignment[\position])=\rhide \lor \op(\alignment[\position])=\lhide\\
    0,& \text{if } \op(\alignment[\position])=\match
\end{cases}
\]
The total cost $\costF$ for an alignment is the sum of function $\syncCostStandard$ for each of its elements:
\[
\costF(\alignment) = \sum_{\position\in1\dots\left|\alignment\right|}\syncCostStandard(\alignment,\position)
\]
\end{definition}
For computing alignments with domain-specific knowledge the cost function could be extended by using multipliers for each activity of the event log or process model.

\newpage
In this work, we rely on Alg. 4 from \cite{s-comps} to compute one optimal alignment for each trace in the event log. The alignments are computed by synchronously traversing the DAFSA and the reachability graph.
Using a DAFSA for the entire log instead of single traces to compute alignments has technical advantages, such as memory considerations and using prefix/suffix memoization when computing alignments as proposed in~\cite{ReissnerCDRA17}.
Algorithm~\ref{alg:fsmAlignments} shows the procedure to compute alignments between two FSMs, as presented in~\cite{s-comps}, further improvements to this algorithm will be introduced in the following sections. 
This algorithm constructs a function $\alignments$ that relates each trace of a given event log to its optimal alignment (lines \ref{ln:beginConstructAlignments}-\ref{ln:endConstructAlignments}). It applies function align (lines \ref{ln:beginAlign}-\ref{ln:endAlign}) to search for the optimal alignment for each trace $t$. The optimal alignment uses Dijkstra's algorithm for finding the shortest sequence of synchronizations with the cost function from Def.~\ref{def:CostFunction}. The algorithm uses an open queue $\open$ that keeps track of the nodes to be investigated in the search. A node in the search is defined as a triplet consisting of the current node in the DAFSA $n_{\dafsa}$, in the reachability graph $n_{\reachGraph}$ and a sequence of synchronizations $\alignment$. Each node is stored in the open list $\open$ paired with the cost of its alignment to speed up the selection of the next candidate node for the search.
The search starts from the source of the DAFSA $\dfaSource$, the source of the reachability graph $\rgSource$ and an empty sequence $\langle\rangle$ (line \ref{ln:startOpen}).
In every iteration, the search removes a node $\actN$ from $\open$ with minimal cost (line \ref{ln:remActN}). If $\actN$ is already a proper alignment, i.e. if both $n_{\dafsa}$ and $n_{\reachGraph}$ are final and if all synchronizations relating to the DAFSA are equal to the trace labels 
(line \ref{ln:condOptimalityAlg1}), the procedure finishes and the alignment is returned. Otherwise, the search continues by inserting new candidate nodes to the open queue (lines \ref{ln:beginNextCandidates}-\ref{ln:endNextCandidates}). The candidate nodes are determined in three steps:
\begin{inparaenum}[(1)] 
    \item the outgoing arc of $n_{\dafsa}$ with the next label of the trace is added with an $\lhide$-synchronization to $\alignment$ shifting $n_{\dafsa}$ to the target of the arc, 
    \item all outgoing arcs of $n_{\reachGraph}$ are added with an $\rhide$-synchronization shifting $n_{\reachGraph}$ to the target of each arc and
    \item for every pair of outgoing arcs from both $n_{\dafsa}$ and $n_{\reachGraph}$ with the next trace label are added with a $\match$-synchronization to $\alignment$ and both $n_{\dafsa}$ and $n_{\reachGraph}$ are shifted along the corresponding arcs.
\end{inparaenum}
Last, the search uses a closed function $\closed$ that links previously visited pairs of $n_{\dafsa}$ and $n_{\reachGraph}$ to the lowest cost achieved so far. The search will only investigate nodes that are visited again, if their cost is lower or equal to the cost of the closed function value (lines \ref{ln:startClosed}-\ref{ln:endClosed}).

\begin{algorithm}[h]
{
    \SetKwInOut{Input}{input}
    \SetKwProg{Fn}{Function}{}{}
    \Input{Event log $\logL$; DAFSA $\dfa$; Reachability Graph $\rg$}
    $\alignments \leftarrow \{\}$\label{ln:beginConstructAlignments}\;
    \For{$t\in\logL$}
    {
        $\alignments \leftarrow \alignments\cup\{t\rightarrow\text{align}(t,\dfa,\rg)\}$\;
    }
    \Return{$\alignments$}\label{ln:endConstructAlignments}\;
    \Fn{align$(t,\dafsa, \reachGraph$)}{\label{ln:beginAlign}
        $\open \leftarrow \{((\dfaSource,\rgSource,\langle\rangle), \costF(\langle\rangle))\}$\label{ln:startOpen}\;
        $\closed\leftarrow\{\}$\;
        \While{$\open \neq \varnothing$}
        {
            $\actN \leftarrow $ remove $(n_{\dafsa},n_{\reachGraph},\alignment)$ from $\open\mid\nexists (n',\rho')$ with $\rho'<\costF(\alignment)$\label{ln:remActN}\;
           \uIf{$\closed(n_{\dafsa},n_{\reachGraph})=\perp\lor\closed(n_{\dafsa},n_{\reachGraph})\geq\costF(\alignment)$\label{ln:startClosed}}{$\closed \leftarrow \closed \cup \{(n_{\dafsa},n_{\reachGraph})\rightarrow\costF(\alignment)\}$\;}\label{ln:endClosed}
           \lElse{Continue} 
           \lIf{$n_{\dafsa} \in \dfaFinStates \land n_{\reachGraph} \in \rgFinStates \land \lbl(\filterSyncDFA)=t$\label{ln:condOptimalityAlg1}}{\Return{$\alignment$}}
           \Else
           {
                $\out_{\dafsa} \leftarrow n_{\dafsa}\shift \mid \lbl(\out_{\dafsa})=t(\left|\filterSyncDFA(\epsilon)\right|+1)$\;\label{ln:beginNextCandidates}
                
                $\open \leftarrow \open \cup \{(\tgt(\out_{\dafsa}),n_{\reachGraph},\alignment\tieconcat(\lhide,\out_{\dafsa},\perp))\}$\;
                \lFor{$\out_{\reachGraph}\in n_{\reachGraph}\shift\mid\lbl(\out_{\reachGraph})=\lbl(\out_{\dafsa})$}
                {$\open\leftarrow\open\cup\{(\tgt(\out_{\dafsa}),\tgt(\out_{\reachGraph}),\alignment\tieconcat(\match,\out_{\dafsa},\out_{\reachGraph}))\}$}                
                
                \lFor{$\out_{\reachGraph}\in n_{\reachGraph}\shift$}{$\open\leftarrow\open\cup\{(n_{\dafsa},\tgt(\out{\reachGraph}),\alignment\tieconcat(\rhide,\perp,\out_{\reachGraph}))\}$\label{ln:endNextCandidates}}
            }
        }\label{ln:endAlign}
    }
    \caption{Computing alignments between two FSMs}\label{alg:fsmAlignments}    
}
\end{algorithm}

\newpage
\subsection{Tandem repeats.}\label{sec:tandemRepeats}

The main contribution of this paper relies on identifying and reducing the repetitive sequences of activity occurrences in the traces, \emph{a.k.a.} tandem repeats, thus compressing each trace. A tandem repeat for a trace $t$ is a triplet$(\start,\TRtype,\reps)$, where $\start$ is the position in the trace where the tandem starts, $\TRtype$ is the repetitive pattern, \emph{a.k.a. repeat type}, and $\reps\geq2$ is the number of repetitions of $\TRtype$ in $t$. 
Given a trace $t$, $\Delta(t)$ is an oracle that retrieves the set of tandem repeats in $t$, such that the repeat type occurs at least twice (in other words, any tandem repeat $(\start,\TRtype,\reps)$ has $\reps \geq 2$).
We use $\TRs(t,\position)$ to refer to the set of tandem repeats in trace $t$ that start at a position in the trace $1\leq\position\leq\left|t\right|$, i.e. $\TRs(t,\position)=\{(\start,\TRtype,\reps)\in\Delta(t)\mid\start=\position\}$.
For the evaluation (Section~\ref{sec:evaluation}), the approach proposed by Gusfield and Stoye~\cite{FindingTRs} was used. The approach uses suffix trees to find tandem repeats in linear time with respect to the length of the input string and defines an order between the tandem repeats by reporting the leftmost occurrences, 
i.e. tandem repeats shifted right by any amount of characters are omitted. 
That technique can be speed up by using suffix arrays as the underlying data structure~\cite{SuffixArrays}. Additionally, the tandem repeats considered in this work are \emph{maximal} and \emph{primitive}~\cite{BoseTRinProcessMining}. A tandem repeat is called maximal if no repetitions of the repeat type occur at the left or right side of the tandem repeat. The tandem repeat is primitive, if the repeat type is not itself a tandem repeat.

\begin{definition}[Maximal and primitive tandem repeat 
with no right shifts
]
Given a trace $t$,
 a tandem repeat $(\start,\TRtype,\reps) \in \Delta(t)$ is \emph{maximal}, 
if neither $(\start-\left|\TRtype\right|,\TRtype,\reps+1)$ nor $(\start,\TRtype,\reps+1)$ is a tandem repeat,
and \emph{primitive} if $\alpha$ is not itself a tandem repeat. 
An operation to shift a tandem repeat right by $x$ characters is defined as $(\start,\TRtype,\reps)\rightshift x=(\start+x,\TRtype[x+1,\left|\TRtype\right|]\oplus \TRtype[1,x],\reps)$ for $1\leq x < \left|\TRtype\right|$. All right-shifts of any tandem repeats are omitted, i.e. $(\start,\TRtype,\reps) \in \Delta(t) \Rightarrow \forall_{1\leq x < \left|\TRtype\right|} (\start,\TRtype,\reps)\rightshift x \notin \Delta(t)$.
\end{definition}

Figure~\ref{fig:runningExampleTandemRepeats} shows the primitive and maximal tandem repeats %
with no right shifts 
for the input event log of the running example of Fig.~\ref{fig:runningExampleLog}. For example, in trace (1), there is one tandem repeat (3, C, 4), that starts on position 3 and the sequence C is repeated 4 times. Another possible tandem repeat for trace (1) is (3, CC, 2), but this is not primitive since CC is itself another tandem repeat %
(3,C,2). 
Another tandem repeat in trace (2) is (2,BDEEF,3), it is primitive because its repeat type is not itself a tandem repeat despite containing the tandem repeat (4,E,2). This kind of tandem repeats, i.e. (4,E,2), is also known as nested tandem repeats. 
In the case of trace (3), (5, BDF, 2) is another tandem repeat, but it is not maximal because it can be extended to the left side by one more repetition.
Last, trace (3) contains another tandem repeat (3,DFB,3), but it is omitted since it is the same as (2,BDF,3) shifted right by one character.
Gusfield and Stoye~\cite{FindingTRs} show how to avoid detecting tandem repeats shifted right by any number of characters.

\vspace{2\baselineskip}
\begin{figure*}[htbp]
\centering
\resizebox{0.6 \textwidth}{!}{
 \tikzstyle{ID} = [draw, rectangle, fill=white, align=center, minimum height=5mm, text width={width("$t  (1)  t$")}, font=\footnotesize]
 \tikzstyle{block} = [draw, rectangle, fill=white, align=left, minimum height=5mm, text width={width("$t$  test Maximal and primitive Tandem Repeat $t$")}, font=\footnotesize]
 \begin{tikzpicture}[>=stealth', node distance=-0.3pt]
  \node[block] (log) {\bf{Maximal and primitive Tandem Repeats}};
  \node[ID, left=of log] (id) {\bf{ID}};
  \node[block, below=of log] (trace1) {$(3, C, 4)$};
  \node[ID, left=of trace1] (id1) {(1)};
  \node[block, below=of trace1] (trace2) {$(2, BDEEF, 3), %
  (4, E, 2),
   (9, E, 2), (14, E, 2)$};
  \node[ID, left=of trace2] (id2) {(2)};
  \node[block, below=of trace2] (trace3) {$(2, BDF, 3)$};
  \node[ID, left=of trace3] (id3) {(3)};
  \node[block, below=of trace3] (trace4) {$(2, BDF, 4)$};
  \node[ID, left=of trace4] (id4) {(4)};
  \node[block, below=of trace4] (trace5) {$(2, BDF, 5)$};
  \node[ID, left=of trace5] (id5) {(5)};
  \node[block, below=of trace5] (trace6) {$(2, DBF, 4)$};
  \node[ID, left=of trace6] (id5) {(6)};
  \end{tikzpicture}
  }
 \caption{Primitive and maximal tandem repeats with no right shifts
  for the event log of Fig.~\ref{fig:runningExampleLog}.}\label{fig:runningExampleTandemRepeats}
\end{figure*}

%% file: tex/Approach.tex
\newcommand{\posExtend}{\mathit{pos}_{ext}}

\newcommand{\copyExtend}{\alignment_{mid}}
\newcommand{\copyFirstExtend}{\alignment_{mid,1}}
\newcommand{\copySecExtend}{\alignment_{mid,2}}
\newcommand{\suff}{\mathit{suff}}
\newcommand{\myceil}[1]{\left \lceil #1 \right \rceil }
\newcommand{\myfloor}[1]{\left \lfloor #1 \right \rfloor }
\newcommand{\lowerB}{\mathit{lo}}
\newcommand{\upperB}{\mathit{up}}
\newcommand{\sortAsc}{\uparrow}
\newcommand{\aligncomplement}{\mathit{TR}_{c}}
\newcommand{\alignTRend}{\mathit{TR}_{e}}
\newcommand{\complemnt}{\mathit{TR}_{c}}
\newcommand{\syncCost}{\mathit{f}}
\newcommand{\minF}{\mathit{min}}
\newcommand{\thirdPos}{\mathit{k}}
\newcommand{\extension}{\mathit{E}}
\newcommand{\applyExtension}{\Gamma}
\newcommand{\alignTRstart}{\mathit{TR}_{s,1}}
\newcommand{\alignTRstartSec}{\mathit{TR}_{s,2}}
\newcommand{\extalignment}{\alignment_{\mathit{ext}}}
\newcommand{\redalignment}{\alignment_{\mathit{red}}}
\newpage
\section{Automata-based Conformance Checking with Tandem Repeats Reductions}\label{sec:approach}

\begin{figure}[h]
\centering
\includegraphics[width=0.8\textwidth]{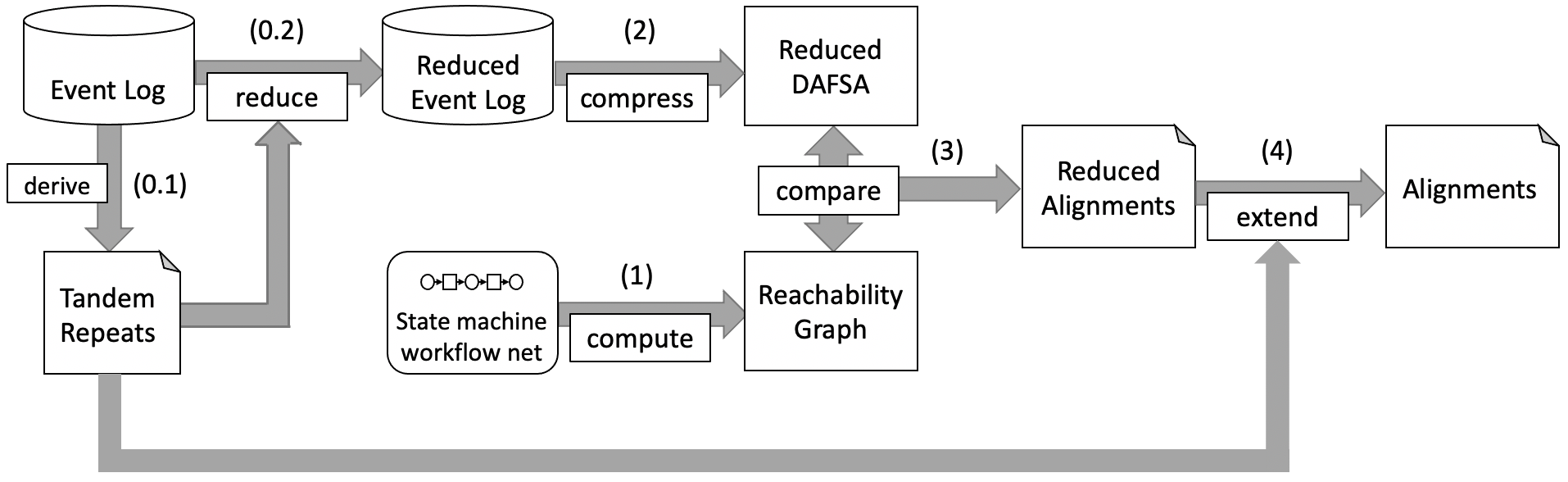}
\vspace{-.5\baselineskip}
\caption{%
Overview of the Tandem Repeats Approach
} \label{TR-approach}
\vspace{-2\baselineskip}
\end{figure}

This section presents a novel approach for computing the differences between an event log and a process model. These differences are expressed in terms of trace alignments. The proposed approach is depicted in Fig.~\ref{TR-approach}.
In order to increase the scalability of the approach, the first step consists in reducing the event log (Step 0.1) by finding patterns of repetition, a.k.a. \emph{tandem repeats}, in each of the event log traces (Step 0.2). Then, the reachability graph of the 
state machine workflow net is computed (Step 1) and, in parallel, the reduced event log is compressed into an automaton (Step 2). Finally, both automata are compared with Dijkstra's algorithm to derive alignments representing the differences and commonalities between the log and the model (Step 3). Given that the computed alignments represent reduced event log traces, the final step (Step 4) expands those alignments to obtain the alignments of the original traces.
Throughout this section we use the input event log from Fig.~\ref{fig:runningExampleLog}, its tandem repeats from Fig.~\ref{fig:runningExampleTandemRepeats} and the input system net from Fig.~\ref{fig:runningExampleRG} as a running example to demonstrate each step of the proposed technique. 


\input{tex/scomp_integration.tex}

\subsection{Determining trace alignments with a reduced event log}

The technique presented in this paper is based on the identification of primitive and maximal tandem repeats within the traces in the event log. These repeats are reduced to two repetitions in each of the traces, producing a reduced version of the log. Then, the alignments are computed between the model and the reduced log. The intuition behind the trace reductions is that, if the two repetitions are matched over the model, then the model is cyclic (the model is uniquely labelled) and we can assume that any additional repetition of the tandem repeat can be matched over the model.
This intuition relies on the fact that if a transition can be repeated twice in a 
state machine workflow net then it can be repeated any number of times, which was proven in \cite{armas2016diagnosing} for free-choice workflow nets, a super class of state machine workflow nets.

For reducing the traces, we use maximal and primitive tandem repeats without right shifts as described in Section~\ref{sec:tandemRepeats}. These tandem repeats can overlap such that it is unclear how the tandem repeats should be reduced. To avoid such overlapping, we propose a deterministic and greedy selection of tandem repeats, i.e. tandem repeats are selected for reduction in the order of the trace. If two tandem repeats occur at the same trace position, the longer one is selected.
The result is a reduced event log $\redLog$ containing a set of reduced traces. 
Figure~\ref{fig:runningExampleLog} shows the order for collapsing tandem repeats for trace (2) in the running example: The first primitive and maximal tandem repeat from left to right in the trace is (2,BDEEF,3). This tandem repeat is reduced to two repetitions in the reduced trace. Next, the trace reductions end, because there are no more primitive and maximal tandem repeats after trace position 17, which is the position where the tandem repeat ends. 
This ordering has the effect that tandem repeats contained in another tandem repeat and overlapping tandem repeats will not be considered for trace reductions. In the case of trace (2), the tandem repeats (4,E,2),(9,E,2) and (14,E,2) are not considered, because they are covered by (2,BDEEF,3).

The reduction operation can collapse different traces into the same reduced trace.
For instance, consider the traces (3) and (4) in Fig.~\ref{fig:runningExampleLog}, which have different number of repetitions for the same repeat type. Both tandem repeats: (2,BDF,3) and (2,BDF,4) in Fig.~\ref{fig:runningExampleTandemRepeats}, will be reduced to only two copies, thus resulting in the reduced trace: $\langle A, $\colorbox{gray!20}{$B, D, F$},\colorbox{gray!20}{$B, D, F$}$, B, D \rangle$, where the greyed-out areas represent the two repetitions of the token repeats. 
The elements in the first copy of the tandem repeats have a corresponding element in the second copy, the i-$th$ element in the first copy is related to the i-$th$ element in the second copy. 
In the example, $\langle A, $\colorbox{gray!20}{$B, D, F$},\colorbox{gray!20}{$B, D, F$}$, B, D \rangle$, $B$ is related to $B$, $D$ with $D$, and $F$ with $F$.
In this way, when both elements are matched, an element in the tandem repeat and its corresponding element in the second copy, then a loop is found in the model. 
Please note that the reduced event log is still an event log according to Def.~\ref{def:log} and it only contains a set of traces. Thus, if a reduced trace is added a second time to a reduced event log, it would be discarded.
Additional information has to be retained after the reduction in order to have a unique identifier relating each reduced trace to its original trace.

The information about the reduction operations applied over a trace will be used later for reconstructing the original trace, thus it is important to preserve the information about the reductions applied. 
In order to do so, a \emph{reduced trace} is represented as a tuple $T= (\redTrace,\additionalCostF,\redReps,\complemnt, \position)$, 
where $\redTrace$ is the trace to reduce, 
$\additionalCostF$ is a function relating each index of each reduced tandem repeat in the reduced trace to the number of reduced repetitions of this tandem repeat,
$\redReps$ is the total number of reduced labels,
$\complemnt$ relates the two repetitions of the tandem repeats: an i$th$ element in the first repetition is related to the i$th$ element in the second repetition, 
and $\position$ is an auxiliary index representing the position in the trace from which tandem repeats can be identified. 
Finally, $\reductions$ relates each trace to its reduced version. Observe that a trace $t$ with no tandem repeats, or prior a reduction, is $(t,\additionalCostF=\{i\rightarrow0\mid1\leq i\leq\left|t\right|\},0,\emptyset, 1)$ where $t$ is a trace and the tandem repeats shall be identified from position 1. Next, we define the \emph{trace reduction} operation.

\begin{definition}[Trace reduction]\label{def:reduction}
Let $T = (t,\additionalCostF,\redReps,\trPositions, \position)$ be a -- possibly reduced -- trace and $(\start,\TRtype,\reps) \in \TRs(t, \position)$ be a maximal tandem repeat, such that $\nexists (\start_i,\TRtype_i,\reps_i)\in\TRs(t,\position) : \left|\TRtype_i\right| * \reps_i > \left|\TRtype\right| * \reps$. The \emph{reduced} $T$ by $k$ is $\reduce(T) = (\redTrace,\additionalCostF',\redReps',\trPositions', \position')$, where:
\begin{compactitem}
 	      \item $\redTrace \leftarrow \pref(t,\position-1) \oplus \TRtype \oplus \TRtype \oplus \suffix(t,\position + (\left|\TRtype\right|*\reps))$,
 	      \item $\additionalCostF' \leftarrow \additionalCostF \cup \{\secondPos\rightarrow\reps-2 \mid \position \leq \secondPos \leq (\position+\left|\TRtype\right|*2-1)\}$,
                 \item $\redReps' \leftarrow \redReps + (\reps-2)*\left|\TRtype\right|$,
                 \item $\complemnt' \leftarrow \complemnt \cup \{\secondPos\rightarrow\secondPos\!+\!\left|
                 \TRtype\right|, \secondPos\!+\!\left|\TRtype\right|\rightarrow\secondPos \mid \position\leq \secondPos \leq (\position\!+\!\left|\TRtype\right|\!-\!1)\}$, and
                 \item $\position' \leftarrow \position + \left|\TRtype\right| * 2$.
\end{compactitem}
Given a possibly reduced trace $T = (t,\additionalCostF,\redReps,\trPositions, \position)$, a trace reduction $\reduce(T)$ will return $T$ if there exists no tandem repeat at the current position $\position$, i.e. $\TRs(t,\position)=\varnothing\Rightarrow \reduce(T)=T$.                
\end{definition}


The reduction of a given event log and its tandem repeats is displayed in Alg.~\ref{alg:reduceLog}. Each trace $t\in\logL$ is reduced from position $\position$ until no more tandem repeats can be found and reduced, and $\position$ reaches the end of the trace $\left|t\right|$. Algorithm~\ref{alg:reduceLog} returns the reduced log $\redLog$ and the reduction information $\reductions$.

\begin{algorithm}[h!]{
    \SetKwInOut{Input}{input}
    \Input{Event log $\logL$; Tandem repeats $\TRs(t)$ for each $t\in\logL$} 
    Set $\redLog \leftarrow \reductions \leftarrow \{\}$\;
    \For{$t\in\logL$}{
    	Set $\position = 1$\;
    	Set $\trReductions = (t,\emptyset,0,\emptyset, 1)$\;
	\While{$\position \leq \left| \trReductions[1] \right|$}{
	    Set $\trReductions = \reduce(rTrace)$\;
	    \lIf{$\position=\trReductions[5]$}{Increase $\position$ by 1}
	    \lElse{Set $\position\leftarrow\trReductions[5]$}
	    Set $\trReductions[5]\leftarrow\position$\;
	}
    	$\redLog \leftarrow \redLog \cup \{\trReductions[1]\}$\;
           $\reductions \leftarrow \reductions \cup \{t\rightarrow \trReductions\}$\;
      }
    \Return{$\redLog$, $\reductions$}\;
    \caption{Reduce event log}\label{alg:reduceLog}
}
\end{algorithm}

Figure~\ref{fig:runningExampleReduction} shows the reduced event log for our running example after applying Alg.~\ref{alg:reduceLog} to the input event log from Fig.~\ref{fig:runningExampleLog}. 
For example, trace (4) was reduced to the trace $\langle A,$\colorbox{gray!20}{$B,D,F$},\colorbox{gray!20}{$B,D,F$}$,B,D\rangle$ by reducing the tandem repeat at position 2 with length 3, the labels reduced are $\redReps=6$, function $\complemnt$ relates ($\leftrightarrow$) positions 2 and 5, 3 and 6, and so forth. 

\begin{figure*}[htbp]
\centering
\resizebox{1\textwidth}{!}{
\tikzstyle{ID_head} = [draw, rectangle, fill=white, align=center, minimum height=9mm, text width={width("$t  (1)  t$")}, font=\footnotesize]
 \tikzstyle{ID} = [draw, rectangle, fill=white, align=center, minimum height=6.5mm, text width={width("$t  (1)  t$")}, font=\footnotesize]
 \tikzstyle{block_head} = [draw, rectangle, fill=white, align=left, minimum height=9mm, text width={width("$ A, B, D, E, E, F, B, D, E, E, F, B, C $")}, font=\footnotesize]
 \tikzstyle{block} = [draw, rectangle, fill=white, align=left, minimum height=6.5mm, text width={width("$ A, B, D, E, E, F, B, D, E, E, F, B, C $")}, font=\footnotesize]
 \tikzstyle{p} = [draw, rectangle, fill=white, align=center, minimum height=6.5mm, text width={width("$t  reduced repetitions  t$")}, font=\footnotesize]
 \tikzstyle{p_head} = [draw, rectangle, fill=white, align=center, minimum height=9mm, text width={width("$t  reduced repetitions  t$")}, font=\footnotesize]
 \tikzstyle{trpos} = [draw, rectangle, fill=white, align=center, minimum height=6.5mm, text width={width("Tandem complement")}, font=\footnotesize]
 \tikzstyle{trpos_head} = [draw, rectangle, fill=white, align=center, minimum height=9mm, text width={width("Tandem complement")}, font=\footnotesize]
 \tikzstyle{kred} = [draw, rectangle, fill=white, align=center, minimum height=6.5mm, text width={width("\#reduced labels")}, font=\footnotesize]
 \tikzstyle{kred_head} = [draw, rectangle, fill=white, align=center, minimum height=9mm, text width={width("\#reduced labels")}, font=\footnotesize]
 \tikzstyle{posH} = [draw, rectangle, fill=white, align=center, minimum height=6.5mm, text width={width("$t  10  t$")}, font=\footnotesize]
 \tikzstyle{posH_head} = [draw, rectangle, fill=white, align=center, minimum height=9mm, text width={width("$t  10  t$")}, font=\footnotesize]
 \setlength{\fboxsep}{2pt}
 \begin{tikzpicture}[>=stealth', node distance=-0.3pt]
  \node[block_head] (log) {\bf{Reduced Trace $\redTrace$}};
  \node[ID_head, left=of log] (id) {\bf{ID}};
  \node[p_head, right=of log] (p) {\bf{reduced repetitions\\p}};
  \node[kred_head, right=of p] (kred) {\bf{\#reduced labels\\k\textsubscript{red}}};
  \node[trpos_head, right=of kred] (trpos) {\textbf{Tandem complement\\TR\textsubscript{c}}};
  \node[posH_head, right=of trpos] (posH) {\bf{pos}};
  \node[block, below=of log] (trace1) {$\langle A, B, $\colorbox{gray!20}{$C$},\colorbox{gray!20}{$C$}$ \rangle$};
  \node[ID, left=of trace1] (id1) {(1)};
  \node[p, right=of trace1] (p1) {$3-4 \rightarrow 2$};
  \node[kred, right=of p1] (kred1) {2};
  \node[trpos, right=of kred1] (trpos1) {$3 \leftrightarrow 4$};
  \node[posH, right= of trpos1] (pos1) {4};
  \node[block, below=of trace1] (trace2) {$\langle A, $\colorbox{gray!20}{$B, D, E, E, F$},\colorbox{gray!20}{$B, D, E, E, F$}$, B, C \rangle$};
  \node[ID, left=of trace2] (id2) {(2)};
  \node[p, right=of trace2] (p2) {$2-11 \rightarrow 1$};
  \node[kred, right=of p2] (kred2) {5};
  \node[trpos, right=of kred2] (trpos2) {$2-6 \leftrightarrow 7-11$};
  \node[posH, right= of trpos2] (pos2) {12};
  \node[block, below=of trace2] (trace3) {$\langle A, $\colorbox{gray!20}{$B, D, F$},\colorbox{gray!20}{$B, D, F$}$, B, D \rangle$};
  \node[ID, left=of trace3] (id3) {(3)};
  \node[p, right=of trace3] (p3) {$2-7 \rightarrow 1$};
  \node[kred, right=of p3] (kred3) {3};
  \node[trpos, right=of kred3] (trpos3) {$2-4 \leftrightarrow 5-7$};
  \node[posH, right= of trpos3] (pos3) {9};
  \node[block, below=of trace3] (trace4) {$\langle A, $\colorbox{gray!20}{$B, D, F$},\colorbox{gray!20}{$B, D, F$}$, B, D \rangle$};
  \node[ID, left=of trace4] (id4) {(4)};
  \node[p, right=of trace4] (p4) {$2-7 \rightarrow 2$};
  \node[kred, right=of p4] (kred4) {6};
  \node[trpos, right=of kred4] (trpos4) {$2-4 \leftrightarrow 5-7$};
  \node[posH, right= of trpos4] (pos4) {9};
  \node[block, below=of trace4] (trace5) {$\langle A, $\colorbox{gray!20}{$B, D, F$},\colorbox{gray!20}{$B, D, F$}$, B, D \rangle$};
  \node[ID, left=of trace5] (id5) {(5)};
  \node[p, right=of trace5] (p5) {$2-7 \rightarrow 3$};
  \node[kred, right=of p5] (kred5) {9};
  \node[trpos, right=of kred5] (trpos5) {$2-4 \leftrightarrow 5-7$};
  \node[posH, right= of trpos5] (pos5) {9};
  \node[block, below=of trace5] (trace6) {$\langle A, $\colorbox{gray!20}{$D, B, F$},\colorbox{gray!20}{$D, B, F$}$, B, D \rangle$};
  \node[ID, left=of trace6] (id6) {(6)};
  \node[p, right=of trace6] (p6) {$2-7 \rightarrow 2$};
  \node[kred, right=of p6] (kred6) {6};
  \node[trpos, right=of kred6] (trpos6) {$2-4 \leftrightarrow 5-7$};
  \node[posH, right= of trpos6] (pos6) {9};
  \end{tikzpicture}
  }
 \caption{Reduced event log after applying Alg.~\ref{alg:reduceLog}.}\label{fig:runningExampleReduction}
\end{figure*}

\newpage
Next, we compute alignments for the reachability graph and the DAFSA of the reduced event log. 
In order to compute the alignments, Alg.~\ref{alg:fsmAlignments} is adapted to deal with reduced traces.
\begin{inparaenum}[]
	\item First, the cost function in Def.~\ref{def:CostFunction} is modified
	to prioritise finding repeatable sequences in the process model.
	\item Second, for improving the computation time, a binary search is implemented for traces reduced to the same reduced trace. 
\end{inparaenum}

\subsubsection{Cost function}

The cost function is modified to consider the amount of reduced tandem repeats. Specifically, even though several traces can have the same reduced trace, their alignment with a path in the reachability graph can have different costs. Consider the case when an element in a tandem repeat needs to be hidden ($\lhide$), and this hiding operation is required in every repetition of the element. Thus, the more it is repeated in a trace (the higher the reduction factor in the reduced trace), the higher the cost for the computed alignment. 
The cost of an alignment involving a reduced trace needs to consider different cases: if a synchronization does not involve an element in a tandem repeat, then the cost is the usual (0 for $\match$ and 1 otherwise); whereas if it involves an element in a tandem repeat, then it is necessary to determine if the element is loopable in the reachability graph and can be synchronized in all repetitions. 

Definition~\ref{def:adjustedCostF} shows the modified cost function. By the abuse of notation, we use $\seq{\start}{end}$ to create a sequence of numbers from $\start$ to $end$ with an increment of 1. Given a sequence $t$, we use $\pref(t,i)$ to refer to the prefix of sequence $t$ from position 1 to $i$, and $\suffix(t,i)$ to refer to the suffix of sequence $t$ from position $i$ to $|t|$. Let $\tpos$ be a function relating each index $i$ of an alignment, where $1 \leq i \leq \left|\alignment\right|$, to the trace position that has been aligned up to, then $\tpos(\alignment,i)=\left|\{\sync\in\pref(\alignment,i)\mid\op(\sync)\neq\rhide\}\right|$. 
For the other direction, we define a function $\alignpos$ that given a trace position $\secondPos$ returns the exact position in an alignment $\alignment$ where the trace label is aligned, i.e. $\alignpos(\alignment,\secondPos)=min\{1\leq \position \leq \left|\alignment\right| : \tpos(\alignment,\position)=\secondPos\}$. 
For assigning the additional cost, we use function $\additionalCostF$ (Def.~\ref{def:reduction}) relating each trace index of a tandem repeat to the number of reduced repetitions. 
We complete the definition of this function by relating all remaining trace indices to 0, i.e. $\additionalCostF \leftarrow \additionalCostF \cup \{\secondPos\rightarrow0\mid 1 \leq \secondPos \leq \left|\redTr\right|\land\secondPos\notin\dom(\additionalCostF)\}$ for every $\redTr\in\redLog$. 

So far the cost function for reduced alignments assigns a value of $1+\additionalCostF(\tpos(\alignment,\position))$ to all synchronizations that are hide operations and 0, otherwise. 
For each complementary pair of positions of a tandem repeat, an additional cost is assigned at most once, even if both labels are aligned with a $\lhide$ operation. This ensures that hiding all labels of a tandem repeat with $\lhide$ operations results in the same cost as if all labels in the extended tandem repeat where hidden with the traditional cost function from Def.~\ref{def:CostFunction}. For implementing this idea, we rely on the complement function $\complemnt$ from Def.~\ref{def:reduction} that links each trace position to its complementary position of its tandem repeat. We extend this function to also apply to alignments (denoted as $\aligncomplement(\alignment,\position)$), which given a position $\position$ in alignment $\alignment$, first retrieves its trace position with function $\tpos$, second retrieves the complementary trace position with function $\complemnt$ and finally retrieves the position of the complement in the alignment with function $\alignpos$, i.e. $\aligncomplement(\alignment,\position) = \alignpos(\complemnt(\tpos(\alignment,\position)))$.
We cover the case of two $\lhide$ operations for two complementary trace labels by only altering the cost of the element in the second copy, i.e. where the trace position is larger than the complement position ($\aligncomplement(\alignment,\position)\!<\!\tpos(\alignment,\position)$). 

\newpage
If both the operation at position $\position$ and at the complementary position in the alignment $\aligncomplement(\alignment,\position)$ are $\lhide$, then the cost of the alignment position $\position$ is reduced to one. 
This means that the penalty cost of $\additionalCostF(\tpos(\alignment,\position))$ is assigned only to the $\lhide$ operation of the trace label in the first copy of the tandem repeat and not the label in the second copy.
Even though we define the cost function based on standard costs, it is possible to set individual costs to activities based on domain knowledge. Given that we only introduce multipliers based on the number of reduced iterations of tandem repeats, the weighting of domain knowledge would be preserved.
Next, we can introduce the cost function of a reduced alignment. 

\begin{definition}[Cost function of a reduced alignment]\label{def:adjustedCostF}
Given an alignment $\alignment$, function $\additionalCostF$ for relating trace indices to the number of reduced repetitions, function $\complemnt$ that links each trace position of a tandem repeat to its complement,
 we define the cost of a position $\position$ within $\alignment$ with function $\syncCost$:
\[
\syncCost(\alignment,\additionalCostF,\complemnt,\position)=
\begin{cases}
    1,& \text{if }  \additionalCostF(\tpos(\alignment,\position))\!\geq 1\!\land \aligncomplement(\alignment,\position)\!<\!\tpos(\alignment,\position)\\
    & \text{and }\op(\alignment[\position])=\lhide \land \op(\alignment[\aligncomplement(\alignment,\position)])=\lhide\\
    1+\additionalCostF(\tpos(\alignment,\position)),& \text{else if } \op(\alignment[\position])=\rhide \lor \op(\alignment[\position])=\lhide\\
    0,& \text{else if } \op(\alignment[\position])=\match
\end{cases}
\]
The total cost $\redCostF$ for a reduced alignment is the sum of $\syncCost$ for each element in the alignment $\alignment$
\[
    \redCostF(\alignment,\additionalCostF, \complemnt) = \sum_{\position\in\seq{1}{\left|\alignment\right|}} \syncCost(\alignment,\additionalCostF, \complemnt,\position)
\]
\end{definition}

In comparison to the standard cost function (Def.~\ref{def:CostFunction}), the cost function of a reduced alignment is an over-approximation of the cost of the original alignment with an optimal cost for complete traces.
Please note that this only holds for alignments of complete traces and not for any partial alignments of trace prefixes. 
The cost function for reduced alignments penalises the cases when both kept repetitions are not aligned with $\match$ operations. This additional cost implies that all hidden repetitions of that label need to be aligned with a hide operation once the alignment is extended. We give the relation of the two cost functions next.

\begin{lemma}\label{lem:costFrelation}
Let $\dfa$ be a DAFSA, $\reachGraph$ be a reachability graph, $t$ be a trace and $t' = (\redTrace,\additionalCostF,\redReps,\complemnt)$ be a trace reduction of $t$. The cost of alignment $\alignment$ of $t$ is always lower or equal than the cost of the reduced alignment $\alignment_r$ of $t'$, i.e. $\costF(\alignment)\leq\redCostF(\alignment_r,\additionalCostF,\complemnt)$, if $\alignment$ is proper for trace $t$ and optimal, i.e. $\nexists \alignment':\costF(\alignment')<\costF(\alignment)$, and if $\alignment_r$ is proper for the reduced trace $\redTrace$ and optimal, i.e. $\nexists\alignment_r':\redCostF(\alignment_r',\additionalCostF,\complemnt)<\redCostF(\alignment_r,\additionalCostF,\complemnt)$.
\end{lemma}

The proof for Lemma~\ref{lem:costFrelation} is given in~\ref{app:proof_costFrelation}.

Figure~\ref{fig:runningExampleAlignment} shows an alignment for the reduced trace (3) in Fig.~\ref{fig:runningExampleReduction} and the computation for the cost function with all its auxiliary functions. The alignment can match all the trace labels of the reduced trace, but has to hide label $E$ with a $\rhide$ operation when traversing the loop $B,D,E,F$ in the process model. The trace position does not move during the $\rhide$ synchronization at alignment position 4, i.e. function $\tpos(\alignment,\position)$ is still at position 3. Since the alignment does not contain any $\lhide$ synchronizations, the complement functions do not influence the cost of this alignment. One point of interest, however, is that the trace complement $\complemnt(\tpos(\alignment,\position)))$ of position 2 points to trace position 5 while the alignment complement $\aligncomplement(\alignment,\position)$ points to the alignment position 6 (because $\rhide(E)$ was aligned in between). Since one repetition has been reduced ($\additionalCostF(\tpos(\alignment,\position))$), the cost for each of the two $\rhide$ synchronizations is 2 because they are contained in a tandem repeat ($\syncCost(\alignment,\additionalCostF,\complemnt,\position)$). The cost of the reduced alignment is 4. Please note that this cost overestimates the optimal cost of the extended alignment, which is 3 for the sequence $\langle\match(B),\match(D),\rhide(E),\match(F)\rangle$ inserted after position 5 and before 6. However, this does not pose a problem since this fact only discourages on overly use of $\rhide$ synchronizations to construct large repetitive sequences while the extension algorithm (presented later) properly constructs the extended alignment with the correct cost.
\vspace{.5\baselineskip}
\begin{figure*}[h]
\centering
\resizebox{1\textwidth}{!}{
 \tikzstyle{row} = [draw, rectangle, fill=white, align=left, minimum height=12mm, text width={width("Synchronization cost "}, font=\normalsize]
 \tikzstyle{box} = [draw, rectangle, fill=white, align=center, minimum height=12mm, text width={width("$t MT(A)  t$")}, font=\normalsize]
 \tikzstyle{result} = [draw, rectangle, fill=white, align=center, minimum height=12mm, text width={\textwidth*1.1}, font=\normalsize]
 \begin{tikzpicture}[>=stealth', node distance=-0.3pt]
  \node[row] (variable1) {Alignment pos $\position$};
  \node[box, right=of variable1] (val1) {1};
  \node[box, right=of val1] (val2) {2};
  \node[box, right=of val2] (val3) {3};
  \node[box, right=of val3] (val4) {4};
  \node[box, right=of val4] (val5) {5};
  \node[box, right=of val5] (val6) {6};
  \node[box, right=of val6] (val7) {7}; 
  \node[box, right=of val7] (val8) {8};
  \node[box, right=of val8] (val9) {9};
  \node[box, right=of val9] (val10) {10};
  \node[box, right=of val10] (val11) {11};
  \node[row, below=of variable1] (variable2) {Alignment $\alignment$};
  \node[box, right= of variable2] (val12) {$\match(A)$};
  \node[box, right= of val12] (val13) {\colorbox{gray!20}{$\match(B)$}};
  \node[box, right= of val13] (val14) {\colorbox{gray!20}{$\match(D)$}};
  \node[box, right= of val14] (val15) {$\rhide(E)$};
  \node[box, right= of val15] (val16) {\colorbox{gray!20}{$\match(F)$}};
  \node[box, right= of val16] (val17) {\colorbox{gray!20}{$\match(B)$}};
  \node[box, right= of val17] (val18) {\colorbox{gray!20}{$\match(D)$}};
  \node[box, right= of val18] (val19) {$\rhide(E)$};
  \node[box, right= of val19] (val20) {\colorbox{gray!20}{$\match(F)$}};
  \node[box, right= of val20] (val21) {$\match(B)$};   
  \node[box, right= of val21] (val22) {$\match(D)$};
  \node[row,  below=of variable2] (variable3) {Reduced trace $\redTr$};
  \node[box, right= of variable3] (val23) {$A$};
  \node[box, right= of val23] (val24) {\colorbox{gray!20}{$B$}};
  \node[box, right= of val24] (val25) {\colorbox{gray!20}{$D$}};
  \node[box, right= of val25] (val26) {};
  \node[box, right= of val26] (val27) {\colorbox{gray!20}{$F$}};
  \node[box, right= of val27] (val28) {\colorbox{gray!20}{$B$}};
  \node[box, right= of val28] (val29) {\colorbox{gray!20}{$D$}};
  \node[box, right= of val29] (val30) {};
  \node[box, right= of val30] (val31) {\colorbox{gray!20}{$F$}};
  \node[box, right= of val31] (val32) {$B$};   
  \node[box, right= of val32] (val33) {$D$};
  \node[row,  below=of variable3] (variable4) {Trace position $\tpos(\alignment,\position)$};
  \node[box, right= of variable4] (val34) {1};
  \node[box, right= of val34] (val35) {2};
  \node[box, right= of val35] (val36) {3};
  \node[box, right= of val36] (val37) {3};
  \node[box, right= of val37] (val38) {4};
  \node[box, right= of val38] (val39) {5};
  \node[box, right= of val39] (val40) {6};
  \node[box, right= of val40] (val41) {6};
  \node[box, right= of val41] (val42) {7};
  \node[box, right= of val42] (val43) {8};   
  \node[box, right= of val43] (val44) {9};
  \node[row,  below=of variable4] (variable5) {Trace complement $\complemnt(\tpos(\alignment,\position)))$};
  \node[box, right= of variable5] (val45) {};
  \node[box, right= of val45] (val46) {5};
  \node[box, right= of val46] (val47) {6};
  \node[box, right= of val47] (val48) {};
  \node[box, right= of val48] (val49) {7};
  \node[box, right= of val49] (val50) {2};
  \node[box, right= of val50] (val51) {3};
  \node[box, right= of val51] (val52) {};
  \node[box, right= of val52] (val53) {4};
  \node[box, right= of val53] (val54) {};   
  \node[box, right= of val54] (val55) {};
  \node[row,  below=of variable5] (variable6) {Alignment complement $\aligncomplement(\alignment,\position)$};
  \node[box, right= of variable6] (val56) {};
  \node[box, right= of val56] (val57) {6};
  \node[box, right= of val57] (val58) {7};
  \node[box, right= of val58] (val59) {};
  \node[box, right= of val59] (val60) {9};
  \node[box, right= of val60] (val61) {2};
  \node[box, right= of val61] (val62) {3};
  \node[box, right= of val62] (val63) {};
  \node[box, right= of val63] (val64) {5};
  \node[box, right= of val64] (val65) {};   
  \node[box, right= of val65] (val66) {};
  \node[row,  below=of variable6] (variable7) {Additional cost $\additionalCostF(\tpos(\alignment,\position))$};
  \node[box, right= of variable7] (val67) {0};
  \node[box, right= of val67] (val68) {1};
  \node[box, right= of val68] (val69) {1};
  \node[box, right= of val69] (val70) {1};
  \node[box, right= of val70] (val71) {1};
  \node[box, right= of val71] (val72) {1};
  \node[box, right= of val72] (val73) {1};
  \node[box, right= of val73] (val74) {1};
  \node[box, right= of val74] (val75) {1};
  \node[box, right= of val75] (val76) {0};   
  \node[box, right= of val76] (val77) {0};
  \node[row,  below=of variable7] (variable7) {Synchronization cost $\syncCost(\alignment,\additionalCostF,\complemnt,\position)$};
  \node[box, right= of variable7] (val67) {0};
  \node[box, right= of val67] (val68) {0};
  \node[box, right= of val68] (val69) {0};
  \node[box, right= of val69] (val70) {2};
  \node[box, right= of val70] (val71) {0};
  \node[box, right= of val71] (val72) {0};
  \node[box, right= of val72] (val73) {0};
  \node[box, right= of val73] (val74) {2};
  \node[box, right= of val74] (val75) {0};
  \node[box, right= of val75] (val76) {0};   
  \node[box, right= of val76] (val77) {0};
  \node[row,  below=of variable7] (variable7) {Alignment cost $\redCostF(\alignment,\additionalCostF, \complemnt)$};
  \node[result, right= of variable7] (res) {4};
  \end{tikzpicture}
  }
 \caption{Cost of the reduced alignment of trace (3) from the running example.}\label{fig:runningExampleAlignment}
\end{figure*}

Different from other approaches, this work uses Dijkstra algorithm to find optimal alignments instead of an $A^*$-search as other approaches. 
The adaptation of this work to an $A^*$-search is left for future work.

\newpage
\subsubsection{Binary search}
As mentioned previously, several traces of an event log can be reduced to the trace when they only differ in the number of repetitions of reduced tandem repeats. Computing an alignment for these reduced traces then only differs in the additional cost assigned by the adjusted cost function, which is the number of repetitions minus two. We can use this property to further speed up the alignment computation as follows: the alignments are computed only for the reduced traces with the lowest and highest numbers of reduced repetitions. If the alignments are the same for both, then all other traces with different repetitions share the same alignment. This relies on the implications of the adjusted cost function from Def.~\ref{def:adjustedCostF}. Specifically, two matches for the same label in the two maintained copies of the reduced tandem repeat imply a loop in the reachability graph, and it can be repeated for any number of reduced repetitions; whereas an $\lhide$ synchronization for a label in one repetition implies it can not be matched in the reduced repetitions. 
If the reduced traces with the smallest and largest numbers of repetitions are not the same, then the tandem repeat has not been aligned to the same loop, and the reduced traces with a number of repetitions in between the two (smallest and largest numbers of repetitions) will not share their alignments. 
In this section, we propose to search for these intervals with a binary search to potentially avoid the computation of some alignments. In a worst case scenario no intervals can be found with the search and every alignment needs to be computed once for each reduced trace.

The binary search starts by taking all original traces that share the same reduced trace, and ordering them in an ascending order with respect to the total number of reduced labels ($\redReps$ from Def.~\ref{def:reduction}), which will define an interval with the reduced traces with lowest and highest number of reduced labels on the extremes. The binary search proceeds by computing the alignments for the reduced traces, it starts by taking the two reduced traces with the lowest and highest number of reduced labels.
The search stops when the alignments for both -- lowest and highest reduced labels -- are equal (i.e. involve the same synchronizations) and, if there is any reduced traced between them w.r.t. the order, then it will get the same alignment.
In case the alignments are not equal, the search continues by splitting the interval into two, investigating one interval from the lowest value of $\redReps$ to the average and one interval from the average to the highest value of $\redReps$ until all alignments have been computed (either implicitly as part of an interval or as explicitly as a border of an interval).

\newpage
In the running example, traces (3) to (5) in Fig.~\ref{fig:runningExampleReduction} are an interval for the binary search as their reduced traces are the same.
The traces are already sorted according to $\redReps$, next the alignments are computed for traces (3) and (5) with the lowest and highest numbers of $\redReps$, respectively. Both reduced traces lead to the same alignment as reported in Fig.~\ref{fig:runningExampleAlignment} and the computation of the alignment for trace (4) can be omitted. 

The binary search is described in Alg.~\ref{alg:BinarySearchAlignments}, it starts by sorting all original trace reductions for a given reduced trace according to their overall number of reduced labels $\redReps$. Please note that we use $\sortAsc x$ to formalize sorting a set into a sequence by using the order of variable $x$ in ascending order. We start with the largest interval from the minimum to the maximum number of repetitions. Then we calculate a reduced alignment for the lower and one for the upper border and store them in a function $\alignments$ relating trace reductions to alignments of reduced traces (to prevent re-computing alignments when the interval needs to be split). If the alignment of the lower equals the alignment of the upper border, then all intermediate trace reductions relate to the same reduced alignment. Otherwise, the binary search continues with the two new intervals, one from the lower to the average and another from the average to the upper number of reduced labels. This binary search continues until all open intervals have been investigated, which in the worst-case computes one reduced alignment for each trace reduction. For the function align, we refer to \cite{s-comps}. In this article, we use the adjusted cost function according to Def.~\ref{def:adjustedCostF}.

\setlength{\textfloatsep}{1\baselineskip}

\begin{algorithm}[h]
{
    \SetKwInOut{Input}{input}
    \SetKwProg{Fn}{Function}{}{}
    \Input{Reduced event log $\redLog$; Reductions $\reductions$;  Reduced DAFSA $\redDafsa$; Reachability Graph $\rg$}
    $\alignments \leftarrow \{\}$\;
    \For{$t\in\redLog$}
    {
        $\trReductions\leftarrow\langle(\redTrace,\additionalCostF,\redReps,\complemnt)\in\val(\reductions)\mid\redTrace=t\mkern6mu\land\sortAsc\redReps\rangle$\;
        $\pairs \leftarrow \{(1,\left|\trReductions\right|)\}$\;        
        \While{$\pairs\neq\varnothing$}{        
            $(\lowerB,\upperB) \leftarrow $ remove an element from $\pairs$\;
            \uIf{$\alignments(\trReductions(\lowerB))=\perp)$}{$\alignments \leftarrow \alignments \cup \{\trReductions(\lowerB)\rightarrow \text{align}(\trReductions(\lowerB),\redDafsa,\reachGraph)\}$\;}            
            \uIf{$\alignments(\trReductions(\upperB))=\perp)$}{$\alignments \leftarrow \alignments \cup \{\trReductions(\upperB)\rightarrow \text{align}(\trReductions(\upperB),\redDafsa,\reachGraph)\}$\;}
            \uIf{$\alignments(\trReductions(\lowerB))=\alignments(\trReductions(\upperB))$}
            {
                \lFor{$i\in\seq{\lowerB}{\upperB}$}{$\alignments \leftarrow \alignments \cup \{\trReductions(i) \rightarrow \alignments(\trReductions(\lowerB))\}$}               
            }
            \lElse{$\pairs \leftarrow \pairs \cup \{(\lowerB,\myfloor{(\lowerB + \upperB) / 2}),(\myceil{(\lowerB + \upperB) / 2},\upperB)\}$}
        }  
    }
    \Return{$\alignments$}\;
    \Fn{align$((t,\additionalCostF,\redReps,\complemnt),\dafsa, \reachGraph$)}{
        $\open \leftarrow \{((\dfaSource,\rgSource,\langle\rangle), \redCostF(\langle\rangle,\additionalCostF,\complemnt))\}$\;
        $\closed\leftarrow\{\}$\;
        \While{$\open \neq \varnothing$}
        {
            $\actN \leftarrow $ remove $(n_{\dafsa},n_{\reachGraph},\alignment)$ from $\open\mid\nexists (n',\rho')$ with $\rho'<\redCostF(\alignment,\additionalCostF,\complemnt)$\;
           \uIf{$\closed(n_{\dafsa},n_{\reachGraph})=\perp\lor\closed(n_{\dafsa},n_{\reachGraph})\geq\redCostF(\alignment,\additionalCostF,\complemnt)$}{$\closed \leftarrow \closed \cup \{(n_{\dafsa},n_{\reachGraph})\rightarrow\redCostF(\alignment,\additionalCostF,\complemnt)\}$\;}
           \lElse{Continue} 
           \lIf{$n_{\dafsa} \in \dfaFinStates \land n_{\reachGraph} \in \rgFinStates \land \lbl(\filterSyncDFA)=t$\label{ln:condOptimality}}{\Return{$\alignment$}}
           \Else
           {
                
                $\out_{\dafsa} \leftarrow n_{\dafsa}\shift \mid \lbl(\out_{\dafsa})=t(\left|\filterSyncDFA(\epsilon)\right|+1)$\;
                
                $\open \leftarrow \open \cup \{(\tgt(\out_{\dafsa}),n_{\reachGraph},\alignment\tieconcat(\lhide,\out_{\dafsa},\perp))\}$\;
                \lFor{$\out_{\reachGraph}\in n_{\reachGraph}\shift\mid\lbl(\out_{\reachGraph})=\lbl(\out_{\dafsa})$}
                {$\open\leftarrow\open\cup\{(\tgt(\out_{\dafsa}),\tgt(\out_{\reachGraph}),\alignment\tieconcat(\match,\out_{\dafsa},\out_{\reachGraph}))\}$}                
                
                \lFor{$\out_{\reachGraph}\in n_{\reachGraph}\shift$}{$\open\leftarrow\open\cup\{(n_{\dafsa},\tgt(\out{\reachGraph}),\alignment\tieconcat(\rhide,\perp,\out_{\reachGraph}))\}$}
            }
        }
    }
    \caption{Binary search for computing reduced alignments}\label{alg:BinarySearchAlignments}    
}
\end{algorithm}

\newpage
\subsection{Extending Reduced Trace Alignments}

This subsection describes how reduced alignments are extended to full and proper alignments, which represent the original traces. Every tandem repeat compressed during the trace reduction step is inserted back into the reduced alignment. They are inserted between the synchronizations of the two copies of the tandem repeats preserved in the reduced trace. In order to be considered a valid alignment, the insertion of the 
missing iterations into the reduced alignment shall also form a valid path in the reachability graph.

The alignment of a reduced trace extends each of the token repeats in order, as observed in the trace, from right to left. 
The start of the first and second copy of a tandem repeat to expand are $\alignTRstart(\alignment,\position)$ and $\alignTRstartSec(\alignment,\position)$, respectively.
Then, the middle copy is inserted between the first and second aligned copies of the tandem. If several repetitions of the tandem repeat have been reduced,
the middle copy is inserted repeatedly in between the two copies until the original trace is reconstructed. 
For finding the middle copy, the leftmost alignment position $\secondPos$ in the first tandem repeat copy is identified, where both $\secondPos$ and its complement in the second copy $\aligncomplement(\alignment,\secondPos)$ have been aligned with a $\match$ operation. 
Since the underlying system net is uniquely labelled, we know that all synchronizations in between these two aligned positions necessarily form a loop, and this sequence can be repeatedly executed on the process model. 
We also know that the sequence involves all trace labels of the tandem repeat since the complement function denotes the first and second occurrence of one specific label of the repeat type and all other labels of the repeat type need to occur in between.
However, the trace labels of the identified repeatable sequence might not be in the order of the tandem repeat. The middle copy is then constructed by rotating the repeatable sequence until the first label of the tandem repeat is at the first position. This is achieved by first taking the sequence from the start of the second copy ($\alignTRstartSec(\alignment,\position)$) up to the end of the repeatable sequence in the second copy ($\aligncomplement(\alignment,\secondPos)$) and then adding the sequence after $\secondPos$ from the first copy up to the end of the first copy ($\alignTRstartSec(\alignment,\position)\!-\!1$).  
If no repeatable sequence can be identified, we construct the middle copy consisting of all $\lhide$ and $\match$ synchronizations from the first copy and change their operation to $\lhide$ to represent all trace labels of the tandem repeat but to not change the path in the process model.
After inserting the middle copies, we update functions $\complemnt$ and $\additionalCostF$ by removing the reference to the extended tandem repeat and move position $\position$ left to one position before the start of the first copy.
The definition of an extended alignment follows next:

\vspace{-.25\baselineskip}
\begin{definition}[Extended alignment]\label{def:extendedAlignment}
Let $\extension = (\alignment,\complemnt,\additionalCostF,\position)$ be a - possibly extended - alignment and the trace position of $\position$ is the last position of a tandem repeat to be extended, i.e. $\tpos(\alignment,\position)\in\dom(\complemnt)$. The two copies of this tandem repeat $\firstCopy$ and $\secondCopy$ can be identified as two sequences in $\alignment$ as follows:
\begin{compactitem}
    \item $\firstCopy = \alignment[\seq{\alignTRstart(\alignment,\position)}{\alignTRstartSec(\alignment,\position)\!-\!1}]$\label{def:firstCopy}
    \item $\secondCopy = \alignment[\seq{\alignTRstartSec(\alignment,\position)}{\position}]$\label{def:secondCopy}
\end{compactitem}
A sequence within $\firstCopy$ and $\secondCopy$ is repeatable by a single position $\thirdPos$ if both alignment positions $\thirdPos$ and its complement $\aligncomplement(\alignment,\thirdPos)$ are $\match$ operations. The left most position of a repeating sequence $\secondPos$ is defined as:
\begin{compactitem}
\item $\secondPos\shorteq\minF\{\thirdPos\in\seq{\alignTRstart(\alignment,\position)}{\aligncomplement(\alignment,\position)} \mid \op(\alignment[\thirdPos])\shorteq\match
\land\op(\alignment[\aligncomplement(\alignment,\thirdPos)])\shorteq\match\}
$
\end{compactitem}
If there exists a repeating sequence ($\secondPos>0$), a middle copy for the extension can be defined as follows:
\begin{compactitem}
\item $
\copyExtend=
\begin{cases}
\alignment[\seq{\alignTRstartSec(\alignment,\position)}{\aligncomplement(\alignment,\secondPos)}]\oplus\alignment[\seq{j\!+\!1}{\alignTRstartSec(\alignment,\position)\!-\!1}], & \text{if } \secondPos>0\\
\langle(\lhide,\fnSyncDFA(\sync),\perp) \mid \sync\in\firstCopy\land\op(\sync)\neq\rhide\rangle, & \text{Otherwise}
\end{cases}
$
\end{compactitem}
Then the extended alignment is $\applyExtension(\extension) = (\extendedAlignment,\complemnt',\additionalCostF',\position')$, where: 
\begin{compactitem}    
\item  $\extendedAlignment\leftarrow\pref(\alignment,\alignTRstart(\alignment,\position)\!-\!1) \oplus \firstCopy\oplus \langle\copyExtend\rangle^{\additionalCostF(\tpos(\alignment,\position))}\oplus\secondCopy\oplus\suff(\alignment,\position)$
\item $\complemnt'\leftarrow\complemnt\setminus\{\thirdPos\rightarrow\complemnt(\thirdPos),\complemnt(\thirdPos)\rightarrow\thirdPos\mid\thirdPos\in\seq{\tpos(\alignment,\alignTRstart(\alignment,\position))}{\complemnt(\tpos(\alignment,\position))}\}$
\item $\additionalCostF'\leftarrow\additionalCostF\oplus\{\thirdPos\rightarrow0\mid\thirdPos\in\seq{\tpos(\alignment,\alignTRstart(\alignment,\position))}{\tpos(\alignment,\position)}\}$
\item $\position' \leftarrow \alignTRstart(\alignment,\position)-1$
\end{compactitem}
\end{definition}

\vspace{-.25\baselineskip}
Algorithm~\ref{alg:ExtendRedAlignments} describes the procedure to fully extend the reduced alignments. The aim is to create a mapping $\extendedAlignments$ that maps each original trace of the event log to its full alignment. Each reduced alignment is extended by moving from right to left with a counter $\position$, and extending each encountered tandem repeat by applying the extension in Def.~\ref{def:extendedAlignment}. If no tandem repeat is to be extended then we decrease the position by 1. This procedure gets repeated until all tandem repeats have been extended and the counter $\position$ reaches the start of the alignment.


\begin{algorithm}[!h]
{
    \SetKwInOut{Input}{input}
    \Input{Event log $\logL$; Reductions $\reductions$;  Reduced alignments $\alignments$;}
    $\extendedAlignments \leftarrow \{\}$\;
    \For{$t\in\logL$}
    {
        $\trReductions = (\redTrace,\additionalCostF,\redReps,\complemnt) \leftarrow \reductions(t)$\;
        $\alignment\leftarrow\alignments(\trReductions)$\;
        $\position\leftarrow \left|\alignment\right|$\;
        $\extension \leftarrow (\alignment,\complemnt,\additionalCostF,\position)$\;\label{line:initExtension}
        \While{$\position\geq1$}
        {
            Set $\extension \leftarrow \applyExtension(\extension)$\;
            \lIf{$\position=\extension[4]$}{Decrease $\position$ by 1}
            \lElse{Set $\position\leftarrow\extension[4]$}
            Set $\extension[4] \leftarrow \position$\;
        }
        $\extendedAlignments \leftarrow \extendedAlignments \cup \{t\rightarrow\extension[1]\}$\;    
    }
    \Return{$\extendedAlignments$}\;
    \caption{Extending reduced alignments}\label{alg:ExtendRedAlignments}       
}
\end{algorithm}

\newpage
Figure~\ref{fig:runningExampleExtension} shows the extension of the reduced alignment $\alignment$ for the reduced trace (3) from Fig.~\ref{fig:runningExampleAlignment}. Algorithm~\ref{alg:ExtendRedAlignments} moves backwards from alignment position 11 to position 9, which is the last position of the reduced tandem repeat $B,D,F$. Next, the positions of the two copies of the tandem repeat (both highlighted with grey background colour in the figure) are identified in the alignment. The first copy starts at position $\alignTRstart(\alignment,\position) = 2$ and continues up to position $\aligncomplement(\alignment,\position)=5$, while the second copy starts at position $\alignTRstartSec(\alignment,\position) = 6$ and ends at position $\position=9$. A repeatable sequence can be identified at position $\secondPos=2$ since both position 2 and its complementary position $\aligncomplement(\alignment,\secondPos) = 6$ are aligned with $\match$ operations. The middle copy then can be constructed from the prefix of the second copy up to and including position 6, which is $\match(B)$, and then adding the suffix from the first copy after position 2, i.e. $\match(D),\rhide(E),\match(F)$. This middle copy is then inserted after the end of the first copy at position 5 to extend the reduced alignment.

\vspace{0.5\baselineskip}
\begin{figure*}[htbp]
\centering\scriptsize
\resizebox{1\textwidth}{!}{
 \tikzstyle{row} = [draw, rectangle, fill=white, align=left, minimum height=13mm, text width={width("ttttttt Repeatable ttttttt"}, font=\large]
 \tikzstyle{box} = [draw, rectangle, fill=white, align=center, minimum height=13mm, text width={width("$t TRs,2(e,i)  t$")}, font=\large]
 \tikzstyle{result} = [draw, rectangle, fill=white, align=center, minimum height=13mm, text width={\textwidth*1.366}, font=\normalsize]
 \begin{tikzpicture}[>=stealth', node distance=-0.3pt]
  \node[row] (variable1) {Position $\position$};
  \node[box, right=of variable1] (val1) {1};
  \node[box, right=of val1] (val2) {2};
  \node[box, right=of val2] (val3) {3};
  \node[box, right=of val3] (val4) {4};
  \node[box, right=of val4] (val5) {5};
  \node[box, right=of val5] (val75) {};
  \node[box, right=of val75] (val76) {};
  \node[box, right=of val76] (val77) {};
  \node[box, right=of val77] (val78) {};
  \node[box, right=of val78] (val6) {6};
  \node[box, right=of val6] (val7) {7}; 
  \node[box, right=of val7] (val8) {8};
  \node[box, right=of val8] (val9) {9};
  \node[box, right=of val9] (val10) {10};
  \node[box, right=of val10] (val11) {11};
  \node[row, below=of variable1] (variable2) {Reduced\\alignment $\alignment$};
  \node[box, right= of variable2] (val12) {$\match(A)$};
  \node[box, right= of val12] (val13) {\colorbox{gray!20}{$\match(B)$}};
  \node[box, right= of val13] (val14) {\colorbox{gray!20}{$\match(D)$}};
  \node[box, right= of val14] (val15) {$\rhide(E)$};
  \node[box, right= of val15] (val16) {\colorbox{gray!20}{$\match(F)$}};
  \node[box, right=of val16] (val79) {};
  \node[box, right=of val79] (val80) {};
  \node[box, right=of val80] (val81) {};
  \node[box, right=of val81] (val82) {};
  \node[box, right= of val82] (val17) {\colorbox{gray!20}{$\match(B)$}};
  \node[box, right= of val17] (val18) {\colorbox{gray!20}{$\match(D)$}};
  \node[box, right= of val18] (val19) {$\rhide(E)$};
  \node[box, right= of val19] (val20) {\colorbox{gray!20}{$\match(F)$}};
  \node[box, right= of val20] (val21) {$\match(B)$};   
  \node[box, right= of val21] (val22) {$\match(D)$};
  \node[row, below=of variable2] (variable3) {Functions};
  \node[box, right=of variable3] (val23) {};
  \node[box, right=of val23] (val24) {$\alignTRstart(\alignment,\position)$};
  \node[box, right=of val24] (val25) {};
  \node[box, right=of val25] (val26) {};
  \node[box, right=of val26] (val27) {$\aligncomplement(\alignment,\position)$};
  \node[box, right=of val27] (val83) {};
  \node[box, right=of val83] (val84) {};
  \node[box, right=of val84] (val85) {};
  \node[box, right=of val85] (val86) {};
  \node[box, right=of val86] (val28) {$\alignTRstartSec(\alignment,\position)$};
  \node[box, right=of val28] (val29) {}; 
  \node[box, right=of val29] (val30) {};
  \node[box, right=of val30] (val31) {$\position$};
  \node[box, right=of val31] (val32) {};
  \node[box, right=of val32] (val33) {};
  \node[row, below=of variable3] (variable4) {Repeatable};
  \node[box, right=of variable4] (val34) {};
  \node[box, right=of val34] (val35) {$\secondPos$};
  \node[box, right=of val35] (val36) {};
  \node[box, right=of val36] (val37) {};
  \node[box, right=of val37] (val38) {};
  \node[box, right=of val38] (val87) {};
  \node[box, right=of val87] (val88) {};
  \node[box, right=of val88] (val89) {};
  \node[box, right=of val89] (val90) {};
  \node[box, right=of val90] (val39) {$\aligncomplement(\alignment,\secondPos)$};
  \node[box, right=of val39] (val40) {}; 
  \node[box, right=of val40] (val41) {};
  \node[box, right=of val41] (val42) {};
  \node[box, right=of val42] (val43) {};
  \node[box, right=of val43] (val44) {};
  \node[row, below=of variable4] (variable5) {Middle copy $\copyExtend$};
  \node[box, right=of variable5] (val45) {};
  \node[box, right=of val45] (val46) {};
  \node[box, right=of val46] (val47) {};
  \node[box, right=of val47] (val48) {};
  \node[box, right=of val48] (val49) {};
  \node[box, right= of val49] (val50) {$\match(B)$};
  \node[box, right= of val50] (val51) {$\match(D)$};
  \node[box, right= of val51] (val52) {$\rhide(E)$};
  \node[box, right= of val52] (val53) {$\match(F)$};
  \node[box, right=of val53] (val54) {};
  \node[box, right=of val54] (val55) {}; 
  \node[box, right=of val55] (val56) {};
  \node[box, right=of val56] (val57) {};
  \node[box, right=of val57] (val58) {};
  \node[box, right=of val58] (val59) {};
  \node[row, below=of variable5] (variable6) {Extended\\alignment $\extendedAlignment$};
  \node[box, right= of variable6] (val60) {$\match(A)$};
  \node[box, right= of val60] (val61) {$\match(B)$};
  \node[box, right= of val61] (val62) {$\match(D)$};
  \node[box, right= of val62] (val63) {$\rhide(E)$};
  \node[box, right= of val63] (val64) {$\match(F)$};
  \node[box, right= of val64] (val65) {$\match(B)$};
  \node[box, right= of val65] (val66) {$\match(D)$};
  \node[box, right= of val66] (val67) {$\rhide(E)$};
  \node[box, right= of val67] (val68) {$\match(F)$};
  \node[box, right= of val68] (val69) {$\match(B)$};
  \node[box, right= of val69] (val70) {$\match(D)$};
  \node[box, right= of val70] (val71) {$\rhide(E)$};
  \node[box, right= of val71] (val72) {$\match(F)$};
  \node[box, right= of val72] (val73) {$\match(B)$};   
  \node[box, right= of val73] (val74) {$\match(D)$};
  \end{tikzpicture}
  }
 \caption{Extending the reduced alignment of trace (3) from the running example.}\label{fig:runningExampleExtension}
\end{figure*}

\vspace{0.5\baselineskip}
The following lemma~\ref{lemma:properAlignment} shows that extended alignments are proper. Its proof can be found in the~\ref{lemma:properAlignment:app}. 

\vspace{-.25\baselineskip}
\begin{lemma}[An extended alignment is a proper alignment]\label{lemma:properAlignment}
Given a trace $t\in\logL$, the alignment $\extendedAlignment$, returned by Alg.~\ref{alg:ExtendRedAlignments}, is a proper alignment. 
Thus, the following two properties hold for $\extendedAlignment$:
\begin{compactenum}
    \item\label{prop:one} the labels in the synchronizations related to the DAFSA represent the trace, i.e. $\lbl(\filterSyncDFA(\extendedAlignment))=t$, and
      \item\label{prop:two} the arcs in the synchronizations of the reachability graph form a path in the reachability graph, i.e., for $\mPath=\fnSyncRG(\filterSyncRG(\extendedAlignment))$ holds $\src(\mPath(1)=\rgSource \land \tgt(\mPath(\left|\mPath\right|))\in\rgFinStates \land \forall \position\in\seq{1}{\left|\mPath\right|-1} : \tgt(\mPath(\position))=\src(\mPath(\position+1))$.
\end{compactenum}
\end{lemma}

\newpage
As a result, extended alignments will never under estimate the minimal cost of an alignment. 
In fact, in some cases the cost of an extended alignment will be minimal.
Lemma~\ref{lemma:casesExactCost} formalizes the cases where an extended alignment achieves minimal cost. The proof of this lemma is included in \ref{app:proof_exact_cost}.

\begin{lemma}[Cases of extended alignments with guaranteed minimal cost]\label{lemma:casesExactCost}
Let $t$ be a trace and $\reachGraph$ a reachability graph. The cost of the optimal alignment $\alignment$ and the extended alignment $\reachGraph$ have the same cost: $\costF(\extalignment)=\costF(\alignment)$ in three cases:
\begin{compactenum}
	\item $t$ contains \emph{no tandem repeats},
	\item $t$ contains only tandem repeats $(i,\alpha,k)$ corresponding to one of the following cases:\label{case:exactCostWithTandemRepeats}
	\begin{compactenum}
	    \item no possible matches could be found for the tandem repeat, or
	    \item no corresponding loop in the model is found for the tandem repeat, or
	    \item the tandem repeat could be fully matched to a corresponding loop in the model.
	\end{compactenum}
\end{compactenum}
\end{lemma}

Finally, the worst-case over-approximation of the minimal cost by an extended alignment is less than two times the length of the repeating sequence $\alpha$ 
i.e. $2*(\left|\alpha\right|-1)-1$. This worst-case over-approximation can occur only if the repetitions $k$ of the tandem repeat are greater or equal to the worst case cost, i.e. $k\geq2*(\left|\alpha\right|-1)-1$. For lower numbers of repetitions the worst case over-approximation is bound by $k$. 
Intuitively, the worst-case cost over-approximation occurs if the extended alignment matches one label of the repeating sequence in every repetition and, to achieve that, it needs to hide all other labels of the repeating sequence and all arcs of the reachability graph related to the labels of the repeating sequence (e.g., because they were in a different order than in the repeating sequence). 
An optimal alignment could achieve a better cost than the extended alignment by hiding the matched label of the extended alignment in one repetition while finding one additional match for every other label in the repeating sequence. Thereby, the optimal alignment can avoid one hide of each trace label of the tandem repeat besides the one matched by the extended alignment, and one hide for each corresponding move in the model while hiding the matched label in the extended alignment once. 
If a trace contains multiple reduced tandem repeats, the worst case cost over-approximation of the extended alignment is the sum of the worst-case over-approximations of each tandem repeat since the tandem repeats are aligned independently. 
Lemma~\ref{lemma:ExtAlignmentCostOverApprox} shows the relation of the cost of an extended alignment to an alignment with minimal cost. Its proof can be found in~\ref{app:proof_cost_overApprox}.

\begin{lemma}[Cost over-approximation of an extended alignment]\label{lemma:ExtAlignmentCostOverApprox}
Let $t$ be a trace and $\reachGraph$ a reachability graph. The worst case cost over-approximation of an extended alignment $\extalignment$ of $t$ is $\costF(\alignment)+\sum_{(i,\alpha,k)\in\Delta(rt)} 2*(\left|\alpha\right|-1)-1$, where $\alignment$ is the optimal alignment of $t$ and $rt$ is the reduced trace of $t$.
\end{lemma}

In the running example, trace (6) of the reduced event log from Fig.~\ref{fig:runningExampleReduction} will lead to an alignment with over-approximated cost. Its reduced alignment is $\langle\match(A),\lhide(D),\match(B),\lhide(F),\match(D),\rhide(E),\lhide(B),\\ \match(F),\match(B),\match(C)\rangle$. No label of the tandem repeat is matched in both copies, i.e. $B$ is matched in the first copy while $D$ and $F$ are matched in the second copy. Hence, the extension algorithm can not identify a repeating sequence and the middle copy consisting of $\lhide$ synchronization for all labels of the tandem, i.e. $\langle \lhide(D),\lhide(B),\lhide(F)\rangle$, will be inserted two times in the reduced alignment after $\lhide(F)$ and before $\match(D)$ resulting in a total cost of 10. A minimal alignment would be 
$\langle \match(A),\lhide(D),\match(B),\lhide(F),\match(D),\\ \lhide(B),\rhide(E),\match(F),\lhide(D),\match(B),\lhide(F),\match(D),\lhide(B),\rhide(E),\match(F),\match(B),\match(C) \rangle$ 
with a cost of 8. The difference in cost results from the fact that the minimal alignment can match activities $B$ and $D$ in return of hiding activity $E$ twice in the middle copies of the tandem, which are hidden to the reduced alignment. However, the optimal alignment can also not match any label of the tandem repeat in every repetition, which would have lead to the identification of a repeating sequence for the reduced alignment. The reduced alignment over-approximates, because it assumes that every label needs to be hidden in the middle copies if it can not be matched in every repetition of a tandem repeat. Hence, the proposed technique provides a trade-off between efficiency and accuracy. In the running example, traces (1)-(5) do not lead to over-approximations despite not fully fitting the behavior of the system net.

%% file: tex/scomp_integration.tex
\subsection{Handling Concurrency in Process Models}

The presented technique can be used in combination with 
a decomposition/recomposition technique, such as the one described in~\cite{s-comps} 
to deal with models with parallelism, i.e. sound free-choice workflow nets.
Parallel activities can disrupt repeatable sequences from an alignment position to its complement. While it is possible to filter parallel activities from repeatable sequences with additional effort, we want to propose an alternative approach for using tandem repeat reductions for computing alignments of concurrent process models.
The decomposition/recomposition technique divides an input sound free-choice workflow net into state machine workflow nets that fully represent the behavior of the input net. The input event log is split into sub-logs, one for each state machine workflow net, by filtering out the activities not present in the corresponding net. Then, we can use the tandem-repeats technique presented in this article (see Fig.~\ref{TR-approach}) on each pair of state-machine workflow net and the corresponding sub-log to compute partial alignments with the improvements of the reductions from tandem repeats. In a final step, the partial alignments are recomposed into proper alignments for the original traces with an approximate cost. This recomposition may fail for some traces of the original log when some partial alignments disagree in their operation for a shared activity. In that case, only the traces with recomposition conflicts need to be aligned with an alternative algorithm such as with Alg.~\ref{alg:fsmAlignments}. 
The decomposition/recomposition technique~\cite{s-comps} used in this article is based on S-Components, please refer to the article for detailed explanations of each of the steps outlined above. Note that the technique presented in this paper can be combined with any other decomposition/recomposition techniques for dividing a process model into state machine workflow nets.


\begin{figure}[h]
\vspace{0.5\baselineskip}
\includegraphics[width=\textwidth]{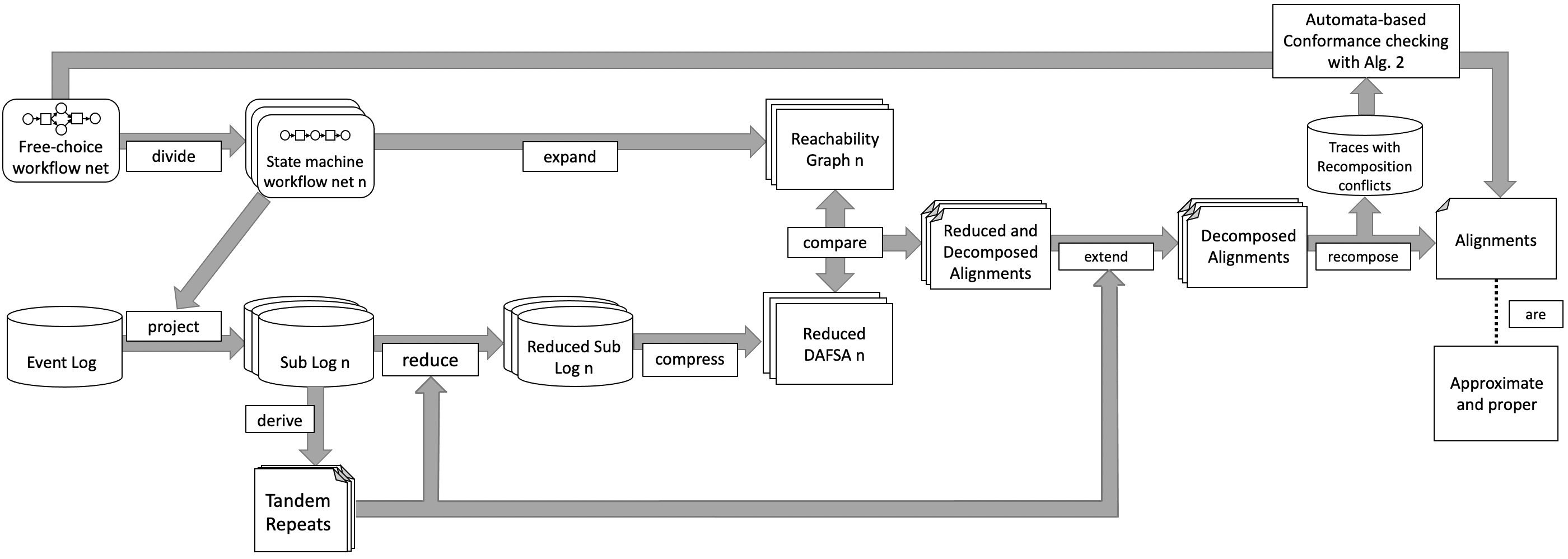}
\vspace{-1\baselineskip}
\caption{%
Integrating tandem repeats with the S-Components approach
} \label{fig:TR-SComp-approach}
\end{figure}

Figure~\ref{fig:TR-SComp-approach} shows the integration between the tandem repeat reductions and S-Component-based approach for the computation of alignments. 
First, 
%
an input sound free-choice workflow net is divided into S-Components, i.e. state machine workflow nets,
and the the sub-logs are created for each S-Component (as specified in~\cite{s-comps}), then given that the S-Components are concurrency-free, we detect and reduce tandem repeats for each sub-log. Thus, Alg.~\ref{alg:BinarySearchAlignments}
 is applied to each pair of S-Component and reduced sub-log to derive reduced and decomposed alignments. These alignments are then extended with Alg.~\ref{alg:ExtendRedAlignments}
and recomposed with the procedure described in~\cite{s-comps}. 
Last, traces with recomposition conflicts are aligned with Alg.~\ref{alg:fsmAlignments}. The resulting alignments are proper and of approximate cost. 

%% file: tex/evaluation.tex
\newpage
\section{Evaluation}\label{sec:evaluation}
We implemented our technique as a standalone open-source command-line tool\footnote{The command line tool is available as \emph{ProConformance 3.0 (automata-based)} at \url{https://apromore.org/platform/tools}; All public logs and models used in the evaluation are available at \url{https://melbourne.figshare.com/articles/dataset/Public_benchmark_data-set_for_Conformance_Checking_in_Process_Mining/8081426};  Source code is available at \url{https://github.com/reissnda/AutomataConformance}} as part of the Apromore process mining environment \cite{Apromore}. Given an event log in XES format and a process model in either BPMN or PNML format, the tool produces various alignment statistics such as fitness and raw fitness costs. Optionally, the tool can output the alignments found or manifest the PSP data structure. It is also possible to specify which extension to the Automata-based approach should be applied, i.e.\
i) base approach without any extension (Automata); ii) with the S-Components extension (SComp); iii) with the S-Components and tandem repeats reduction (TR-SComp); or iv) a hybrid approach that tries to automatically select the most suitable extension based on the characteristics of the input model and log (Hybrid).

Using this tool, we evaluated the time performance and accuracy of our technique in a series of experiments, against two internal baselines (Automata and SComp), and four external baselines. For the external baselines, we chose three exact and one approximate approach for computing trace alignments: 
\begin{inparaenum}[(1)]
    \item the newest version of the trace alignment technique presented in \cite{ILP-Alignment} using the \emph{ILP} marking equation and implemented in the PNetReplayer package of ProM\footnote{Available at \url{http://www.promtools.org}} (ILP);
    \item the trace alignment technique presented in \cite{ILP-Alignment} using the \emph{LP} marking equation and implemented in the PNetReplayer package of ProM\footnote{Available at \url{http://www.promtools.org}} (LP);
    \item the extended version of the trace alignment approach presented in \cite{BVD-Alignment},  using the extended marking equation, which is also implemented in the PNetReplayer package of ProM (eMEQ); and
    \item an approximate approach using local search to compute alignments of large instances presented in \cite{ALI_TOSEM} (ALI). 
\end{inparaenum}
Please note, that we only consider baseline approaches that compute trace alignments and hence we omit other conformance checking techniques such as the token-based replay~\cite{TokenBasedReplay}.
While the approximate approach ALI is only implemented as a Python prototype,\footnote{Available at \url{https://www.cs.upc.edu/~taymouri/tool.html}} the authors previously compared it with the two exact baselines for optimal alignments computation, and showed to outperform these on a synthetic dataset \cite{ALI_TOSEM}. We conducted the experiments for ALI with the commercial ILP solver Gurobi.
We use each technique with default parameter settings for a common application scenario. Table~\ref{tb:eval_inputAssumptions} summarizes the different assumptions on the process model of all techniques. Notably, some baseline techniques are applicable to a wider class of process models. Currently there exist no other techniques to compute alignments specifically for the class of free-choice sound workflow nets as used by the proposed technique for a better comparison.

\begin{table}[htbp!] 
{\footnotesize{           
\setlength{\tabcolsep}{3pt}   
\centering{           
\begin{tabular}{|c|c|}
\hline
\bf{Technique} & \bf{Class of input models} \\ \hline
ILP, LP, eMEQ & easy sound Petri nets \\ \hline
ALI & sound Petri nets\\ \hline
Automata, SComp, TR-Scomp & free-choice sound workflow nets\\ \hline
\end{tabular}
}
\vspace{.5\baselineskip}           
\caption{Assumptions on input models of evaluated techniques}\label{tb:eval_inputAssumptions}           
\vspace{.5\baselineskip}           
}}  
\end{table}

\subsection{Setup}

For the purpose of measuring time performance, we recorded the execution time of our technique and of the four baselines by computing the alignments of a range of model-log pairs. 
The execution times exclude the time to load the datasets but include any preprocessing times such as computing concurrency-free nets via S-Components or tandem repeats for the proposed technique. 
Including the preprocessing times ensures a fair comparison to the baseline techniques as reduced processing times come at the cost of increased preprocessing times.

Each experiment was repeated five times and we reported the average execution time of runs \#2--\#4 to exclude the influence of the Java class loader and to reduce variance. For practical reasons, we set a time bound of ten minutes for each measurement taking into account the worst-case exponential time complexity of computing alignments. We note that previous experiments reported that in certain cases the computation of an alignment may take over a dozen hours \cite{Munoz-GamaCA14}. However, setting such large time bound would have rendered this evaluation impractical, given the very large number of model-log pairs.

As for accuracy, we measured the alignment cost per trace (cf. Def.~\ref{def:CostFunction}) for each model-log pair. This allows us to assess the degree of optimality of the different approximate techniques. We chose alignment cost over other conformance measures such as fitness as cost allows us to better pinpoint over-approximation. 
We conducted these experiments on a single-threaded 22-core Intel Xeon CPU E5-2699 v4 with 2.30GHz and with 128GB of RAM running JVM 8.

\subsection{Datasets}

In terms of datasets, we used a range of public and private log-model pairs from a recent benchmark on automated process discovery~\cite{PD-Discovery-BM}. The publicly available dataset consists of twelve event logs, which originate from the 4TU Centre for Research Data.\footnote{\url{https://data.4tu.nl/repository/collection:event\_logs\_real}} It consists of logs from the Business Process Intelligence challenge (BPIC) series, i.e.\ BPIC12 \cite{BPIC12}, BPIC13\textsubscript{cp} \cite{BPIC13cp}, BPIC13\textsubscript{inc} \cite{BPIC13inc}, BPIC14 \cite{BPIC14}, BPIC15 \cite{BPIC15}, BPIC17 \cite{BPIC17}, the Road Traffic Fines Management process log (RTFMP) \cite{RTFMP} and the SEPSIS Cases log (SEPSIS) \cite{SEPSIS}. The BPIC logs from  years 2011 and 2016 (BPIC11 and BPIC16) were excluded since they do not represent real business processes. 

We extended this dataset with the BPIC logs from the years 2018 (BPIC18) and 2019 (BPIC19), which were published after the benchmark paper. As suggested by the description of BPIC19, we split this log into four sublogs according to the attribute item category since the log captures four different types of processes. Hence, the public dataset was extended to a total of 17 event logs.
These public logs cover process executions from different domains such as finance, healthcare, government and IT service management. 

The private dataset from \cite{PD-Discovery-BM} encompasses eight proprietary logs. These originate from several organizations around the world, including healthcare, banking, insurance and software vendors. 

The authors of the benchmark in~\cite{PD-Discovery-BM} could not discover process models for two of the public event logs (BPIC15 and BPIC17) since the majority of discovery techniques used in this benchmark exceeded the allotted memory 
To overcome this problem, they applied the filtering technique described in~\cite{Noise-filtering} to filter infrequent behavior. 
We retained this filtering step to ensure compatibility with the dataset used in the benchmark paper (in the results, these filtered logs are annotated with ``$_f$''). As for the logs BPIC18 and BPIC19, we were also unable to discover process models using the unfiltered logs. In contrast to the benchmark paper, also the filtering technique could not be applied in a reasonable time frame. Thus, we decided to apply a naive filter to remove all traces that occur only once in order to obtain a process model for these two logs. When comparing the logs with the discovered process models for alignment computation, however, we retained the unfiltered logs, so as to detect more complex alignments.

In Table~\ref{tb:log_stats} we report the log characteristics as well as a range of statistics related to the application of the tandem repeats reduction. The size of the logs differs in terms of total number of events (5.9K to 2.5M) and traces (681 to 788K). Particularly relevant for the computation of alignments are the number of unique (``Unq.'') events (4--74), the number of unique traces (128--28K) and the average (``$\varnothing$'') and maximum (``Max'') trace length (``$\left|\text{Trace}\right|$'', average 3.4--64, maximum 9--2,973). These statistics are closely linked with the complexity of computing alignments since they determine the number and length of alignments to be computed. These logs thus feature a wide range of characteristics, including both simple and complex logs. 

To quantify the degree of repeated behavior, we also report various measures related to tandem repeats in Table~\ref{tb:log_stats}. Specifically, we report the average (``$\varnothing$'') number of identified tandem repeat tuples per trace (``\#TRs'', 0--2.47); the average number of repetitions per tandem repeat (``Reps'', 0--8.11); and the average length of the repeating sequence (length, 0--1.72). Next, we show the results of reducing the event logs also using our reduction algorithm (cf. Alg.~\ref{alg:reduceLog}). 

Specifically, we report the average and maximum trace length before (``$\left|\text{Trace}\right|$'') and after reduction (``$\left|\text{Trace TR}\right|$''). We can observe that on average each trace is reduced by 2.3 events across all logs.\footnote{We observe that this statistic includes both traces with and traces without repeated events.} In particular, for the BPIC12 log, we had the highest average reduction per trace (14 events) with the longest trace being reduced from 175 to 117 events (``Max''). One extreme case can be observed in BPIC18, where the longest trace of 2,973 events was reduced to 1,535 events (about half its length). 
Another effect of the tandem repeat reduction is that some reduced traces can be mapped to the same unique trace resulting in a smaller number of unique traces (``Unq. TR''), and so less alignments to compute. On average across all event logs around 500 unique traces were reduced. Two interesting examples are BPIC12 and BPIC17\textsubscript{f}, where the number of unique traces could be reduced by 2,086 and respectively 6,119 traces, highlighting the benefits of the binary search (cf. Alg.~\ref{alg:BinarySearchAlignments}).

\input{tex/TableLogStats}
The automated process discovery benchmark~\cite{PD-Discovery-BM} used four state-of-the-art automated discovery methods, namely: Inductive Miner (IM) \cite{InductiveMiner}, Split Miner (SM) \cite{SplitMiner}, Structured Heuristics Miner \cite{StructuredHeuristicsMiner} and Fodina \cite{Fodina}.
We decided to conduct experiments only on the process models discovered by the Inductive Miner and Split Miner algorithms, as the benchmark in~\cite{PD-Discovery-BM} shows that these two algorithms strike the best performance in terms of fitness, precision and simplicity. Moreover, these two algorithms have been embedded in commercial tools (e.g.\ Apromore, Celonis, Minit, MyInvenio) and hence the models discovered by them represent common use cases for conformance checking. In addition, these models fulfil all assumptions of the Tandem Repeat technique, i.e.\ they are sound or deadlock-free, free-choice and have uniquely-labelled transitions.
This resulted in a total of 40 log-model pairs from the benchmark dataset. We then discovered two process models for BPIC18 and eight process models for the four sublogs of BPIC19, using the latest version of Split Miner and Inductive Miner, giving rise to ten additional models. This resulted in a total of 50 model-log pairs for our evaluation.
\newpage
\input{tex/TableModelStats}

In Table~\ref{tb:model_stats} we provide the characteristics of the process models discovered by Inductive (IM) and Split Miner (SM). For these models, we report the overall size as the sum of places, transitions and arcs (``Size''), as well as the number of transitions (``Trns.''), choices (``XOR'') and parallel splits (``AND''), and the size of the reachability graph generated from each model (``RG Size''). The complexity of computing alignments is mainly linked to the size of the reachability graph of the process model, which is worst-case exponential on the size of the model. For example, we can observe that the size of the reachability graph of PRT2 (IM) is around 5.5M nodes and arcs, while the corresponding process model has a size of 175 nodes and arcs. 

\newpage
In Table~\ref{tb:model_stats} we also report the number of S-Components identified per model (``\#Scomp'') and the average size of the reachability graph (``$\varnothing$ RG Size'') after applying the S-Components decomposition (empty cells indicate concurrency-free models -- the number of S-Components being 1). We can observe that this size is usually smaller than that of the original reachability graph. For example, for PRT2 (IM) the size is reduced to an average of 15 nodes and arcs. Sometimes, this reduction does not lead to a smaller state space, e.g.\ for BPIC12 (SM) the size reduces from 95 to 90 per S-Component, which leads to a total state space of 180 nodes and arcs for all S-Components, which is larger than the size of the original model.

The two discovery methods (IM and SM) pose different challenges to conformance checking. 
Inductive Miner often uses nested structures in process models and hence their reachability graphs exhibit large state spaces. In the  benchmark in~\cite{PD-Discovery-BM}, it was shown that this algorithm systematically produces process models with high fitness values.
Split Miner strikes a trade-off between fitness and precision by filtering the directly-follows graph of the log before discovering the model. The models produced by Split Miner will have a smaller state space but may lead to a higher number of fitness mismatches
. Altogether, the models obtained by these two methods present two different scenarios for conformance checking: the models discovered by Inductive Miner require a larger state space to be traversed with a low to medium number of mismatches per trace, while the models discovered by Split Miner have a smaller state space with a medium to high number of mismatches per trace.

\subsection{Results}
In Table~\ref{tb:eval_results} we show the time performance in milliseconds (ms) of all approaches against the 50 model-log pairs including any preprocessing times. The best result for each dataset is highlighted in bold and timeout cases are recorded with ``t/out''. The table also shows the number of S-Components (``\#SComp'') of the model and the average trace reduction length (``$\varnothing$Red.''), computed as the average trace lengths minus the average length of the reduced traces from Table~\ref{tb:log_stats}. We included these model-log properties as they have explanatory value for the time results.

\input{tex/TableEvalResults}

\textbf{Analyzing the overall performance.} 
The Automata-based approach (Automata) outperforms the other datasets in 32 of 50 cases; its S-Components extension (SComp) performs best in eight out of 50 cases while the use of the tandem repeats reduction on top of the S-Components extension (TR-SComp), performs best in seven out of 50 cases. Both eMEQ and ALI outperform the other approaches in one case only, while the ILP as well as the LP approach never outperforms any other approach. When including the number of times an approach was ranked second, in order to reduce small variations, the results between the Automata-based variants homogenize. The base approach increases to 36 out of 50 cases, the SComp extension increases to 29 of 50 and the TR-SComp variant increases to 27 out of 50. The LP approach increases to 3 out of 50 datasets. The results of ILP, eMEQ and ALI do not change. When considering the total time spent across all 50 datasets (excluding timed out cases), the TR-SComp approach is the fastest with 251 seconds, followed by the SComp approach at around 300 seconds, the Automata approach at 540 seconds, eMEQ at 840 seconds, ILP and LP at 1,250 seconds and last ALI at 1,600 seconds. The difference between the total execution time (TR-SComp ranks first) and the ranking in individual datasets (TR-SComp ranks third) indicates that the tandem repeats approach reduces execution times significantly but only for certain datasets, i.e.\ the cases where repetitions can be observed in the log.

\textbf{Investigating timeout cases.}
All approaches time out for dataset BPIC18 for the model discovered from IM. This log has a very high number of unique traces (28,457) and nested parallel structures in the process model, resulting in 72 S-Components. The second most difficult case is PRT2 with IM. This log has a very large underlying state space (RG size: 5M), and only ALI can compute alignments. The S-Component approach can compute alignments quickly for most traces in this log (the average size of the reachability graph of the S-Components is only 15), though this approach times out when some conflicting traces need to be aligned on the original reachability graph. Our TR-SComp approach is able to compute alignments quickly for other difficult cases (i.e.\ cases where more than 3 other approaches time out) such as BPIC19\textsubscript{2} (IM) or BPIC18 (SM). In total, the TR-SComp and SComp approaches have the lowest number of timeouts (two cases) tied with the LP technique, followed by ILP and Automata with three, ALI with four and eMEQ with twelve.

\textbf{Improvements of the tandem repeats approach.}
The TR-SComp approach outperforms the other approaches significantly when the input event log is reduced on average by at least two events (the average trace reduction length $\varnothing$Red. is greater than two). For example, in BPIC12 (IM), TR-SComp outperforms S-Components by a factor of two, ALI by five and Automata as well as ILP by an order of magnitude. In the dataset BPIC18 (SM), TR-SComp improves over all the Automata-based variants by 20 seconds, while other approaches could not compute alignments for this model-log pair. In the case of BPIC14\textsubscript{f}, TR-Scomp outperforms the other approaches even though the average reduction is low (0.4 events per trace). This could be due to the high number of unique traces (circa 15K) that overall a high number of events is reduced.

\textbf{Problematic cases of the tandem repeats approach.}
Some datasets like the BPIC15 logs do not contain any repeating events. In these cases, TR-SComp performs similarly as the SComp approach, which in turn will fall back to the performance of the Automata-based approach when the process model is concurrency-free (i.e.\ \#SComps=1). In some cases, the log contains a relevant number of repetitions ($\varnothing$Red.>2), but TR-SComp is not faster than SComp, e.g.\ in BPIC13\textsubscript{inc} (IM) or BPIC19\textsubscript{1} (SM). We attribute this to the fact that these model-log pairs are quite small and thus the processing of tandem repeats creates a computational overhead compared to not applying the reduction altogether.

\textbf{Hybrid approach: definition and performance.}
Since the TR-SComp technique improves computation time only for a specific type of input event logs, we decided to define a hybrid approach that only applies the tandem repeat reduction if the traces can be reduced on average by at least two events per trace. In addition, we preserve the hybrid rule from~\cite{s-comps}, that is, we only apply the S-Component extension if the sum of the reachability graph sizes of all S-Components is not larger than the size of the original reachability graph. If the S-Component extension is not applied to a process model with concurrency, then the tandem repeats reduction is also not applied. We report the results of this Hybrid approach in Table~\ref{tb:eval_results}. In total, this approach gains eight seconds over TR-SComp, outperforming all other approaches in 32 cases out of 50 and is first- or second-ranked in 39 cases. This is the highest result compared to all other approaches. 

\textbf{Analyzing the distribution of preprocessing and processing times.}
The proposed TR-SComp technique includes many preprocessing steps such as computing concurrency free subnets via S-Components for the input model as described in~\cite{s-comps} and the tandem repeats for the event log. While the evaluation so far compared overall execution times including the pre-processing times, it is also interesting to investigate its breakdown into pre-processing and processing times. The table in \ref{app:preprocessing} reports on the preprocessing times of the TR-SComp technique as well as the percentage at the overall processing times. The pre-processing fraction ranges from 2-52\% for IM models and 5-72\% for SM models. It is significantly larger for small datasets with low processing times, i.e. the 72\% is a part of 62ms processing time for BPIC19\textsubscript{4} (SM). For larger datasets the preprocessing fraction decreases significantly, i.e. for BPIC19\textsubscript{2} (IM) the preprocessing takes only 2\% of 62 seconds.

\input{tex/TableEvalCost}

\newpage
\textbf{Analyzing cost over-approximations.}
Table~\ref{tb:eval_cost} shows the cost and over-approximation for all datasets where the TR-SComp approach over-approximates the minimal cost. The table with the cost and over-approximations for all datasets can be found in \ref{app:cost_comparison}. In total, TR-SComp over-approximates in ten cases out of 50 and computes the minimal cost in all other datasets. In detail, the degree of over-approximation ranges between 0.02\% to 2.47\% for nine cases and 32\% for PRT6 (IM). The over-approximation of the latter dataset may seem high, but the overall fitness of this dataset is very low such that any small variation on the minimal cost will cause a high degree of over-approximation. When drilling down the over-approximation for this model-log pair, we found that the cost was on average higher than the minimal by 1.05 in 2.7\% of the unique traces. 
In comparison, ALI, as a representative approach for approximate alignment computation, over-approximates the minimal cost in all datasets ranging from 17\% to 11 times the value of the optimal cost. More specifically, ALI over-approximates the optimal cost by more than 100\% in 38 cases out of 50. Both TR-SComp and ALI never under-estimate the optimal cost.

\textbf{Investigating causes of over-approximation.}
The first reason for over-approximation is that the TR-SComp approach is applied on top of the SComp approach and as such it carries the over-approximation induced by this latter approach~\cite{s-comps}. Specifically, the cost will be over-approximated if a trace contains an activity that is after a parallel block in the process model, but appears before the activities of the parallel block in the trace, i.e.\ it is misplaced. This was the most common cause of over-approximation.

\begin{figure}[h!]
\centering
\includegraphics[width=0.5\textwidth]{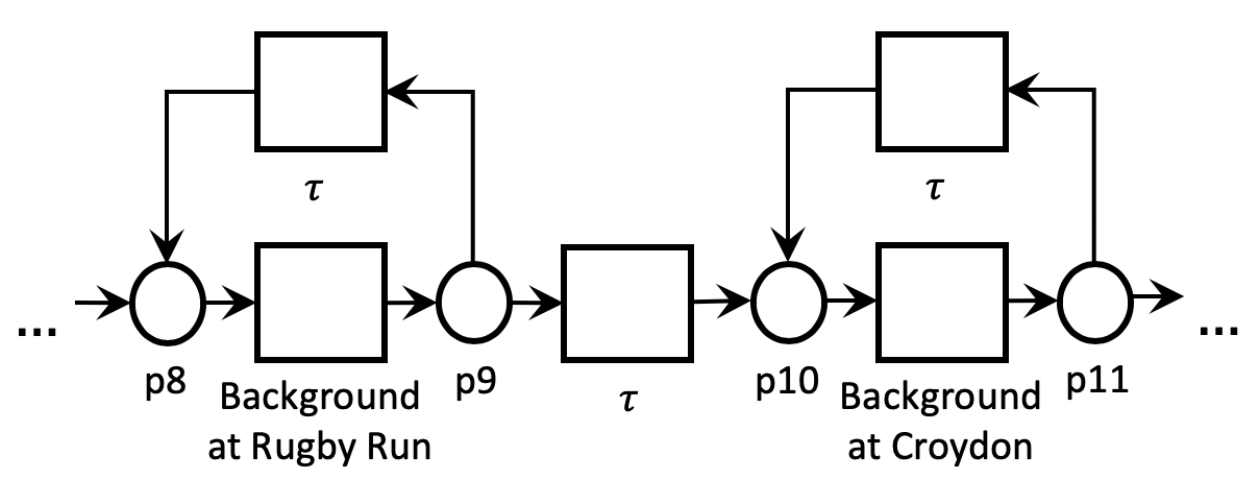}
\caption{Snippet of the process model PRT4 (SM), where an over-approximation occurs}\label{fig:OverApprox}
\end{figure}

As a second reason, TR-SComp induces over-approximation when a tandem repeat involves exactly three repetitions and no repeating sequence can be found in the returned reduced alignment of Alg.~\ref{alg:BinarySearchAlignments}. The over-approximation occurs if there exists another reduced alignment with the same cost that could contain a repeating sequence. In that case, the extended alignment would include a middle copy of the tandem repeat with all $\lhide$-operations while some of the events could actually be matched. We encountered this problem for example in the dataset PRT4 (SM). The snippet of the problematic part of the process model is shown in Fig.~\ref{fig:OverApprox}. Here an activity ``Background at Rugby Run'' can be repeated any number of times and is then followed by activity ``Background at Croyden'', which can also be repeated. 

The problematic tandem repeat is $(2, \text{Background at Croyden}, \text{Background at Rugby Run})$, 3), i.e.\ the repeat type ``Background at Croyden, Background at Rugby Run'' is repeated three times at the trace position two. The reduced tandem repeat containing only two copies is then aligned with $\langle (\lhide,\text{Background at Croyden}),$ $(\match,\text{Background at Rugby Run}),(\match,\text{Background at Croyden}),$ $(\lhide,\text{Background at Rugby Run})\rangle$. When trying to extend this tandem repeat with Alg.~\ref{alg:ExtendRedAlignments}, we can not identify a repeating sequence within this tandem repeat since neither of the two activities has been aligned with a $\match$ in both copies of the tandem repeat. Hence, the extended alignment would include both activities with $\lhide$ operations and incur a cost of two. However, there exists another alignment for the reduced tandem repeat with the same cost, where a repeating sequence can be found, i.e.\ %
$\langle (\lhide,\text{Background at Croyden}),$ $(\match,\text{Background at Rugby Run}),$ $(\lhide,\text{Background at Croyden}),$ $(\match,\text{Background at Rugby Run}),$  $(\rhide,\text{Background at Croyden})\rangle$
. For that alignment activity ``Background at Rugby Run'' can now be identified as a repeating sequence and will be included with a $\match$ operation in the middle copy such that only a cost of one is added. The problem of over-approximation occurs because we find the former alignment and not the latter since they have the same cost and we only find one optimal alignment with Alg.~\ref{alg:BinarySearchAlignments}. 

\newpage
The reason why the two alignments have the same cost is that the two activities in the trace are reversed in the order of the process model such that only one activity can be repeated and the other needs to be hidden in both the trace and the model, or both activities can be matched and then no repetitions can be matched resulting in an overall cost of four in both cases. We observe that this situation cannot happen for higher numbers of repetitions since the alignment that matches the two activities without any repetitions incurs an additional cost from the collapsed repetitions. Specifically, the former alignment would have a cost of six while the latter alignment would have a cost of five. Thus, the correct alignment would be returned by Alg.~\ref{alg:BinarySearchAlignments}, avoiding the over-approximation. 
To avoid such cases of over-approximation, it is possible to discard tandem repeats of three repetitions during the reduction, to increase the precision of the tandem repeats approach. However, the amount of over-approximation caused by this issue was low in our experiments and thus we argue that trading small amounts of accuracy for potential gains in performance is worthwhile.

\input{tex/TableEvalMemory.tex}

\textbf{Analyzing memory requirements.}
Table~\ref{tb:eval_memory} reports on heap memory measurements in megabytes (MB) of TR-SComp and of the LP alignments technique for comparison to investigate the requirements of the proposed technique. On average the TR-SComp uses 50 MB with a maximum of 550 MB, while the LP baseline only uses 17 MB on average with a maximum of 139 MB. The priority of the proposed technique lies on reducing the processing time requiring slightly more memory. The memory requirements of TR-SComp are acceptable for most common computing devices such as laptops or PCs, but show improvement potential.
In particular, using a Dijkstra search for alignments uses more memory than an A\textsuperscript{*} search used in LP since more nodes need to be investigated and stored. Hence, the memory requirements of the technique could be further improved by embedding the technique in an A\textsuperscript{*} search, which we consider as future work.

\newpage
\subsection{Threats to validity}
The selection of datasets is a threat to validity. We decided to use two datasets of real-life log-model pairs from a recent benchmark \cite{PD-Discovery-BM}, and enriched this with the BPIC logs from the last two years to keep the dataset up to date. These model-log pairs exhibit a wide range of structural characteristics and originate from different industry domains, so they provide a good representation of reality. However, some of the event logs such as the BPIC15 series do not contain any repeating events. This calls for further experiments with event logs with a higher degree of repetitions, and  with very large real-life log-model pairs. Such datasets are not publicly available at the time of writing. This problem could be alleviated with artificial datasets~\cite{SyntheticDataset}. 

The selection of the baseline approaches is another threat to validity. The technique proposed in this article, as well as the Automata-based technique and its S-Components extension, are applicable to a specific subclass of Petri nets, namely sound free-choice workflow nets, while the two exact approaches chosen as baselines (ILP and eMEQ) are applicable to a wider class of Petri nets, namely easy sound Petri nets, and the approximate technique (ALI) is applicable to sound Petri nets. To the best of our knowledge, however, there are no conformance checking techniques available that target the specific subclass of Petri nets addressed by our technique for a better comparison. In addition, this specific class of Petri nets has relevance to the the field of process mining and more widely to the field of business process management, since BPMN models can be translated to this class and several mining algorithms such as Split Miner \cite{SplitMiner}, Inductive Miner \cite{InductiveMiner} or Fodina \cite{Fodina} produce Petri nets of this class. However, the BPMN models also need to have uniquely-labeled activities to be used by the proposed technique. This is fulfilled by many mining algorithms such as Split miner and Inductive miner. Additionally, the models need to be sound, which is sometimes violated by the Split miner algorithm. All process models used in the evaluation of this article were tested to be sound to avoid this issue.

A final threat to validity is posed by the number of methods used for automated process discovery (only two). Potentially we could have chosen a larger number of methods. The choice of Split Miner and Inductive Miner was determined by both pragmatic reasons (other methods such as Structured Heuristics Miner return models with duplicate events which we cannot handle, or lead to models for which fitness could not be computed) as well as by the need to test two extreme cases: models with large state spaces versus event logs with large degrees of repeated activities. Moreover, they are the best performing automated discovery methods according to the benchmark in \cite{PD-Discovery-BM}. So, all considered, they constitute a sufficiently representative set of automated discovery methods. 

%% file: tex/TableLogStats.tex
\begin{table}[h]                           
{\footnotesize{                           
\setlength{\tabcolsep}{3pt}                           
\centering{                           
\begin{tabular}{|l|r r|r r|r r|r r r||r|r r|}                           
\hline                           
\multirow{2}{*}{\textbf{Log}} & \multicolumn{2}{c|}{\bf{\#Events}}   & \multicolumn{2}{c|}{\bf{\#Traces}}   & \multicolumn{2}{c|}{\textbf{$\left|\text{Trace}\right|$}}   & \multicolumn{3}{c||}{\textbf{Tandem repeats ($\varnothing$)}}     & \bf{\#Traces} & \multicolumn{2}{c|}{\textbf{$\left|\text{Trace TR}\right|$}}   \\ \cline{2-13} 
 & \bf{Overall} & \bf{Unq.} & \bf{Overall} & \bf{Unq.} & \bf{$\varnothing$} & \bf{Max} & \bf{\#TRs} & \bf{Reps} & \bf{length} & \bf{Unq. TR} & \bf{$\varnothing$} & \bf{Max} \\ \hline 
BPIC12 & 262,200 & 24 & 13,087 & 4,366 & 41.8 & 175 & 2.27 & 8.11 & 1.05 & 2,280 & 27.8 & 117 \\ \hline 
BPIC13\textsubscript{cp} & 6,660 & 4 & 1,487 & 183 & 9.9 & 35 & 0.63 & 4.3 & 1.25 & 120 & 8.2 & 25 \\ \hline 
BPIC13\textsubscript{inc} & 65,533 & 4 & 7,554 & 1,511 & 19.2 & 123 & 1.19 & 4.71 & 1.27 & 992 & 15.4 & 104 \\ \hline 
BPIC14\textsubscript{f} & 369,485 & 9 & 41,353 & 14,948 & 15.5 & 167 & 0.22 & 3.8 & 1.19 & 14,477 & 15.1 & 165 \\ \hline 
BPIC15\textsubscript{1f} & 21,656 & 70 & 902 & 295 & 35.1 & 50 & 0.0 & 0.0 & 0.0 & 295 & 35.1 & 50 \\ \hline 
BPIC15\textsubscript{2f} & 24,678 & 82 & 681 & 420 & 42.4 & 63 & 0.0 & 0.0 & 0.0 & 420 & 42.4 & 63 \\ \hline 
BPIC15\textsubscript{3f} & 43,786 & 62 & 1,369 & 826 & 34.6 & 54 & 0.0 & 0.0 & 0.0 & 826 & 34.6 & 54 \\ \hline 
BPIC15\textsubscript{4f} & 29,403 & 65 & 860 & 451 & 36.6 & 54 & 0.0 & 0.0 & 0.0 & 451 & 36.6 & 54 \\ \hline 
BPIC15\textsubscript{5f} & 30,030 & 74 & 975 & 446 & 42.1 & 61 & 0.0 & 0.0 & 0.0 & 446 & 42.1 & 61 \\ \hline 
BPIC17\textsubscript{f} & 714,198 & 18 & 21,861 & 8,767 & 38.1 & 113 & 2.47 & 5.41 & 1.04 & 2,648 & 29.5 & 73 \\ \hline 
RTFMP & 561,470 & 11 & 150,370 & 231 & 8.2 & 20 & 0.15 & 6.32 & 1.0 & 202 & 7.5 & 11 \\ \hline 
SEPSIS & 15,214 & 16 & 1,050 & 846 & 16.3 & 185 & 0.15 & 3.58 & 1.54 & 846 & 15.9 & 184 \\ \hline 
BPIC18 & 2,514,266 & 41 & 43,809 & 28,457 & 64.0 & 2,973 & 1.61 & 5.65 & 1.5 & 27,279 & 55.7 & 1,535 \\ \hline 
BPIC19\textsubscript{1} & 5,898 & 11 & 1,044 & 148 & 10.9 & 21 & 0.78 & 6.02 & 1.0 & 109 & 7.8 & 15 \\ \hline 
BPIC19\textsubscript{2} & 319,233 & 38 & 15,182 & 4,228 & 60.5 & 990 & 1.82 & 6.67 & 1.19 & 3,579 & 51.6 & 843 \\ \hline 
BPIC19\textsubscript{3} & 1,234,708 & 39 & 221,010 & 7,832 & 10.5 & 179 & 0.14 & 7.15 & 1.1 & 7,404 & 9.8 & 63 \\ \hline 
BPIC19\textsubscript{4} & 36,084 & 15 & 14,498 & 281 & 5.7 & 17 & 0.23 & 4.48 & 1.0 & 238 & 5.2 & 12 \\ \hline \hline
PRT1 & 75,353 & 9 & 12,720 & 1,026 & 13.1 & 64 & 0.5 & 4.11 & 1.55 & 784 & 11.6 & 56 \\ \hline 
PRT2 & 46,282 & 9 & 1,182 & 1,153 & 39.8 & 276 & 1.04 & 3.96 & 1.3 & 1,153 & 37.2 & 252 \\ \hline 
PRT3 & 13,720 & 15 & 1,600 & 318 & 8.6 & 9 & 0.0 & 0.0 & 0.0 & 318 & 8.6 & 9 \\ \hline 
PRT4 & 166,282 & 11 & 20,000 & 5,941 & 11.5 & 36 & 0.22 & 3.23 & 1.03 & 5,551 & 11.2 & 36 \\ \hline 
PRT6 & 6,011 & 9 & 744 & 167 & 10.1 & 21 & 0.01 & 3.0 & 2.0 & 166 & 10.1 & 21 \\ \hline 
PRT7 & 16,353 & 13 & 2,000 & 128 & 9.1 & 11 & 0.0 & 0.0 & 0.0 & 128 & 9.1 & 11 \\ \hline 
PRT9 & 1,808,706 & 8 & 787,657 & 909 & 10.5 & 58 & 0.4 & 4.36 & 1.72 & 777 & 8.8 & 50 \\ \hline 
PRT10 & 78,864 & 19 & 43,514 & 172 & 3.4 & 15 & 0.15 & 5.42 & 1.0 & 151 & 2.9 & 8 \\ \hline 
\end{tabular}                           
}                           
\caption{Logs and reduction statistics}\label{tb:log_stats}                           
\vspace{-\baselineskip}                           
}}                           
\end{table}                                                     

%% file: tex/TableModelStats.tex
\begin{table}[htbp!]                    
{\footnotesize{                    
\setlength{\tabcolsep}{3pt}                    
\centering{                    
\begin{tabular}{|c|c|c|c c c c c|c c|}                    
\hline                    
\bf{Miner} & \bf{Domain} & \bf{Dataset} & \bf{Size} & \bf{Trns} & \bf{XOR} & \bf{AND} & \bf{RG Size} & \bf{\#Scomp} & \bf{$\varnothing$ RG Size} \\ \hline
 \multirow{17}{*}{IM} &  \multirow{17}{*}{public} & BPIC12 & 177 & 45 & 16 & 2 & 1,997 & 10 & 58.3 \\ \cline{3-10}
 &  & BPIC13\textsubscript{cp} & 31 & 8 & 2 & 0 & 9 & 1 &  \\ \cline{3-10}
 &  & BPIC13\textsubscript{inc} & 56 & 13 & 3 & 1 & 121 & 3 & 14.0 \\ \cline{3-10}
 &  & BPIC14\textsubscript{f} & 124 & 29 & 8 & 2 & 4,383 & 10 & 26.1 \\ \cline{3-10}
 &  & BPIC15\textsubscript{1f} & 449 & 127 & 48 & 0 & 719 & 1 &  \\ \cline{3-10}
 &  & BPIC15\textsubscript{2f} & 537 & 150 & 55 & 1 & 1,019 & 2 & 232.0 \\ \cline{3-10}
 &  & BPIC15\textsubscript{3f} & 464 & 128 & 47 & 3 & 875 & 8 & 191.5 \\ \cline{3-10}
 &  & BPIC15\textsubscript{4f} & 469 & 131 & 51 & 1 & 1,019 & 2 & 202.0 \\ \cline{3-10}
 &  & BPIC15\textsubscript{5f} & 381 & 111 & 31 & 0 & 429 & 1 &  \\ \cline{3-10}
 &  & BPIC17\textsubscript{f} & 121 & 33 & 8 & 0 & 59 & 1 &  \\ \cline{3-10}
 &  & RTFMP & 111 & 26 & 9 & 2 & 2,394 & 6 & 25.0 \\ \cline{3-10}
 &  & SEPSIS & 145 & 37 & 13 & 3 & 2,274 & 8 & 44.0 \\ \cline{3-10}
 &  & BPIC18 & 235 & 57 & 18 & 6 & 1,057 & 72 & 62.9 \\ \cline{3-10}
 &  & BPIC19\textsubscript{1} & 44 & 11 & 4 & 1 & 49 & 2 & 13.5 \\ \cline{3-10}
 &  & BPIC19\textsubscript{2} & 186 & 47 & 13 & 4 & 20,264 & 8 & 37.3 \\ \cline{3-10}
 &  & BPIC19\textsubscript{3} & 279 & 70 & 23 & 7 & 6,450 & 44 & 83.4 \\ \cline{3-10}
 &  & BPIC19\textsubscript{4} & 85 & 23 & 8 & 1 & 86 & 2 & 29.5 \\ \cline{2-10}
 &  \multirow{8}{*}{private} & PRT1 & 70 & 16 & 4 & 1 & 195 & 4 & 19.3 \\ \cline{3-10}
 &  & PRT2 & 175 & 43 & 16 & 1 & 5,515,357 & 7 & 23.0 \\ \cline{3-10}
 &  & PRT3 & 111 & 27 & 8 & 2 & 167 & 8 & 33.0 \\ \cline{3-10}
 &  & PRT4 & 91 & 21 & 5 & 2 & 154 & 8 & 26.0 \\ \cline{3-10}
 &  & PRT6 & 86 & 20 & 4 & 2 & 65 & 6 & 29.0 \\ \cline{3-10}
 &  & PRT7 & 99 & 23 & 5 & 2 & 158 & 8 & 30.0 \\ \cline{3-10}
 &  & PRT9 & 96 & 21 & 7 & 2 & 9,121 & 7 & 13.9 \\ \cline{3-10}
 &  & PRT10 & 124 & 35 & 8 & 1 & 184 & 2 & 48.5 \\ \hline
 \multirow{25}{*}{SM} &  \multirow{17}{*}{public} & BPIC12 & 315 & 85 & 29 & 1 & 95 & 2 & 140.0\\ \cline{3-10}
 &  & BPIC13\textsubscript{cp} & 49 & 13 & 4 & 0 & 13 & 1 & \\ \cline{3-10}
 &  & BPIC13\textsubscript{inc} & 56 & 15 & 5 & 0 & 17 & 1 &  \\ \cline{3-10}
 &  & BPIC14\textsubscript{f} & 88 & 24 & 9 & 0 & 24 & 1 &  \\ \cline{3-10}
 &  & BPIC15\textsubscript{1f} & 368 & 98 & 25 & 0 & 156 & 1 &  \\ \cline{3-10}
 &  & BPIC15\textsubscript{2f} & 444 & 117 & 25 & 0 & 186 & 1 &  \\ \cline{3-10}
 &  & BPIC15\textsubscript{3f} & 296 & 78 & 17 & 0 & 136 & 1 &  \\ \cline{3-10}
 &  & BPIC15\textsubscript{4f} & 323 & 85 & 18 & 0 & 141 & 1 &  \\ \cline{3-10}
 &  & BPIC15\textsubscript{5f} & 359 & 94 & 18 & 0 & 159 & 1 &  \\ \cline{3-10}
 &  & BPIC17\textsubscript{f} & 149 & 40 & 12 & 0 & 54 & 1 &  \\ \cline{3-10}
 &  & RTFMP & 102 & 28 & 11 & 0 & 37 & 1 &  \\ \cline{3-10}
 &  & SEPSIS & 162 & 44 & 15 & 0 & 41 & 1 &  \\ \cline{3-10}
 &  & BPIC18 & 251 & 72 & 16 & 0 & 87 & 1 &  \\ \cline{3-10}
 &  & BPIC19\textsubscript{1} & 63 & 18 & 4 & 0 & 29 & 1 &  \\ \cline{3-10}
 &  & BPIC19\textsubscript{2} & 232 & 68 & 14 & 0 & 105 & 1 &  \\ \cline{3-10}
 &  & BPIC19\textsubscript{3} & 378 & 112 & 20 & 0 & 172 & 1 &  \\ \cline{3-10}
 &  & BPIC19\textsubscript{4} & 106 & 31 & 8 & 0 & 51 & 1 &  \\ \cline{2-10}
 &  \multirow{8}{*}{private} & PRT1 & 104 & 28 & 9 & 0 & 28 & 1 &  \\ \cline{3-10}
 &  & PRT2 & 166 & 45 & 15 & 0 & 37 & 1 &  \\ \cline{3-10}
 &  & PRT3 & 96 & 25 & 8 & 1 & 34 & 2 & 41.0 \\ \cline{3-10}
 &  & PRT4 & 126 & 33 & 10 & 1 & 34 & 2 & 55.0 \\ \cline{3-10}
 &  & PRT6 & 46 & 11 & 2 & 1 & 20 & 2 & 19.0 \\ \cline{3-10}
 &  & PRT7 & 86 & 19 & 3 & 5 & 39 & 6 & 27.7 \\ \cline{3-10}
 &  & PRT9 & 107 & 29 & 10 & 0 & 32 & 1 &  \\ \cline{3-10}
 &  & PRT10 & 327 & 90 & 34 & 0 & 92 & 1 &  \\ \hline
\end{tabular}                    
}                    
\caption{Models and S-Components statistics}\label{tb:model_stats}                    
}}                    
\end{table}                                  

%% file: tex/TableEvalResults.tex
\begin{table}[htbp!]                         
{\footnotesize{                         
\setlength{\tabcolsep}{3pt}                         
\centering{                         
\begin{tabular}{|c|c|c c|c c c c c c|c c|}                         
\hline                         
 \multirow{2}{*}{\bf{Miner}} &  \multirow{2}{*}{\bf{Dataset}} &  \multirow{2}{*}{\bf{\#SComps}} &  \multirow{2}{*}{\bf{$\varnothing$Red.}} & \multicolumn{6}{c|}{\bf{Baselines}}           & \multicolumn{2}{c|}{\bf{Our approaches}}    \\ 
 &  &  &  & \bf{ILP} & \bf{LP} & \bf{eMEQ} & \bf{ALI} & \bf{Automata} & \bf{SComp} & \bf{TR-SComp} & \bf{Hybrid}  \\ \hline
 \multirow{25}{*}{IM} & BPIC12 & 10 & 14.0 & 135,911 & 42,685 & t/out &  49,018  & 104,523 & 21,516 &\bf{ 11,537 }&\bf{ 11,537 } \\ 
 & BPIC13\textsubscript{cp} & 1 & 1.7 & 123 & 61 & 707 &  3,836  &\bf{ 23 }& 32 & 38 &\bf{ 23 } \\ 
 & BPIC13\textsubscript{inc} & 3 & 3.8 & 1,708 & 1,852 & 59,068 &  27,850  & 1,385 &\bf{ 137 }& 242 & 242  \\ 
 & BPIC14\textsubscript{f} & 10 & 0.4 & 101,837 & 233,452 & t/out &  100,351  & 199,488 &\bf{ 2,048 }& 11,185 &\bf{ 2,048 } \\ 
 & BPIC15\textsubscript{1f} & 1 & 0.0 & 3,501 & 1,501 & 3,041 &  19,651  &\bf{ 126 }& 165 & 153 &\bf{ 126 } \\ 
 & BPIC15\textsubscript{2f} & 2 & 0.0 & 20,529 & 7,675 & 20,759 &  43,863  & 1,187 & 1,739 & 1,521 & 1,739  \\ 
 & BPIC15\textsubscript{3f} & 8 & 0.0 & 32,328 & 12,564 & 60,310 &  47,525  &\bf{ 2,292 }& 4,330 & 4,118 &\bf{ 2,292 } \\ 
 & BPIC15\textsubscript{4f} & 2 & 0.0 & 12,184 & 5,728 & 21,942 &  33,958  &\bf{ 686 }& 1,047 & 887 & 1,047  \\ 
 & BPIC15\textsubscript{5f} & 1 & 0.0 & 8,877 & 2,600 & 11,059 &  23,239  &\bf{ 348 }& 431 & 427 &\bf{ 348 } \\ 
 & BPIC17\textsubscript{f} & 1 & 8.6 & 21,982 & 8,340 & 167,013 &  60,621  & 4,436 & 4,996 & 1,314 & 1,314  \\ 
 & RTFMP & 6 & 0.6 & 2,449 & 2,002 & 1,162 &  5,884  &\bf{ 428 }& 674 & 678 & 674  \\ 
 & SEPSIS & 8 & 0.4 & 7,109 & 2,212 & 17,848 &  23,842  & 3,063 &\bf{ 1,461 }& 1,590 &\bf{ 1,461 } \\ 
 & BPIC18 & 72 & 8.3 & t/out & t/out & t/out &   t/out   & t/out & t/out & t/out & t/out  \\ 
 & BPIC19\textsubscript{1} & 2 & 3.1 & 118 & 57 & 542 &  7,006  &\bf{ 38 }& 57 & 54 & 54  \\ 
 & BPIC19\textsubscript{2} & 8 & 9.0 & 179,802 & 183,685 & t/out &   t/out   & t/out & 66,751 &\bf{ 61,967 }&\bf{ 61,967 } \\ 
 & BPIC19\textsubscript{3} & 44 & 0.8 & 113,305 & 145,917 &\bf{ 35,354 }&  71,759  & 74,565 & 53,635 & 54,283 & 53,635  \\ 
 & BPIC19\textsubscript{4} & 2 & 0.6 & 463 & 301 & 676 &  7,563  &\bf{ 52 }& 95 & 105 & 95  \\ \cline{2-11}
 & PRT1 & 4 & 1.5 & 1,022 & 1,447 & 1,822 &  10,812  & 749 &\bf{ 144 }& 255 &\bf{ 144 } \\ 
 & PRT2 & 7 & 2.6 & t/out & t/out & t/out &\bf{  17,296  }& t/out & t/out & t/out & t/out  \\ 
 & PRT3 & 8 & 0.0 & 214 & 114 & 473 &  6,500  &\bf{ 58 }& 84 & 88 &\bf{ 58 } \\ 
 & PRT4 & 8 & 0.3 & 4,895 & 5,913 & 7,824 &  32,811  & 2,731 &\bf{ 486 }& 1,815 & 2,731  \\ 
 & PRT6 & 6 & 0.0 & 142 & 60 & 347 &  6,223  &\bf{ 33 }& 58 & 64 &\bf{ 33 } \\ 
 & PRT7 & 8 & 0.0 & 125 & 57 & 214 &  5,898  &\bf{ 29 }& 71 & 73 &\bf{ 29 } \\ 
 & PRT9 & 7 & 1.7 & 34,949 & 49,374 & 7,763 &  10,738  & 21,297 &\bf{ 1,890 }& 4,745 &\bf{ 1,890 } \\ 
 & PRT10 & 2 & 0.5 & 721 & 444 & 564 &  8,677  &\bf{ 37 }& 131 & 150 & 131  \\ \hline\hline
 \multirow{25}{*}{SM} & BPIC12 & 2 & 14.0 & 199,209 & 55,275 & t/out & 118,053 & 6,096 & 5,783 & 4,112 & 6,096  \\ 
 & BPIC13\textsubscript{cp} & 1 & 1.7 & 153 & 71 & 1,609 & 4,351 &\bf{ 22 }& 28 & 35 &\bf{ 22 } \\ 
 & BPIC13\textsubscript{inc} & 1 & 3.8 & 1,239 & 595 & 84,507 & 18,906 &\bf{ 210 }& 274 & 244 & 244  \\ 
 & BPIC14\textsubscript{f} & 1 & 0.4 & 41,956 & 14,290 & t/out & 157,733 & 10,936 & 11,498 &\bf{ 3,421 }& 10,936  \\ 
 & BPIC15\textsubscript{1f} & 1 & 0.0 & 2,995 & 1,391 & 1,362 & 12,986 &\bf{ 146 }& 190 & 196 &\bf{ 146 } \\ 
 & BPIC15\textsubscript{2f} & 1 & 0.0 & 9,591 & 11,946 & 5,319 & 26,114 &\bf{ 721 }& 805 & 869 &\bf{ 721 } \\ 
 & BPIC15\textsubscript{3f} & 1 & 0.0 & 7,644 & 5,160 & 8,576 & 25,067 & 784 &\bf{ 744 }& 755 & 784  \\ 
 & BPIC15\textsubscript{4f} & 1 & 0.0 & 7,508 & 4,681 & 5,300 & 19,893 &\bf{ 494 }& 571 & 568 &\bf{ 494 } \\ 
 & BPIC15\textsubscript{5f} & 1 & 0.0 & 9,148 & 6,682 & 4,312 & 24,608 & 729 &\bf{ 623 }& 658 & 729  \\ 
 & BPIC17\textsubscript{f} & 1 & 8.6 & 28,410 & 9,918 & 53,545 & 265,169 & 4,646 & 5,094 & 1,306 & 1,306  \\ 
 & RTFMP & 1 & 0.6 & 2,421 & 1,885 & 1,514 & 7,020 &\bf{ 71 }& 374 & 382 &\bf{ 71 } \\ 
 & SEPSIS & 1 & 0.4 & 5,126 & 960 & t/out & 18,821 &\bf{ 224 }& 262 & 243 &\bf{ 224 } \\ 
 & BPIC18 & 1 & 8.3 & t/out & 302,964 & t/out & t/out & 88,274 & 96,902 &\bf{ 62,026 }&\bf{ 62,026 } \\ 
 & BPIC19\textsubscript{1} & 1 & 3.1 & 185 & 63 & 1,136 & 6,570 &\bf{ 35 }& 40 & 41 & 41  \\ 
 & BPIC19\textsubscript{2} & 1 & 9.0 & 70,222 & 13,283 & t/out & t/out &\bf{ 6,497 }& 8,218 & 8,831 & 8,831  \\ 
 & BPIC19\textsubscript{3} & 1 & 0.8 & 91,644 & 69,158 & t/out & 104,798 &\bf{ 4,136 }& 6,500 & 4,518 &\bf{ 4,136 } \\ 
 & BPIC19\textsubscript{4} & 1 & 0.6 & 490 & 274 & 1,163 & 7,501 &\bf{ 33 }& 59 & 67 &\bf{ 33 } \\ \cline{2-11}
 & PRT1 & 1 & 1.5 & 2,243 & 780 & 43,426 &  10,788  &\bf{ 132 }& 271 & 172 &\bf{ 132 } \\ 
 & PRT2 & 1 & 2.6 & 39,025 & 4,566 & t/out &  91,186  &\bf{ 910 }& 1,040 & 1,048 & 1,048  \\ 
 & PRT3 & 2 & 0.0 & 203 & 117 & 620 &  6,439  &\bf{ 51 }& 93 & 88 &\bf{ 51 } \\ 
 & PRT4 & 2 & 0.3 & 8,556 & 4,748 & 129,444 &  47,776  &\bf{ 1,733 }& 1,938 & 2,432 & 1,733  \\ 
 & PRT6 & 2 & 0.0 & 83 & 42 & 288 &  5,997  &\bf{ 24 }& 54 & 59 &\bf{ 24 } \\ 
 & PRT7 & 6 & 0.0 & 111 & 57 & 209 &  5,887  &\bf{ 25 }& 66 & 69 &\bf{ 25 } \\ 
 & PRT9 & 1 & 1.7 & 43,094 & 37,747 & 82,628 &  18,165  &\bf{ 336 }& 1,194 & 1,533 &\bf{ 336 } \\ 
 & PRT10 & 1 & 0.5 & 887 & 495 & 1,363 &  9,374  &\bf{ 48 }& 89 & 91 &\bf{ 48 } \\ \hline\hline
\multicolumn{4}{|c|}{Total time spent (ms):}       & 1,256,434 & 1,255,187 & 844,811 & 1,638,124 & 544,886 & 304,688 & 251,052 &\bf{ 243,855 } \\ \hline
\multicolumn{4}{|c|}{Total outperforming:}       & 0 & 0 & 1 & 1 &\bf{ 32 }& 8 & 7 & 30  \\ \hline
\multicolumn{4}{|c|}{Total outperforming and second:}       & 0 & 3 & 1 & 1 & 36 & 29 & 27 &\bf{ 39 } \\ \hline
\multicolumn{4}{|c|}{\#Timeouts:}       & 3 & \bf{2} & 12 & 4 & 3 &\bf{ 2 }&\bf{ 2 }&\bf{ 2 } \\ \hline
\end{tabular}                         
}                         
\vspace{.5\baselineskip}                         
\caption{Time performance}\label{tb:eval_results}                         
\vspace{.5\baselineskip}                         
}}                         
\end{table}                                        

%% file: tex/TableEvalCost.tex
\begin{table}[htbp!]                      
{\footnotesize{                      
\setlength{\tabcolsep}{3pt}                      
\centering{                      
\begin{tabular}{|c|c|c c c c c c|c c c|}                      
\hline                      
 \multirow{2}{*}{\bf{Miner}} &  \multirow{2}{*}{\bf{Dataset}} & \multicolumn{6}{c|}{\bf{Cost}}           & \multicolumn{3}{c|}{\bf{Over-Approximation}}     \\ 
 &  & \bf{ILP} & \bf{eMEQ} & \bf{ALI} & \bf{Automata} & \bf{SComp} & \bf{TR-SComp} & \bf{$\Delta$ ALI} & \bf{$\Delta$ SComp} & \bf{$\Delta$ TR-SComp} \\ \hline
 \multirow{5}{*}{IM} & BPIC15\textsubscript{2f} & 2.02 & 2.02 & 26.73 & 2.02 & 2.07 & 2.07 & 24.706 & 0.050 & 0.050 \\ 
 & SEPSIS & 0.12 & 0.12 & 12.83 & 0.12 & 0.12 & 0.12 & 12.714 & 0.004 & 0.002 \\ 
 & BPIC19\textsubscript{2} & 0.18 & t/out & t/out & t/out & 0.18 & 0.18 & t/out & 0.001 & 0.001 \\ 
 & BPIC19\textsubscript{3} & 1.00 & 1.00 & 6.56 & 1.00 & 1.00 & 1.01 & 5.562 & 0.003 & 0.008 \\ 
 & PRT6 & 0.09 & 0.09 & 2.31 & 0.09 & 0.12 & 0.12 & 2.216 & 0.028 & 0.028 \\ \hline
 \multirow{5}{*}{SM} & BPIC12 & 1.29 & t/out & 17.87 & 1.29 & 1.31 & 1.31 & 16.583 & 0.021 & 0.020 \\ 
 & BPIC18 & t/out & t/out & t/out & 7.60 & 7.60 & 7.60 & t/out & 0.000 & 0.001 \\ 
 & PRT4 & 1.91 & 1.91 & 3.59 & 1.91 & 1.91 & 1.91 & 1.680 & 0.000 & 0.001 \\ 
 & PRT7 & 1.40 & 1.40 & 2.22 & 1.40 & 1.40 & 1.40 & 0.826 & 0.003 & 0.003 \\ 
 & PRT9 & 0.35 & 0.35 & 4.44 & 0.35 & 0.35 & 0.35 & 4.094 & 0.000 & 0.001 \\ \hline
\end{tabular}                      
}                      
\caption{Cost and over-approximation}\label{tb:eval_cost}                      
\vspace{-.5\baselineskip}                      
}}                      
\end{table}                      

%% file: tex/TableEvalMemory.tex
\begin{table}[!h]          
{\footnotesize{          
\setlength{\tabcolsep}{3pt}          
\centering{          
\begin{tabular}{|c|c c|c c|}          
\cline{2-5}          
\multicolumn{1}{c|}{} & \multicolumn{4}{c|}{\bf{Memory (MB)}}       \\ \hline
\bf{Technique} & \multicolumn{2}{c|}{\bf{LP}}   & \multicolumn{2}{c|}{\bf{TR-Scomp}}   \\ 
\bf{Miner} & \bf{IM} & \bf{SM} & \bf{IM} & \bf{SM} \\ \hline
BPIC12 & 26 & 18 & 215 & 74 \\ 
BPIC13\textsubscript{cp} & 1 & 1 & 1 & 1 \\ 
BPIC13\textsubscript{inc} & 8 & 8 & 4 & 10 \\ 
BPIC14\textsubscript{f} & 30 & 31 & 44 & 100 \\ 
BPIC15\textsubscript{1f} & 0 & 0 & 2 & 2 \\ 
BPIC15\textsubscript{2f} & 1 & 1 & 9 & 4 \\ 
BPIC15\textsubscript{3f} & 2 & 2 & 46 & 5 \\ 
BPIC15\textsubscript{4f} & 1 & 1 & 7 & 3 \\ 
BPIC15\textsubscript{5f} & 1 & 1 & 3 & 3 \\ 
BPIC17\textsubscript{f} & 28 & 30 & 69 & 69 \\ 
RTFMP & 29 & 29 & 38 & 15 \\ 
SEPSIS & 2 & 2 & 12 & 5 \\ 
BPIC18 & 0 & 104 & 0 & 495 \\ 
BPIC19\textsubscript{1} & 0 & 0 & 0 & 0 \\ 
BPIC19\textsubscript{2} & 17 & 15 & 288 & 93 \\ 
BPIC19\textsubscript{3} & 59 & 59 & 555 & 53 \\ 
BPIC19\textsubscript{4} & 2 & 2 & 3 & 2 \\ \hline
PRT1 & 4 & 4 & 3 & 6 \\ 
PRT2 & 0 & 3 & 0 & 21 \\ 
PRT3 & 0 & 0 & 1 & 2 \\ 
PRT4 & 12 & 12 & 16 & 58 \\ 
PRT6 & 0 & 0 & 1 & 1 \\ 
PRT7 & 0 & 0 & 0 & 0 \\ 
PRT9 & 139 & 139 & 84 & 85 \\ 
PRT10 & 9 & 9 & 7 & 4 \\ \hline\hline
\bf{AVG:} & \multicolumn{2}{c|}{17}   & \multicolumn{2}{c|}{50}   \\ 
\bf{MAX:} & \multicolumn{2}{c|}{139}   & \multicolumn{2}{c|}{555}   \\ \hline
\end{tabular}          
}          
\vspace{.5\baselineskip}          
\caption{Memory requirements in megabytes (MB)}\label{tb:eval_memory}          
\vspace{.5\baselineskip}          
}}          
\end{table}                             

%% file: tex/conclusion.tex
\section{Conclusion}\label{sec:conclusion}
This article contributes a technique for the efficient computation of alignments with approximate cost in the field of conformance checking. The technique revolves around two key optimizations. First, we show how to use a specific type of repeat pattern, namely tandem repeats, to reduce an event log by collapsing repetitive behavior. We compute alignments on this reduced log and later extend these alignments to work on the original event log. We  prove that these two operations (reduction and later expansion) lead to proper alignments in the original log, in the context of concurrency-free process models. While this first optimization helps us reduce the computation time of each alignment in the presence of repeated behavior in the log, as a second optimization, we use  a binary search to identify reduced traces that map to the same unique trace in the original log. This allows us to reduce the overall number of alignments to be computed, hence further improving the computation time.
Finally, we prove that the worst case over-approximation for an alignment computed by the proposed technique is less than the cost of two repetitions of each reduced tandem repeat. However, the evaluation shows that cost over-approximations rarely occurred and were much lower than its worst case cost.


The proposed technique is applicable to concurrency-free process models. However, we show how the technique can be integrated in a decomposition framework to also work in the context of models that exhibit concurrency. In this article, we propose the use of the S-Component decomposition \cite{s-comps} since it naturally outputs concurrency free process models. However, the technique is not specifically tight to this particular decomposition approach, i.e.\ other decomposition approaches may be used so long as they produce concurrency-free process models. 

Our technique builds on top of our previous approach for Automata-based alignments computation \cite{ReissnerCDRA17,s-comps}. However, the optimizations proposed by this technique are independent from the selected approach and could also be used in conjunction with other trace alignment approaches, e.g.\ those approaches that align one trace at a time, such as \cite{ILP-Alignment} and \cite{BVD-Alignment}. Adapting the technique to work on top of these approaches is an avenue for future work.

In an extensive evaluation using 50 real-life model-log pairs, we showed that our technique used on top of the Automata-based approach systematically outperforms five baseline approaches, in the presence of significant repetitive behavior in the log. 
Our technique exploits the characteristics of sound free-choice and uniquely labeled Petri nets to speed up the computation of alignments; whereas the techniques used as baseline have longer execution times but are applicable to more general classes of Petri nets. While more restrictive, the class of nets considered in this paper are extensively used. In particular, BPMN models can be translated to this class of nets and can be discovered by various mining algorithms -- such as Split Miner \cite{SplitMiner}, Inductive Miner \cite{InductiveMiner} and Fodina \cite{Fodina}.
We also showed that the over-approximation induced by our technique, when it occurs, is negligible.
To benefit from the reduction only when this is really needed, and avoid applying it when not needed, we presented a hybrid approach that selects which extension of the Automata-based conformance checking approach should be applied, based on characteristics of the input log and model. We derived these criteria empirically based on the evaluation results. More research could be conducted on finding finer-grained rules for applying our technique based on more specific model-log characteristics. 
Given the increasing number of optimizations for computing alignments, an avenue for future work is to extend the hybrid approach presented in this article, to exploit a wider range of such optimizations, based on the characteristics of the input model and log.

This article tackled the problem of identifying unfitting log behavior. 
In the course of this research, we found that the traditional definition of alignments for traces with repeated trace labels either try to avoid loop constructs in order to increase the amount of matches, or overly force aligning the repeated labels and hence mismatching surrounding trace labels. Both problems are not intuitive for users that expect repeated trace labels to be matched with a single loop structure in the process model.
Therefore, another interesting idea for future work is to find an alternative definition of alignments that considers patterns such as repetitions and concurrency in a trace and assigns them a specific weight, such that they can be aligned independently and relate better to the model structure -- e.g., considering loops or parallel blocks.
The ideas investigated in this article could also be applied to the problem of identifying how well model structures generalize patterns of an event log \cite{GarciaL17}. This latter problem is related to that of measuring the precision of a process model relative to an event log, which is an open problem in the field of process mining~\cite{PrecisionMeasures}. Another avenue for future work is to investigate the application of tandem repeats to the problem of identifying and measuring additional process model behavior that generalizes the behavior in the log.

\medskip
\noindent \emph{Acknowledgements.} This research is partly funded by the Australian Research Council (grant DP180102839).
\vspace{-1\baselineskip}

%% file: tex/Appendix.tex
\newpage
\section{The cost function of reduced alignments over-approximates the standard cost function}\label{app:proof_costFrelation}
\input{tex/ProofCostFunctionOverApproximates.tex}

\vspace{-2\baselineskip}
\section{Extended Alignments are Proper Alignments}\label{app:proof}
\input{tex/ProofExtendedAlignmentsAreProperAlignments.tex}

\newpage
\section{Cases of extended alignments with guaranteed minimal cost}\label{app:proof_exact_cost}
\input{tex/ProofExactExtendedAlignments.tex}

\section{Worst case cost over-approximation of an extended alignment}\label{app:proof_cost_overApprox}
\input{tex/ProofExtendedAlignmentCostOverApprox.tex}

\newpage
\section{Complete cost comparison and order of approximation}\label{app:cost_comparison}
\input{tex/tableCostFull.tex}

\newpage
\section{Distribution of preprocessing and processing times}\label{app:preprocessing}
\input{tex/TablePreprocessing.tex}

%% file: tex/ProofCostFunctionOverApproximates.tex
\begin{replemma}{lem:costFrelation}
Let $\dfa$ be a DAFSA, $\reachGraph$ be a reachability graph, $t$ be a trace and $t' = (\redTrace,\additionalCostF,\redReps,\complemnt)$ be a trace reduction of $t$. The cost of alignment $\alignment$ of $t$ is always lower or equal than the cost of the reduced alignment $\alignment_r$ of $t'$, i.e. $\costF(\alignment)\leq\redCostF(\alignment_r,\additionalCostF,\complemnt)$, if $\alignment$ is proper for trace $t$ and optimal, i.e. $\nexists \alignment':\costF(\alignment')<\costF(\alignment)$, and if $\alignment_r$ is proper for the reduced trace $\redTrace$ and optimal, i.e. $\nexists\alignment_r':\redCostF(\alignment_r',\additionalCostF,\complemnt)<\redCostF(\alignment_r,\additionalCostF,\complemnt)$.
\end{replemma}

\begin{proof}
We aim to prove Lemma~\ref{lem:costFrelation} by contradiction. Let $\alignment$ be an optimal alignment for a trace $t$, and $\alignment_r$ be the reduced alignment for a reduced trace $t'= (\redTrace,\additionalCostF,\redReps,\complemnt)$ of $t$.
Let us assume the cost of $\alignment_r$, according to the reduced cost function from Def.~\ref{def:adjustedCostF}, for a reduced trace $\redTrace$ is lower than the cost of $\alignment$, i.e. $\redCostF(\alignment_r,\additionalCostF,\complemnt)<\costF(\alignment)$. We consider the following cases:
\begin{enumerate}[leftmargin=*,parsep=0pt]
\item\label{first} If the original trace $t$ did not contain any tandem repeats, i.e. $\TRs(t)=\emptyset$, then the reduced trace $\redTrace$ is the same as the original trace $t$, i.e. $\redTrace=t$ and $\additionalCostF=\{i\rightarrow0\mid1\leq i\leq \left|t\right|\}$. So, the reduced cost function $\redCostF$ returns the same cost for $\redTrace$ as the standard cost function $\costF$ for $t$, because the synchronization operations have the same costs in both functions ($\lhide$ and $\rhide$ have a cost of 1, and $\match$ a cost of 0). Observe that, if the cost of the reduced alignment is lower than the cost of the original trace, we would reach a contradiction about the optimality of $\alignment$, because $\redTrace$ is proper. Then, the only case is if $\redCostF(\alignment_r,\additionalCostF,\complemnt)=\costF(\alignment)$, which is also a contradiction to the assumption $\redCostF(\alignment_r,\additionalCostF,\complemnt)<\costF(\alignment)$.

\item\label{second} If the original trace contains tandem repeats, i.e. $\TRs(t)\neq\varnothing$, then each tandem repeat is reduced to only two copies in the reduced trace. Note that, according to Def.~\ref{def:reduction}, $\additionalCostF$ keeps track of the numbers of reduced labels for each position of the two remaining copies of each tandem repeat. Observe that the cost for synchronizations, which are not part of a tandem repeat, have the same cost in both functions $\redCostF$ and $\costF$: $\match$ has a cost of 0, and $\lhide$ and $\rhide$ have a cost of 1. 
The following cases investigates three sub-cases: when all the labels in the tandem repeats are matched, when all the labels in the tandem repeats are hidden and when there is a mix of matches and hides.
\begin{enumerate}[leftmargin=0pt, parsep=0pt]
\item\label{allMatch} A reduced alignment $\alignment_r$ matching all labels of the reduced tandem repeats has a cost of $0$. In this case, the only possibility where $\alignment_r$ has a lower cost than $\alignment$ is when $\alignment$ cannot match all the repetitions associated to the tandem repeat. However, since $\alignment_r$ is proper, there is a path in the reachability graph with all the labels in the reduced trace. 
Thus, as proven in~\cite{armas2016diagnosing}, given that all pairs of copies preserved in the reduced trace were matched, they can be repeated any number of times as part of a looping behavior. This is a contradiction since there exist an $\alignment$ where all reduced tandem repeat labels can be matched and the cost would be 0.
\item\label{allHide} Consider the case when $\alignment_r$ hides all the labels related to tandem repeats. Then, the reduced cost function $\redCostF$ adds the amount of reduced labels with $\additionalCostF$ for each label of a tandem repeat. As a result, the reduced cost function $\redCostF$ will assign $\left|t\right|$ cost if all labels of $\redTrace$ were hidden, i.e. $\left|\redTrace\right| + (\sum_{i\in\dom(\additionalCostF)\mid\additionalCostF(i)\geq1}\additionalCostF(i))/2 = \left|t\right|$. Thus, this is the worst case scenario and $\alignment$ can only have equal or lower value than $\alignment_r$, which is a contradiction to the premise.
\item\label{comlementCase} The challenging case is when a tandem repeat has a mix of $\match$, $\lhide$ and $\rhide$ operations. In this case we consider the possible combinations of such operations between the two copies preserved in the reduced trace. Note that every label of a tandem repeat is linked between the two copies in $\redTrace$ via function $\complemnt$. 
We build the proof for arbitrary $\match$, $\lhide$ and $\rhide$ synchronizations to align tandem repeats by first investigating two simpler sub-cases. First, we consider the case when exactly one iteration of a tandem repeat can be matched and the other iteration needs to be hidden. Second, we investigate the case when each tandem repeat is aligned only with arbitrary $\match$ and $\lhide$ synchronizations before we finally prove the general case of arbitrary $\match$, $\lhide$ and $\rhide$ synchronizations.
\newpage
\begin{enumerate}[leftmargin=0pt]
	\item\label{comlementCase:1} Exactly one iteration of a tandem repeat can be matched. This case occurs when a reduced alignment cannot find a loop for a tandem repeat. 
	Thus, every label in one copy of the tandem repeat is matched, while their complementary labels in the other copy are hidden.
	Given the length of the tandem repeat $l$ and its number of iterations $i$, the reduced cost function $\redCostF$ adds a cost of $l*(i-1)$ to $\alignment_r$ for the two copies of the tandem repeat, because exactly one iteration of the tandem repeat can be matched assigning a cost of $0$ to $l$ synchronizations and a cost of $1+\additionalCostF(j)$ to $l$ amount of synchronizations, where $j$ are the corresponding trace positions of the tandem repeat. Further, $l*(1+\additionalCostF(j))=l*(i-1)$ since function $\additionalCostF$ assigns a cost of $i-2$ to any synchronization with a trace position in a tandem repeat.
	$\alignment_r$ can only have a lower cost than $\alignment$, if $\alignment$ contains more than $l*(i-1)$ $\lhide$ synchronizations for all iterations of the tandem repeat reduced in $\alignment_r$. 
However, an optimal alignment $\alignment$ can always achieve the same cost as the reduced alignment $\alignment_r$ by matching all labels of one iteration of each tandem repeat and hiding the labels of all other iterations. 
This is possible, because $\alignment_r$ is proper and hence the iteration of the tandem repeat that is matched in $\alignment_r$ forms a path in the reachability graph that can also be used to match one iteration of the tandem repeat in $\alignment$. Hiding all other iterations of the tandem repeat of the original trace in $\alignment$ is trivial and leads to a total cost of $l*(i-1)$. That is exactly the same cost as returned by $\redCostF$ for $\alignment_r$.
Here we arrive at a contradiction and there cannot exist a reduced alignment $\alignment_r$ with a lower cost than $\alignment$.

	\item\label{arbitraryLhides} Arbitrary $\match$ and $\lhide$ synchronizations. 
	This case covers a wide variety of cases, where some complementary pairs of labels of a tandem repeat are both aligned with $\match$ synchronizations, some with both $\lhide$ synchronizations and others with one $\lhide$ and one $\match$ synchronization.
	As shown in the cases \ref{allHide} and \ref{comlementCase:1}, the reduced cost function $\redCostF$ will add the reduced labels via function $\additionalCostF$ if one or both labels of a complementary pair are aligned with an $\lhide$ synchronization.
	Let $l$ and $i$ be the length and the number of iterations of a tandem repeat, respectively. Further, let $x_{m}$, $x_{l}$ and $x_{lm}$ be the numbers of complementary pairs of labels that were aligned with both $\match$, both $\lhide$ and one $\match$ and one $\lhide$ synchronization in $\alignment_r$, respectively, such that $x_m + x_l + x_{lm}=l$. The cost for the tandem repeat returned by function $\redCostF$ then is $x_l * i + x_{lm} * (i-1)$. $\alignment$ can aim to achieve the same cost by applying the following strategy: include a $\lhide$ synchronization for all iterations $i$ for all complementary pairs of $x_l$, include $\match$ synchronizations for $i$ iterations for $x_m$ and include one $\match$ and $i-1$ $\lhide$ synchronizations for all $x_{lm}$ complementary pairs. As a result, function $\costF$ will assign a cost of $x_l * i + x_{lm} * (i-1)$, which is the same cost as $\alignment_r$. 
By the assumptions, given that $\alignment_r$ is proper, then this strategy also forms a proper alignment $\alignment$ following the previous construction. 
Specifically, the introduced $\lhide$ synchronizations just need to be inserted in the order of the tandem repeat and the introduced $\match$ synchronizations, however, need to also form a path in the reachability graph.
	Since the reduced alignment $\alignment_r$ is proper, it follows that all its synchronizations must form a path. Further, if a complementary pair of labels is matched in both copies of the reduced alignment, both $\match$ synchronizations relate to the same transition in the workflow net since it is uniquely labelled and thus all $\match$ synchronizations in between the two complementary labels form a loop. In addition, the input workflow net is sound and free-choice and hence this loop can be executed any number of times. It follows that an optimal alignment $\alignment$ can always find $\match$ synchronizations for the reduced labels in the missing iterations of a tandem repeat by repeating the loopable sequence discovered in $\alignment_r$ for the missing iterations. Following this strategy, $\alignment$ can always achieve the same cost and sometimes a lower cost than $\alignment_r$, which is a contradiction to the assumptions.
	
	
	\item\label{arbitraryLhidesAndRhides} Arbitrary $\match$, $\lhide$ and $\rhide$ synchronizations. Last, we investigate the cases where a reduced proper alignment $\alignment_r$ aligns the reduced tandem repeats with arbitrary $\match$, $\lhide$ and $\rhide$ synchronizations outside the previous cases. An optimal alignment $\alignment$ has a higher cost than $\alignment_r$ either if some added $\rhide$ synchronization in $\alignment_r$ does not allow for a repeatable sequence as in case~\ref{comlementCase:1} or if $\alignment$ needs to use a larger amount of $\rhide$ synchronizations in the reduced iterations of each tandem repeat than the reduced cost function $\redCostF$ assigns to $\alignment_r$. As shown in case~\ref{arbitraryLhides}, for any complementary pair of labels of a tandem repeat that has been aligned with $\match$ synchronizations the sequence in between the complementary labels is repeatable. Adding $\rhide$ synchronizations into the repeatable sequence will again form a repeatable sequence since the reduced alignment $\alignment_r$ has to be proper, i.e. all arcs of the reachability graph need to form a path from the initial marking to the final marking. Hence, an optimal alignment can always repeat the sequence to find all matched complementary label pairs. 
	
	An optimal alignment $\alignment$ can only have a higher cost than $\alignment_r$, if it has a higher amount of $\rhide$ synchronizations than the reduced iterations of each tandem. 
	Let $i$ be the iterations of a tandem repeat. Further, let $y_1$ and $y_2$ be the number of $\rhide$ synchronizations in the first and second copy of the tandem repeat of $\alignment_r$, respectively. The reduced cost function $\redCostF$ then returns a cost of $(y_1+y_2)*(i-1)$ for all $\rhide$ synchronizations of the tandem repeat. However, if $\alignment$ repeats the repeatable sequence found in $\alignment_r$, it will add $\rhide$ synchronizations to the additional iterations that are hidden in $\redTrace$ and will achieve a cost of $(y_1+y_2)*(i-1)$  in the worst case. Thus, a smaller cost can not be obtained and it is a contradiction.
\end{enumerate}


\end{enumerate}
It follows that the initial assumption was false and there exists no proper and reduced alignment with a lower cost for $\redCostF$ than an optimal alignment $\alignment$ of the original trace $t$ according to cost function $\costF$.
\end{enumerate}
\end{proof}

%% file: tex/ProofExtendedAlignmentsAreProperAlignments.tex
\begin{replemma}{lemma:properAlignment}[An extended alignment is a proper alignment]
Given a trace $t\in\logL$, the alignment $\extendedAlignment$, returned by Alg.~\ref{alg:ExtendRedAlignments}, is a proper alignment. 
Thus, the following two properties hold for $\extendedAlignment$:
\begin{compactenum}
    \item\label{prop:one} the labels in the synchronizations related to the DAFSA represent the trace, i.e. $\lbl(\filterSyncDFA(\extendedAlignment))=t$, and
      \item\label{prop:two} the arcs in the synchronizations related to the reachability graph forms a path through the reachability graph, i.e., for $\mPath=\fnSyncRG(\filterSyncRG(\extendedAlignment))$ holds $\src(\mPath(1)=\rgSource \land \tgt(\mPath(\left|\mPath\right|))\in\rgFinStates \land \forall \position\in\seq{1}{\left|\mPath\right|-1} : \tgt(\mPath(\position))=\src(\mPath(\position+1))$.
\end{compactenum}
\end{replemma}


For the proof of lemma~\ref{lemma:properAlignment}, the extended alignment $\extendedAlignment$ are related to the reduced alignment $\alignment$ linked with trace $t$ via its reduced trace, i.e. $\alignment = \alignments(\reductions(t)) =(\redTrace,\additionalCostF,\redReps,\trPositions)$. 
It has been shown in~\cite{s-comps} that the reduced alignment $\alignment$ represents the trace labels of the reduced trace $\redTrace$ as well as forms a path through the reachability graph.

The difference between $t$ and $\redTrace$ is recorded in functions $\complemnt$ and $\additionalCostF$ in Alg.~\ref{alg:reduceLog} as every tandem repeat starting at position $\position\in\dom(\complemnt)$ is reduced to only two copies present in $\redTrace$. These two copies are aligned in $\alignment$ with two subsequences of the alignment $\firstCopy$and $\secondCopy$ acc. to Def~\ref{def:firstCopy}.
The additional cost function $\additionalCostF$ then links each tandem repeat to the amount of reduced repetitions. 

In Alg.~\ref{alg:ExtendRedAlignments}, we insert a sequence of synchronizations $\copyExtend$ after the first aligned copy of each tandem repeat, i.e. after $\firstCopy$, for $k$-amount of times. 
Thus for proving the two properties of lemma~\ref{lemma:properAlignment}, we only need to investigate, if sequence $\copyExtend$ adds the same trace labels as the tandem repeat type, i.e. if $\lbl(\filterSyncDFA(\copyExtend))=\redTrace(\seq{\position}{\complemnt(\position)\!-\!1})$, and if the added arcs of the reachability graph form a valid path.

The construction of $\copyExtend$ depends on whether the aligned two copies of the tandem repeat in $\alignment$ contains a trace label of both copies that is aligned with a $\match$ synchronization ($\secondPos>0$ in Def.~\ref{def:extendedAlignment}). 
If such a subsequence is found, we create the middle copy as all synchronizations from the start of the second copy to the position of the matched label, adding all synchronizations after the position of the matched label in the first copy up to the end of the first copy.
If no label of the tandem repeat can be matched in both copies, $\copyExtend$ consists of all $\lhide$-synchronizations for each trace label in the sequence of the first tandem repeat.

\begin{proof}
We can now prove the two properties of Lemma~\ref{lemma:properAlignment} below:
	\begin{enumerate} 
		\item 
		Property~\ref{prop:one} holds by construction.
		The extended alignment $\extendedAlignment$ adds all synchronizations from the reduced alignment $\alignment$ as shown in line~\ref{line:initExtension} thus it holds that $\lbl(\fnSyncDFA(\extendedAlignment))=\redTrace$. In addition, Alg.~\ref{alg:ExtendRedAlignments} applies the extension defined in Def.~\ref{def:extendedAlignment}
		by adding $\additionalCostF(\position)$ times a middle copy for each last position $\position$ of a tandem repeat.
		This reconstructs the original trace $t$ iff the trace labels of the middle copy equals the labels of the tandem repeat.
		If no repeating subsequence is found, this is trivially true since $\copyExtend$ is constructed of all $\lhide$-synchronizations for the first aligned copy $\firstCopy$ of the tandem repeat, which corresponds to the trace labels of the tandem repeat.
		In case a repeatable subsequence exists beginning at $\secondPos$ as defined in Def.~\ref{def:extendedAlignment}, $\copyExtend$ consists of the prefix of the second copy $\secondCopy$ up to index $\aligncomplement(\alignment,\secondPos)$ and then adds the suffix of the first copy after $\secondPos$. It holds that $\lbl(\filterSyncDFA(\firstCopy))=\lbl(\filterSyncDFA(\secondCopy))=\redTrace(\seq{\complemnt(\position)\!-\!1}{\position})$ since both copies of the tandem repeat contain the same labels. 
		The constructed middle copy from Def.~\ref{def:extendedAlignment} then contains all trace labels of the tandem repeat since it adds the prefix of the trace labels of the second copy and the suffix of the trace labels of the first copy of the same position of the repeat type of the tandem repeat, i.e. $\aligncomplement(\alignment,\secondPos)$ and $\secondPos$ respectively.
		\item
		We will prove that property~\ref{prop:two} holds, i.e. that the insertion of a middle copy is still a path in the reachability graph. The general case will follow by induction. First, note that a reduced alignment forms a path in the reachability graph. 
		
		Alg.~\ref{alg:ExtendRedAlignments} applies the extension defined in Def.~\ref{def:extendedAlignment}
		by adding $\additionalCostF(\position)$ times a middle copy for each last position $\position$ of a tandem repeat.
		We now need to show that after adding this middle copy $\copyExtend$ that $\fnSyncRG(\filterSyncRG(\extendedAlignment))$ again forms a path through the reachability graph.
		
                      In the case that no repeating subsequence is found, this is trivially true since $\copyExtend$ is constructed of all $\lhide$-synchronizations for the first aligned copy of the tandem repeat. It follows that no new arcs in the reachability graph are added to $\extendedAlignment$ and the path of $\alignment$ is not changed.		
		
		In the case of a repeatable subsequence at a position $\secondPos$, $\copyExtend$ consists of the prefix of the second copy up to index $\aligncomplement(\alignment,\secondPos)$ and then adds the suffix of the first copy after $\secondPos$.
		Take the last element in the synchronizations of the first copy of the tandem repeat, the first element in the middle copy (as defined in Def.~\ref{def:extendedAlignment}) is the same as the initial element of the second copy, thus its insertion is also a path in the reachability graph for the first element. All elements until and including $\aligncomplement(\alignment,\secondPos)$ will also form a path since they are equal to the synchronizations of the second copy. The next element being inserted is in the first copy at position $\secondPos+1$. From Def.~\ref{def:extendedAlignment}, we know that both $\match $synchronizations for the same label.
		Since in this work we only consider uniquely-labelled workflow nets, it follows that both $\match$-synchronizations at $\alignment[\secondPos]$ and $\alignment[\aligncomplement(\alignment,\secondPos)]$ relate to the same transition and thus the arcs of the reachability graph in between the two matches form a loop. Hence, inserting the element of the first copy at position $\secondPos\!+\!1$ will form a path since it has the same execution state of the process model after the synchronization at position $\secondPos$. All remaining elements of the middle copy after $\secondPos\!+\!1$ from a path since they are equal to the suffix of $\firstCopy$. In the case that only on copy is inserted the proof is complete since the last element of the middle copy is equal to the last element in the first copy which then forms a path to the first element of the second copy. In case multiple middle copies are inserted, the last element of the middle copy also needs to form a path to its first element. This holds, because the first element of the middle copy is the same as the first element in the second copy and the last element of the middle copy is the last element of the first copy, which also form a path in the reduced alignment. It follows that inserting any number of repetitions of the middle copy as per Def.~\ref{def:extendedAlignment} will again form a path through the reachability graph.
		This holds especially for Petri nets that are parallelism free.
	\end{enumerate}
\end{proof}

%% file: tex/ProofExactExtendedAlignments.tex
\begin{replemma}{lemma:casesExactCost}
Let $t$ be a trace and $\reachGraph$ a reachability graph. The cost of the optimal alignment $\alignment$ and the extended alignment $\reachGraph$ have the same cost: $\costF(\extalignment)=\costF(\alignment)$ in three cases:
\begin{compactenum}
	\item $t$ contains \emph{no tandem repeats},
	\item $t$ contains only tandem repeats $(i,\alpha,k)$ corresponding to one of the following cases:\label{case:exactCostWithTandemRepeats}
	\begin{compactenum}
	    \item no possible matches could be found for the tandem repeat, or
	    \item no corresponding loop in the model is found for the tandem repeat, or
	    \item the tandem repeat could be fully matched to a corresponding loop in the model.
	\end{compactenum}
\end{compactenum}
\end{replemma}
\vspace{-.5\baselineskip}

\begin{proof}[Proof for lemma~\ref{lemma:casesExactCost} (sketch)]\label{proof:CasesExactCost}
We aim to prove lemma~\ref{lemma:casesExactCost} with a direct proof.
\begin{enumerate}[topsep=2pt, partopsep=2pt,itemsep=2pt,parsep=2pt]
\item First, let $t$ contain \emph{no tandem repeats}. In that case, after applying Alg.~\ref{alg:reduceLog}, the reduced trace $rt$ will be the same as trace $t$. The algorithms for computing alignments for reduced traces (Alg.~\ref{alg:BinarySearchAlignments}) and to compute optimal alignments (Alg.~\ref{alg:fsmAlignments}) differ only in their cost function. The reduced cost function (Def.~\ref{def:adjustedCostF}) assigns different costs to alignment synchronizations corresponding reduced tandem repeats. Hence, without tandem repeats the cost functions are  the same. Hence, it follows that an extended alignment for $t$ without tandem repeats will achieve the same cost as an optimal alignment, i.e. $\costF(\extalignment)=\costF(\alignment)$.

\item For proving case~\ref{case:exactCostWithTandemRepeats}, we first only consider $t$ containing a single tandem repeat $(i,\alpha,k)$ and then generalize the cases for multiple tandem repeats.
The tandem repeat will be reduced to two copies (Alg.~\ref{alg:reduceLog}) in a reduced trace $rt$. The main difference for computing the alignments for $rt$ lies in the selection of synchronizations with the use of the adjusted cost function from Def.~\ref{def:adjustedCostF}. 
The selection of synchronizations for a tandem repeat does not change the selection of synchronizations outside the tandem repeat since the accumulated cost of aligning a tandem repeat according to the adjusted cost function is the same as aligning the tandem repeat in $t$ with Alg.~\ref{alg:fsmAlignments}, which has been shown in the proof in \ref{app:proof_costFrelation}. 
Hence, we only need to investigate whether aligning the tandem repeat and its extension will lead to an over-approximation of the alignment cost of $\extalignment$.

\begin{enumerate}[topsep=2pt, partopsep=2pt,itemsep=2pt,parsep=2pt]
	\item First, we consider a \emph{tandem repeat with no possible matches}, i.e. the reachability graph $\reachGraph$ does not contain any arcs with labels of $\alpha$ such that $l\in\alpha\Rightarrow(n_s,l,n_t)\notin\rgArcs$. In that case, an optimal alignment $\alignment$ will hide all trace labels related to the tandem repeat. A reduced alignment $\redalignment$ will hide all trace labels of the two copies of the reduced tandem. The extension from Alg.~\ref{alg:ExtendRedAlignments} will identify no label that has been aligned with a $\match$ operation in both copies and hence the middle copy will consist of all hide operations. The middle copy will be used to recreate the $k$ repetitions and as a result $\extalignment$ will have the same cost as $\alignment$ with no over-approximation, i.e. $\costF(\extalignment)=\costF(\alignment)$.
	
	\item Second, we consider a \emph{tandem repeat with no corresponding loop in the model}. It follows that every label of the repeating sequence $\alpha$ can be matched at most once since a label matching in two repetitions would establish a loop in $\reachGraph$.  An optimal alignment $\alignment$ could then match each of the labels of the repeating sequence once in arbitrary repetitions. The reduced alignment $\redalignment$ can also match  each label of the repeating sequence once in one of the two maintained repetitions. The middle copy would again consist of all hide operations and hence the extended alignment $\extalignment$ can also achieve the same cost as $\alignment$ without over-approximation, i.e. $\costF(\extalignment)=\costF(\alignment)$.
	
	\item Next, we consider a \emph{tandem repeat with all matches}, i.e. its labels can be perfectly matched to a loop in the reachability graph. In that case, the optimal alignment $\alignment$ will align all trace labels corresponding to the tandem repeat with $\match$ operations. The reduced alignment $\redalignment$ can similarly match both repetitions of the reduced tandem repeat since there exists a loop in the reachability graph with the labels of the tandem repeat. The extension algorithm~\ref{alg:ExtendRedAlignments} will identify a repeating sequence at the first label of the repeating sequence with a $\match$ operation in both repetitions. The middle copy inserted to recreate $k$ repetitions then consists of all $\match$ synchronizations since both repetitions of the tandem repeat could be matched in $\redalignment$. Hence, $\extalignment$ will consist of all $\match$ synchronizations for the tandem repeat and has the same cost as $\alignment$ with no over-approximation, i.e. $\costF(\extalignment)=\costF(\alignment)$.
\end{enumerate}

Last, we consider the case where $t$ contains \emph{several tandem repeats}. As discussed before, the selection of synchronizations for a tandem repeat does not change the selection of synchronizations outside the tandem repeat. Furthermore, the reduced tandem repeats can not overlap (as per Alg.~\ref{alg:reduceLog}) since, after reducing a tandem repeat, the next tandem repeat is selected at a trace position strictly after the last position of the previously reduced tandem repeat. It follows then that the tandem repeats will be aligned independently from each other. 
Hence, if $t$ contains multiple tandem repeats that only correspond to any case of item~\ref{case:exactCostWithTandemRepeats}, then the cost of $\extalignment$ will be minimal because all tandem repeats can also be aligned with minimal cost.
\end{enumerate}

\end{proof}

%% file: tex/ProofExtendedAlignmentCostOverApprox.tex

\begin{figure}[h]
\centering
\includegraphics[width=0.9\textwidth]{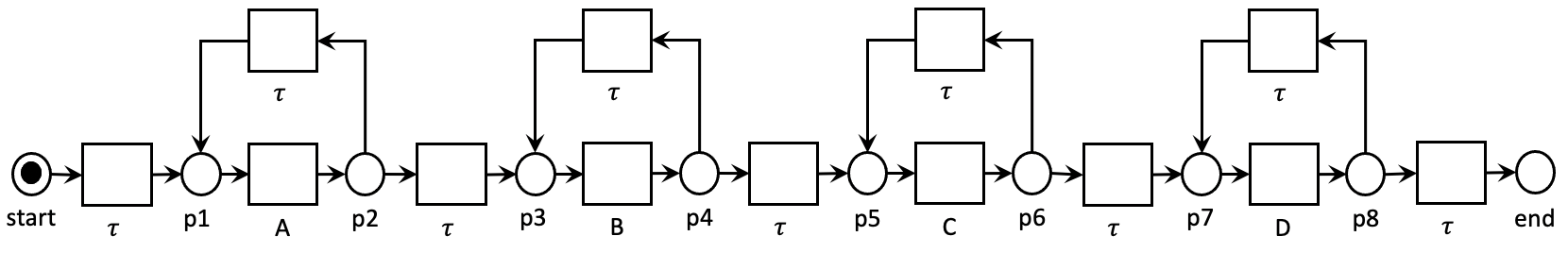}
\vspace{-.5\baselineskip}
\caption{
System net for the worst case cost over-approximation of an extended alignment.
} \label{fig:WorstCaseModel}
\end{figure}

Fig.~\ref{fig:WorstCaseModel} demonstrates the worst case cost over-approximation of an extended alignment with a system net that includes four activities $A,B,C$ and $D$. These activitiesneed to be executed in order and every activity can be executed at least once. The worst case occurs for the specific trace $\langle(1,BCDA,7)\rangle$. The trace is reduced with Alg.~\ref{alg:reduceLog} to $\langle B,C,D,A,B,C,D,A\rangle$, where the penalty cost $\additionalCostF$ for every position is 6. The reduced alignment computed by Alg.~\ref{alg:BinarySearchAlignments} is  $\langle\lhide(B),\lhide(C),\lhide(D),\match(A),\lhide(B),\\ \lhide(C),\lhide(D),\match(A),\rhide(B),\rhide(C),\rhide(D)\rangle$. 
The extension algorithm~\ref{alg:ExtendRedAlignments} will identify the middle copy $\lhide(B),\lhide(C),\lhide(D),\match(A)$ and extend the reduced alignment to $\langle(1,\lhide(B),\lhide(C),\lhide(D),\match(A),7),\\ \rhide(B),\rhide(C),\rhide(D\rangle$. The extended alignment hence will have a total cost of 24 according to the standard cost function (Def.~\ref{def:CostFunction}). However, Alg.~\ref{alg:fsmAlignments} can find the following alignment with minimal cost $\langle\lhide(B),\lhide(C),\lhide(D),\match(A),\match(B),\match(C),\match(D),\lhide(A),(9,\lhide(B),\lhide(C),\match(D),\lhide(A),5)\rangle$. The optimal alignment can achieve a cost of 19. The alignment with minimal cost can achieve a lower cost than the extended alignment by not matching one label in every repetition while hiding it once to match every other label of the repeating sequence one more time than the extended alignment. 
For every match synchronization, it can save one $\lhide$ and one $\rhide$ synchronization.  Hence, the worst case cost over approximation of the extended alignment is two times the length of the repeating sequence of a tandem repeat minus one label, since the extended alignment will match this label in every repetition, and minus one since the minimal alignment needs to hide the matched label of the extended alignment once.
The alignment with minimal cost can never find more than one additional match for a label of the repeating sequence since two match synchronizations for the same activity induce the existence of a loop in the model for this activity that could have been matched in every repetition of the tandem repeat instead. If the example trace would be extended to 100 repetitions, i.e. $\langle(1,BCDA,100)\rangle$, the worst case cost over-approximation is still 5 since it is only related to the length of the repeating sequence. 
Next, we prove the worst-case cost formally.

\begin{proof}[Proof of Lemma~\ref{lemma:ExtAlignmentCostOverApprox} (Sketch)]

\begin{replemma}{lemma:ExtAlignmentCostOverApprox}
Let $t$ be a trace and $\reachGraph$ a reachability graph. The worst cost over-approximation of an extended alignment $\extalignment$ of $t$ is $costF(\alignment)+\sum_{(i,\alpha,k)\in\Delta(rt)} 2*(\left|\alpha\right|-1)-1$, where $\alignment$ is the optimal alignment of $t$.
\end{replemma}

We aim to prove Lemma~\ref{lemma:ExtAlignmentCostOverApprox} by contradiction.
Let $\alignment$ be an alignment with minimal cost computed with Alg.~\ref{alg:fsmAlignments} for trace $t$ and reachability graph $\reachGraph$. We now assume there exists an extended alignment $\extalignment$ with a cost higher than the worst case cost of an extended alignment,
i.e. $\costF(\extalignment)>\costF(\alignment)+\sum_{(i,\alpha,k)\in\Delta(rt)} 2*(\left|\alpha\right|-1)-1$ where $rt$ is the reduced trace of $t$. 

Note the following observations: 
\begin{inparaenum}[a)]
	\item as noted in the case 2 in the proof in~\ref{proof:CasesExactCost}, different selection of synchronizations between $\extalignment$ and $\alignment$ can only occur within the trace positions of a tandem repeat,
	\item if the reduced trace $rt$ of $t$ contains multiple tandem repeats, every tandem repeat is aligned independently and hence it is enough to show that $\extalignment$ does not exist if $t$ contains a single tandem repeat, and 
	\item the tandem repeat cannot fall into any of the cases of Lemma~\ref{lemma:casesExactCost}, as they guarantee minimal cost, i.e. $\extalignment$ can not exist in these cases.
\end{inparaenum}

Hence, $\extalignment$ can only exist if a tandem repeat $(i,\alpha,k)\in\Delta(t)$ can only be partially aligned to a loop in $\reachGraph$. In that case, at least one label $\ell\in\alpha$ of the repeating sequence can be matched in every repetition by $\extalignment$ to the loop in $\reachGraph$. The label $\ell$ will also be matched by $\alignment$ in every repetition.
We now investigate an extended alignment $\extalignment$ with a cost greater than the worst case over-approximation for a single tandem repeat $(i,\alpha,k)$, i.e. $\costF(\extalignment)>\costF(\alignment)+2*(\left|\alpha\right|-1)-1$.
It can only exist if $\alignment$ can achieve at least one more match synchronization for every label of the repeating sequence, i.e. $\left|\alpha\right|$ more $\match$ synchronizations,  since for every $\match$ the cost can be reduced by two, i.e. one $\lhide$ and one $\rhide$ synchronization. In that case, $\extalignment$ would have a cost of 1 more than the worst case cost of Lemma~\ref{lemma:ExtAlignmentCostOverApprox}, i.e. $\costF(\extalignment)=\costF(\alignment)+2*(\left|\alpha\right|-1)$.
It cannot be $\ell$ because it is matched in $\extalignment$ in every repetition and $\alignment$ can only achieve $\left|\alpha\right|$ more matches by matching another label $\ell \neq x \in\alpha$ of the repeating sequence twice. If a label is matched twice, it induces a loop in the system net for label x as it is uniquely labelled. Then, label $x$ would also have been matched in every other repetition of the tandem repeat instead in both $\extalignment$ and $\alignment$ since matching label $x$ in every repetition would have a lower cost than hiding it.
Nevertheless, since label $x$ is now matched in all repetitions of $\extalignment$ and $\alignment$, $\alignment$ can no longer achieve $\left|\alpha\right|$ more $\match$ synchronizations than $\extalignment$, which is a contradiction. The only conclusion left is that there can not exist $\extalignment$ with a cost higher than $\costF(\alignment)+2*(\left|\alpha\right|-1)-1$ for $t$ containing a single tandem repeat.
Furthermore, since $\extalignment$ can not achieve a higher cost than $2*(\left|\alpha\right|-1)-1$ per tandem repeat, there can also not exist an extended alignment $\extalignment$ with a higher cost than $\costF(\alignment)+\sum_{(i,\alpha,k)\in\Delta(rt)} 2*(\left|\alpha\right|-1)-1$ for a trace with multiple tandem repeats, because the tandem repeats can be aligned independently and hence the cost cannot be higher than the sum of worst case cost over approximations of all reduced tandem repeats. Then thee worst case cost over approximation of an extended alignment is $\costF(\extalignment)\leq\costF(\alignment) + \sum_{(i,\alpha,k)\in\Delta(rt)} 2*(\left|\alpha\right|-1)-1$ and the proof is complete.
\end{proof}

%% file: tex/tableCostFull.tex
\begin{table}[htbp!]                      
{\footnotesize{                      
\setlength{\tabcolsep}{3pt}                      
\centering{                      
\begin{tabular}{|c|c|c c c c c c|c c c|}                      
\hline                      
 \multirow{2}{*}{\bf{Miner}} &  \multirow{2}{*}{\bf{Dataset}} & \multicolumn{6}{c|}{\bf{Cost}}           & \multicolumn{3}{c|}{\bf{Over-Approximation}}     \\ 
 \multirow{25}{*}{IM} &  & \bf{ILP} & \bf{eMEQ} & \bf{ALI} & \bf{Automata} & \bf{Scomp} & \bf{TR-Scomp} & \bf{$\Delta$ ALI} & \bf{$\Delta$ Scomp} & \bf{$\Delta$ TR-Scomp} \\ \hline
 & BPIC12 & 0.87 & t/out & 3.38 & 0.87 & 0.87 & 0.87 & 2.51 & 0.00 & 0.00 \\ 
 & BPIC13\textsubscript{cp} & 1.46 & 1.46 & 2.36 & 1.46 & 1.46 & 1.46 & 0.89 & 0.00 & 0.00 \\ 
 & BPIC13\textsubscript{inc} & 0.84 & 0.84 & 4.14 & 0.84 & 0.84 & 0.84 & 3.31 & 0.00 & 0.00 \\ 
 & BPIC14\textsubscript{f} & 1.94 & 0.00 & 3.58 & 1.94 & 1.94 & 1.94 & 1.65 & 0.00 & 0.00 \\ 
 & BPIC15\textsubscript{1f} & 0.50 & 0.50 & 15.01 & 0.50 & 0.50 & 0.50 & 14.51 & 0.00 & 0.00 \\ 
 & BPIC15\textsubscript{2f} & 2.02 & 2.02 & 26.73 & 2.02 & 2.07 & 2.07 & 24.71 & 0.05 & 0.05 \\ 
 & BPIC15\textsubscript{3f} & 1.70 & 1.70 & 25.67 & 1.70 & 1.70 & 1.70 & 23.97 & 0.00 & 0.00 \\ 
 & BPIC15\textsubscript{4f} & 1.14 & 1.14 & 22.84 & 1.14 & 1.14 & 1.14 & 21.71 & 0.00 & 0.00 \\ 
 & BPIC15\textsubscript{5f} & 1.20 & 1.20 & 24.90 & 1.20 & 1.20 & 1.20 & 23.70 & 0.00 & 0.00 \\ 
 & BPIC17\textsubscript{f} & 0.83 & 0.83 & 11.53 & 0.83 & 0.83 & 0.83 & 10.70 & 0.00 & 0.00 \\ 
 & RTFMP & 0.06 & 0.06 & 2.19 & 0.06 & 0.06 & 0.06 & 2.13 & 0.00 & 0.00 \\ 
 & SEPSIS & 0.12 & 0.12 & 12.83 & 0.12 & 0.12 & 0.12 & 12.71 & 0.00 & 0.00 \\ 
 & BPIC18 & 0.00 & t/out & t/out & 0.00 & t/out & 0.00 & t/out & t/out & 0.00 \\ 
 & BPIC19\textsubscript{1} & 0.30 & 0.30 & 3.13 & 0.30 & 0.30 & 0.30 & 2.83 & 0.00 & 0.00 \\ 
 & BPIC19\textsubscript{2} & 0.18 & t/out & t/out & t/out & 0.18 & 0.18 & t/out & 0.00 & 0.00 \\ 
 & BPIC19\textsubscript{3} & 1.00 & 1.00 & 6.56 & 1.00 & 1.00 & 1.01 & 5.56 & 0.00 & 0.01 \\ 
 & BPIC19\textsubscript{4} & 0.27 & 0.27 & 3.12 & 0.27 & 0.27 & 0.27 & 2.85 & 0.00 & 0.00 \\ 
 & PRT1 & 1.43 & 1.43 & 4.42 & 1.43 & 1.43 & 1.43 & 3.00 & 0.00 & 0.00 \\ 
 & PRT2 & 0.00 & t/out & 38.16 & 0.00 & 0.00 & 0.00 & 38.16 & 0.00 & 0.00 \\ 
 & PRT3 & 0.23 & 0.23 & 3.84 & 0.23 & 0.23 & 0.23 & 3.61 & 0.00 & 0.00 \\ 
 & PRT4 & 1.22 & 1.22 & 4.79 & 1.22 & 1.22 & 1.22 & 3.56 & 0.00 & 0.00 \\ 
 & PRT6 & 0.09 & 0.09 & 2.31 & 0.09 & 0.12 & 0.12 & 2.22 & 0.03 & 0.03 \\ 
 & PRT7 & 0.00 & 0.00 & 2.10 & 0.00 & 0.00 & 0.00 & 2.10 & 0.00 & 0.00 \\ 
 & PRT9 & 0.38 & 0.38 & 3.09 & 0.38 & 0.41 & 0.38 & 2.71 & 0.03 & 0.00 \\ 
 \multirow{25}{*}{SM} & PRT10 & 0.06 & 0.06 & 3.16 & 0.06 & 0.06 & 0.06 & 3.10 & 0.00 & 0.00 \\ \hline\hline
 & BPIC12 & 1.29 & t/out & 17.87 & 1.29 & 1.31 & 1.31 & 16.58 & 0.02 & 0.02 \\ 
 & BPIC13\textsubscript{cp} & 0.09 & 0.09 & 2.19 & 0.09 & 0.09 & 0.09 & 2.10 & 0.00 & 0.00 \\ 
 & BPIC13\textsubscript{inc} & 0.24 & 0.24 & 3.77 & 0.24 & 0.24 & 0.24 & 3.54 & 0.00 & 0.00 \\ 
 & BPIC14\textsubscript{f} & 2.91 & t/out & 5.83 & 2.91 & 2.91 & 2.91 & 2.92 & 0.00 & 0.00 \\ 
 & BPIC15\textsubscript{1f} & 3.20 & 3.20 & 19.38 & 3.20 & 3.20 & 3.20 & 16.18 & 0.00 & 0.00 \\ 
 & BPIC15\textsubscript{2f} & 10.22 & 10.22 & 27.44 & 10.22 & 10.22 & 10.22 & 17.21 & 0.00 & 0.00 \\ 
 & BPIC15\textsubscript{3f} & 9.70 & 9.70 & 13.75 & 9.70 & 9.70 & 9.70 & 4.06 & 0.00 & 0.00 \\ 
 & BPIC15\textsubscript{4f} & 10.42 & 10.42 & 22.23 & 10.42 & 10.42 & 10.42 & 11.81 & 0.00 & 0.00 \\ 
 & BPIC15\textsubscript{5f} & 8.10 & 8.10 & 17.04 & 8.10 & 8.10 & 8.10 & 8.94 & 0.00 & 0.00 \\ 
 & BPIC17\textsubscript{f} & 1.47 & 1.47 & 18.94 & 1.47 & 1.47 & 1.47 & 17.47 & 0.00 & 0.00 \\ 
 & RTFMP & 0.03 & 0.03 & 2.77 & 0.03 & 0.03 & 0.03 & 2.73 & 0.00 & 0.00 \\ 
 & SEPSIS & 4.72 & t/out & 12.23 & 4.72 & 4.72 & 4.72 & 7.51 & 0.00 & 0.00 \\ 
 & BPIC18 & t/out & t/out & t/out & 7.60 & 7.60 & 7.60 & t/out & 0.00 & 0.00 \\ 
 & BPIC19\textsubscript{1} & 0.65 & 0.65 & 3.23 & 0.65 & 0.65 & 0.65 & 2.58 & 0.00 & 0.00 \\ 
 & BPIC19\textsubscript{2} & 1.02 & t/out & t/out & 1.02 & 1.02 & 1.02 & t/out & 0.00 & 0.00 \\ 
 & BPIC19\textsubscript{3} & 0.09 & t/out & 7.64 & 0.09 & 0.09 & 0.09 & 7.55 & 0.00 & 0.00 \\ 
 & BPIC19\textsubscript{4} & 0.03 & 0.03 & 2.56 & 0.03 & 0.03 & 0.03 & 2.53 & 0.00 & 0.00 \\ 
 & PRT1 & 0.29 & 0.29 & 4.13 & 0.29 & 0.29 & 0.29 & 3.83 & 0.00 & 0.00 \\ 
 & PRT2 & 8.32 & t/out & 34.53 & 8.32 & 8.32 & 8.32 & 26.22 & 0.00 & 0.00 \\ 
 & PRT3 & 2.54 & 2.54 & 6.50 & 2.54 & 2.54 & 2.54 & 3.96 & 0.00 & 0.00 \\ 
 & PRT4 & 1.91 & 1.91 & 3.59 & 1.91 & 1.91 & 1.91 & 1.68 & 0.00 & 0.01 \\ 
 & PRT6 & 1.08 & 1.08 & 1.26 & 1.08 & 1.08 & 1.08 & 0.19 & 0.00 & 0.00 \\ 
 & PRT7 & 1.40 & 1.40 & 2.22 & 1.40 & 1.40 & 1.40 & 0.83 & 0.00 & 0.00 \\ 
 & PRT9 & 0.35 & 0.35 & 4.44 & 0.35 & 0.35 & 0.35 & 4.09 & 0.00 & 0.01 \\ 
 & PRT10 & 0.10 & 0.10 & 3.08 & 0.10 & 0.10 & 0.10 & 2.98 & 0.00 & 0.00 \\ \hline
\end{tabular}                      
}                      
\vspace{.5\baselineskip}                      
\caption{Full cost comparison}\label{ap:cost_comparison}                      
\vspace{.5\baselineskip}                      
}}                      
\end{table}                                          

%% file: tex/TablePreprocessing.tex
\begin{table}[htbp!]              
{\footnotesize{              
\setlength{\tabcolsep}{3pt}              
\centering{              
\begin{tabular}{|c|c c c|c c c|}              
\cline{2-7}              
\multicolumn{1}{c|}{} & \multicolumn{6}{c|}{\bf{TR-Scomp time performance (ms)}}           \\ \hline
\bf{Miner} & \multicolumn{3}{c|}{\bf{IM}}     & \multicolumn{3}{c|}{\bf{SM}}     \\ 
\bf{Dataset} & \bf{Preprocessing} & \bf{Total} & \bf{\%Preprocessing} & \bf{Preprocessing} & \bf{Total} & \bf{\%Preprocessing} \\ \hline
BPIC12 &  687  &  11,537  & 6\% &  362  &  4,112  & 9\% \\ 
BPIC13\textsubscript{cp} &  15  &  38  & 40\% &  16  &  35  & 46\% \\ 
BPIC13\textsubscript{inc} &  47  &  242  & 19\% &  66  &  244  & 27\% \\ 
BPIC14\textsubscript{f} &  232  &  11,185  & 2\% &  446  &  3,421  & 13\% \\ 
BPIC15\textsubscript{1f} &  35  &  153  & 23\% &  33  &  196  & 17\% \\ 
BPIC15\textsubscript{2f} &  176  &  1,521  & 12\% &  44  &  869  & 5\% \\ 
BPIC15\textsubscript{3f} &  303  &  4,118  & 7\% &  53  &  755  & 7\% \\ 
BPIC15\textsubscript{4f} &  137  &  887  & 15\% &  39  &  568  & 7\% \\ 
BPIC15\textsubscript{5f} &  42  &  427  & 10\% &  41  &  658  & 6\% \\ 
BPIC17\textsubscript{f} &  383  &  1,314  & 29\% &  380  &  1,306  & 29\% \\ 
RTFMP &  341  &  678  & 50\% &  231  &  382  & 61\% \\ 
SEPSIS &  72  &  1,590  & 5\% &  38  &  243  & 16\% \\ 
BPIC18 &  \multicolumn{3}{c|}{t/out} &  2,980  &  62,026  & 5\% \\ 
BPIC19\textsubscript{1} &  17  &  54  & 32\% &  14  &  41  & 35\% \\ 
BPIC19\textsubscript{2} &  1,173  &  61,967  & 2\% &  478  &  8,831  & 5\% \\ 
BPIC19\textsubscript{3} &  3,045  &  54,283  & 6\% &  590  &  4,518  & 13\% \\ 
BPIC19\textsubscript{4} &  60  &  105  & 57\% &  48  &  67  & 72\% \\ \hline
PRT1 &  62  &  255  & 24\% &  61  &  172  & 36\% \\ 
PRT2 &  \multicolumn{3}{c|}{t/out} &  107  &  1,048  & 10\% \\ 
PRT3 &  34  &  88  & 39\% &  27  &  88  & 30\% \\ 
PRT4 &  104  &  1,815  & 6\% &  206  &  2,432  & 8\% \\ 
PRT6 &  27  &  64  & 42\% &  20  &  59  & 35\% \\ 
PRT7 &  31  &  73  & 42\% &  27  &  69  & 39\% \\ 
PRT9 &  1,129  &  4,745  & 24\% &  793  &  1,533  & 52\% \\ 
PRT10 &  68  &  150  & 46\% &  52  &  91  & 57\% \\ \hline
\end{tabular}              
}              
\vspace{.5\baselineskip}              
\caption{Distribution of preprocessing and processing times}\label{tb:eval_preprocessing}              
\vspace{.5\baselineskip}              
}}              
\end{table}              